\pdfoutput=1
\documentclass[PAPER=true,texlive=2023,atlasdraft=false,cernpreprint,UKenglish, texmf, orcidlogo]{atlasdoc}
 
\usepackage{atlaspackage}
\usepackage{atlasbiblatex}

\usepackage{atlasphysics}
\usepackage{atlasbsm}

\graphicspath{{logos/}{figures/}}
 
\usepackage{ANA-SUSY-2020-15-PAPER-defs}

\AtlasTitle{ATLAS Run~2 searches for electroweak production of supersymmetric particles interpreted within the pMSSM}

\AtlasAbstract{

A summary of the constraints from searches performed by the ATLAS Collaboration for the electroweak production of charginos and neutralinos is presented.  Results from eight separate ATLAS searches are considered, each using 140 fb$^{-1}$ of proton--proton data at a centre-of-mass energy of $\sqrt{s}=13$ TeV collected at the Large Hadron Collider during its second data-taking run. The results are interpreted in the context of the 19-parameter phenomenological minimal supersymmetric standard
model, where $R$-parity conservation is assumed and the lightest supersymmetric particle is assumed to be the lightest neutralino. Constraints from previous electroweak, flavour and dark matter related measurements are also considered.  The results are presented in terms of constraints on supersymmetric particle masses and are compared with limits from simplified models. Also shown is the impact of ATLAS searches on parameters such as the dark matter relic density and the spin-dependent and
spin-independent scattering cross-sections targeted by direct dark matter detection experiments. The Higgs boson and $Z$ boson `funnel regions', where a low-mass neutralino would not oversaturate the dark matter relic abundance, are almost completely excluded by the considered constraints. Example spectra for non-excluded supersymmetric models with light charginos and neutralinos are also presented.
}
 
\author{The ATLAS Collaboration}
 
\AtlasRefCode{SUSY-2020-15}
 
\PreprintIdNumber{CERN-EP-2024-021}

\AtlasJournal{JHEP}
\AtlasJournalRef{\JHEP 2024 (2024) 106}
\AtlasDOI{10.1007/JHEP05(2024)106}

\hypersetup{pdftitle={ATLAS document},pdfauthor={The ATLAS Collaboration}}
 
\begin{document}
 
\maketitle

\section{Introduction}
\label{sec:intro}

The second data-taking run (Run~2) of the Large Hadron Collider (LHC) provided sensitivity to a variety of new-physics models well beyond that of LHC Run~1. This is due to the increase in both integrated luminosity and centre of mass energy -- Run~1 corresponded to 4.6 and 20.3 fb$^{-1}$ of proton--proton collisions at centre of mass energies of 7 and 8~\TeV, respectively, whilst Run~2 searches benefitted from 140~fb$^{-1}$ at 13~\TeV.
This in turn has enabled the ATLAS Collaboration~\cite{PERF-2007-01} to develop a suite of new analysis techniques sensitive to the production of electroweakly interacting particles that extends sensitivity into a new realm. Weak scale supersymmetry (SUSY)~\cite{Golfand:1971iw,Volkov:1973ix,Wess:1974tw,Wess:1974jb,Ferrara:1974pu,Salam:1974ig} is a theoretical extension to the Standard Model (SM) that adds a new fermion/boson SUSY partner to each boson/fermion in the SM\@. If fulfilled in nature, this could help to solve the fine-tuning problem~\cite{Weinberg:1975gm,Gildener:1976ai,Weinberg:1979bn,Susskind:1978ms}. In SUSY models that conserve $R$-parity~\cite{Farrar:1978xj}, SUSY particles (sparticles) must be produced in pairs and the lightest SUSY particle (LSP) is stable. If the LSP is weakly interacting it constitutes a viable candidate for dark matter (DM)~\cite{Goldberg:1983nd,Ellis:1983ew}.
Due to its weakly interacting nature, any LSP produced at the LHC would escape detection and lead to missing transverse momentum ($\MET$).
 
LHC sparticle production cross-sections are highly dependent on the sparticle masses. The coloured sparticles (squarks and gluinos) are strongly produced and have significantly larger production \linebreak \mbox{cross-sections} than non-coloured sparticles of equal masses, such as the sleptons (superpartners of the SM leptons) and the electroweakinos. The superpartners of the SM Higgs boson and the electroweak gauge bosons, known as higgsinos, winos and binos are collectively known as electroweakinos. They mix to form chargino ($\tilde{\chi}_i^\pm, i=1,2$) and neutralino ($\tilde{\chi}_j^{0}, j=1,2,3,4$) mass eigenstates (states are ordered by increasing values of their mass).
If gluino and squark masses were much heavier than \mbox{low-mass} electroweakinos,
then SUSY production at the LHC would be dominated by direct electroweak production. The latest limits from the ATLAS Collaboration on squark and gluino production~\cite{SUSY-2019-08,SUSY-2018-17,SUSY-2018-22,SUSY-2018-10} extend well beyond the \TeV\ scale, thus making electroweak production of sparticles a promising and important probe to search for SUSY at the LHC.
 
Throughout the first two data-taking runs of the LHC the ATLAS and CMS collaborations have performed extensive searches for the production of electroweakinos, and in the absence of any significant excesses over the SM prediction, exclusion limits were set on SUSY model parameters~\cite{SUSY-2018-32,SUSY-2018-16,SUSY-2018-06,SUSY-2018-10,SUSY-2018-02, SUSY-2019-09, SUSY-2018-41,SUSY-2018-05,SUSY-2019-02,SUSY-2018-19}. To simplify the design and interpretation of analyses, `simplified
models'~\cite{Alwall:2008ag} are often used such that the masses of relevant sparticles (often the $\Ch{1}$, $\N{2}$ and $\N{1}$) are the only free parameters. Whilst exclusion limits using simplified models provide an easily interpretable picture of the sensitivity of analyses to specific areas of parameter space in the minimal SUSY extension to the SM (MSSM), they are far from an exhaustive exploration of the MSSM\@.  Without any assumed mechanism for SUSY-breaking the MSSM has over a hundred
parameters describing the sparticle masses and their decays. This large number of parameters can be reduced by considering the phenomenological MSSM (pMSSM)~\cite{Djouadi:1998di,Berger:2008cq}. This is based on the most general $CP$-conserving MSSM, meaning no additional $CP$-violating contributions, and assumes $R$-parity conservation, and minimal flavour violation. The first two generations of sfermions are also required to be mass degenerate and to have negligible Yukawa
couplings. This leaves 19 independent weak scale parameters to consider: ten sfermion masses (five for the degenerate first two generations and five for the third generation), three trilinear couplings $A_{\tau,t,b}$ that give the couplings between the Higgs field and the third generation sfermions, the bino, wino and gluino mass parameters $M_{1,2,3}$, the bilinear Higgs mass parameter $\mu$, the ratio of the vacuum expectation values of the Higgs fields $\tan{\beta}$, and the mass parameter of the pseudoscalar Higgs boson $M_{A}$.  These parameters are summarised in Table~\ref{tab:pmssmparams}.
 
\begin{table}[h!]
\caption{Summary of the 19 pMSSM parameters relevant to this study.}
\label{tab:pmssmparams}
\begin{tabular}{ l >{\raggedright\arraybackslash}p{0.75\textwidth} }
\toprule
pMSSM Parameter & Meaning \\
\midrule
\midrule
 
$\tan\beta$                            &  Ratio of the Higgs vacuum expectation values for the two doublets \\
$M_A$                                  &  Pseudoscalar ($CP$-odd) Higgs boson mass parameter \\
$\mu $ & Higgsino mass parameter \\
$M_1$, $M_2$, $M_3$ & Bino, wino and gluino mass parameters \\
$A_t$, $A_b$, $A_{\tau}$ & Third generation trilinear couplings \\
$M_{\tilde{q}}$, $M_{\tilde{u}_{R}}$, $M_{\tilde{d}_{R}}$, $M_{\tilde{l}}$, $M_{\tilde{e}_{R}}$ & First/second generation sfermion mass parameters \\
$M_{\tilde{Q}}$, $M_{\tilde{t}_{R}}$, $M_{\tilde{b}_{R}}$, $M_{\tilde{L}}$, $M_{\tilde{\tau}_{R}}$ & Third generation sfermion mass parameters \\

\bottomrule
\end{tabular}
 
\end{table}
 
At the end of Run~1 a reinterpretation of the ATLAS SUSY searches was performed using 300\,000 pMSSM models in this 19-dimensional parameter space~\cite{SUSY-2014-08}. The strongest direct constraints on sparticle production were found to come from searches for squarks and gluinos. A subsequent interpretation was performed restricting attention to a five-dimensional sub-space of the pMSSM to specifically assess the impact of the ATLAS Run~1 searches for the electroweak production of SUSY particles, and their corresponding constraints on DM~\cite{SUSY-2015-12}.  The CMS Collaboration also performed a reinterpetation of their Run~1 searches in the 19-dimensional pMSSM~\cite{CMS:2016lcl} using a global Bayesian analysis, and additional reinterpretations of LHC SUSY searches have been performed outside the LHC collaborations. These include results obtained using \textsc{GAMBIT}~\cite{GAMBIT:2017yxo,GAMBIT:2017zdo} and \textsc{MasterCode}~\cite{Bagnaschi:2017tru} that incorporate LHC searches into global likelihood fits in the 19-dimensional pMSSM or subsets of it.
This paper presents the sensitivity of the searches performed with the ATLAS detector in Run~2 of the LHC to the electroweak production of SUSY particles in the pMSSM\@. All ATLAS results used in this paper benefit from an extensive software suite~\cite{ATL-SOFT-PUB-2021-001} that is used in data simulation, in the reconstruction and analysis of real and simulated data, in detector operations, and in the trigger and data acquisition systems of the experiment.
Relative to the Run~1 reinterpretation, this analysis benefits from a broader range of search channels that were targeted throughout Run~2 due to the increased statistical precision and centre of mass energy.
It also takes advantage of new tools for reinterpretation developed by the ATLAS Collaboration, such as particle-level analysis evaluation using the \textsc{SimpleAnalysis} framework~\cite{ATL-PHYS-PUB-2022-017}, preserved profile likelihood fits~\cite{ATL-PHYS-PUB-2019-029} and a new full analysis reinterpretation framework, RECAST~\cite{Cranmer:2010hk,ATL-PHYS-PUB-2019-032}, as well as utilising the REANA reproducible data analysis platform~\cite{Simko:2021gqg}.
 
The work is structured as follows: Section~\ref{sec:analysis} describes the analysis strategy, including the pMSSM model samples and ATLAS results considered. Section~\ref{sec:result} then presents the results, with conclusions provided in Section~\ref{sec:conclusion}.


\FloatBarrier
\clearpage
 
\section{Analysis strategy}
\label{sec:analysis}
 
This section describes the analysis strategy used. This includes a description of the framework used to scan over the pMSSM parameters, the external constraints considered, the simulated samples produced for selected models, the choice of the scan ranges for the pMSSM parameters, and the ATLAS searches and measurements considered.
 
\subsection{Scanning framework}
\label{sec:framework}

\begin{figure}
\centering
\includegraphics[trim=2cm 5cm 0 4cm,clip,width=0.98\textwidth]{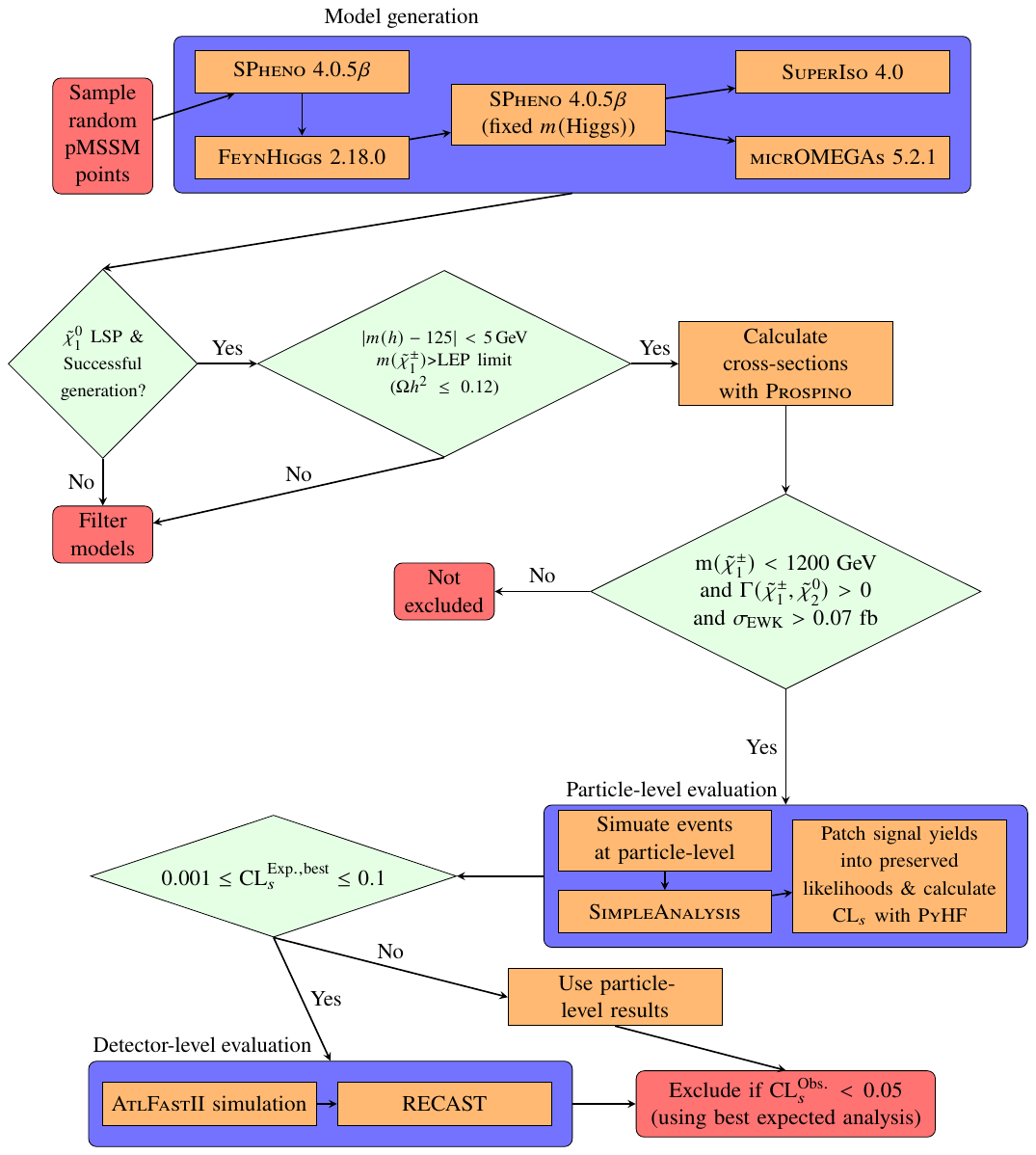}
 
\caption{Workflow for the electroweak pMSSM scan, starting from sampling the pMSSM parameter space and ending at the determination of whether each model is excluded or not.}
\label{workflow}
\end{figure}

A portion of the pMSSM parameter space is randomly sampled to produce a set of models. Two separate samplings (scans) are used as described in Section~\ref{sec:scanconfig}. In both scans a flat prior is chosen as the relevant parameter ranges are relatively narrow ($<2~\tev$) with most of the interest in the range of O(50~\gev) to O(1~\tev). These models are then evaluated using a workflow chain summarised in Figure~\ref{workflow}.
 
The models are first passed through several programs to calculate various observables:
\begin{itemize}
\item \textsc{SPheno} 4.0.5~\cite{Porod:2003um,Porod:2011nf} is used to calculate the mass spectra and decays of the SUSY particles.
\item \textsc{FeynHiggs} 2.18~\cite{Bahl:2018qog,Bahl:2017aev,Bahl:2016brp,Hahn:2013ria,Frank:2006yh,Degrassi:2002fi,Heinemeyer:1998np,Heinemeyer:1998yj} is used to calculate relevant Higgs sector variables, such as the masses and branching fractions of the SUSY and SM Higgs bosons. As the masses of the SUSY particles strongly impact the Higgs boson masses, FeynHiggs automatically takes these into account. The changes to the Higgs sector masses, in turn, influence the branching fractions of SUSY particles. Hence, the resulting spectrum determined with FeynHiggs is reprocessed by \textsc{SPheno}, which allows the calculation of SUSY branching fractions while keeping both the SUSY and Higgs sector masses constant.
\item \textsc{MicrOMEGAS} 5.2.1~\cite{Barducci:2016pcb,Belanger:2020gnr} is used to calculate the predicted DM relic density, annihilation cross-sections, and spin (in)dependent weakly interacting massive particle (WIMP)--nucleon cross-sections. Limits on WIMP--nucleon cross-sections from direct detection experiments assume a DM candidate that saturates the observed relic density $\Omega h^2=0.12$ whereas the majority of pMSSM models considered here under-predict the observed relic density. Therefore all WIMP--nucleon cross-sections are scaled by $(\Omega h^2 / 0.12)$ for each model below the observed relic density to correct for this assuming that a second DM component makes up the remaining relic density without contributing to the DM--nucleon scattering cross-sections.
\item \textsc{SuperISO} 4.0~\cite{Mahmoudi:2008tp} is used to calculate a variety of flavour observables.
 
\end{itemize}

Models that fail to be processed properly by one or more of these programs or contain unphysical spectra are removed. Models where the LSP is not the lightest neutralino (\N{1}) are removed as are models with charginos that are excluded by LEP,
i.e., $m(\tilde{\chi}_{1}^{\pm})<103\,\GeV$ for $\Delta\,m (\tilde{\chi}_{1}^{\pm}, \tilde{\chi}_{1}^{0})\geq 3\,\gev$~\cite{LEPSUSY1} and
$m(\tilde{\chi}_{1}^{\pm})<91.9\,\GeV$ for $\Delta\,m (\tilde{\chi}_{1}^{\pm}, \tilde{\chi}_{1}^{0})< 3\,\gev$~\cite{LEPSUSY2}. Finally, a loose bound is applied to the predicted mass of the SM Higgs boson: $120\,\GeV<m(h)<130\,\GeV$.
The bound on the Higgs boson mass is wider than the mass measurement precision due to the larger theoretical uncertainties in the calculation for the MSSM\@. A loose bound also improves the efficiency of the model generation,
while the bound keeps simulated Higgs boson decays consistent with other experimental constraints.
When presenting the final results, additional external constraints complementary to the ATLAS search constraints are considered based on the observables calculated by these programs. These include constraining the mass of the \Wboson boson to a window around its measured values, spin (in)dependent LSP--nucleon cross-section limits, constraints on electroweak precision observables and constraints on $B$-physics observables. These constraints are summarised in Table~\ref{tab:externalconstraints}.
 
\newenvironment{DIFnomarkup}{}{}
\begin{DIFnomarkup}

\begin{table}
\caption{Constraints from electroweak precision measurements, flavour physics observables and direct-detection DM searches. When used in the results in Section~\ref{sec:result}, the flavour and precision electroweak constraints together correspond to the `non-DM' external constraints, whilst `all external constraints' includes the flavour, precision electroweak and DM constraints. In addition to these constraints, unless otherwise stated, all models considered in this paper include the LEP constraint on the chargino mass and a Higgs boson mass constraint as described in the text.}
\label{tab:externalconstraints}
\centering
\scriptsize
\begin{tabular}{l  l l l l } \toprule
Category & Constraint & Lower bound & Upper bound & Notes \\ \midrule
Flavour & $\mathcal{B}$($b\rightarrow s\gamma$) & $3.11 \times 10^{-4}$ & $3.87 \times 10^{-4}$ & 2022 PDG average ($2\sigma$ window)~\cite{Workman:2022ynf}. \\
& $\mathcal{B}$($B_{s}\rightarrow \mu\mu$) & $1.87 \times 10^{-9}$ & $4.31 \times 10^{-9}$ & Most recent LHCb result ($2\sigma$ window)~\cite{LHCb:2021awg}. \\
& $\mathcal{B}$($B^{+}\rightarrow\tau\nu$) & $6.10 \times 10^{-5}$ & $1.57 \times 10^{-4}$ & 2022 PDG average ($2\sigma$ window)~\cite{Workman:2022ynf}. \\ \midrule
Precision & $\Delta\rho$ & $-0.0004$ & $0.0018$ & Updated global electroweak fit by \textsc{GFitter} group~\cite{Haller:2018nnx} \\
electroweak & & & & (not including CDF $W$ mass measurement~\cite{CDF:2022hxs}). \\
& $\Gamma^{\textup{BSM}}_{\textup{inv}}(Z)$ & -- & 2 \MeV & Beyond-the-Standard Model contributions to precision electroweak \\
& & & & measurements on the $Z$-resonance from experiments at the SLC and \\
& & & &  LEP colliders~\cite{ALEPH:2005ab}. \\
& $m(W)$ & 80.347 \GeV & 80.407 \GeV & 2022 PDG result (excluding CDF $W$ mass measurement~\cite{CDF:2022hxs})~\cite{Workman:2022ynf} \\
& & & &  but with the 2$\sigma$ window expanded by 6 \MeV to allow for uncertainty \\
& & & &   due to the top-quark mass in the MSSM Higgs calculation~\cite{Heinemeyer:2013dia}. \\
\midrule
DM & Relic density  & -- &0.12 & Latest bound from Planck~\cite{Planck:2018vyg}. \\
& $\sigma_{\textup{Spin-independent}}$ & & & Exclusion contour on direct detection of DM from the \\
& & & & LZ Collaboration~\cite{LUX-ZEPLIN:2022xrq}. \\
& $\sigma_{\textup{Spin-dependent}}$ & & & Exclusion contour on direct detection of DM from PICO-60~\cite{PICO:2019vsc}. \\
\midrule
\end{tabular}
\end{table}
 
\end{DIFnomarkup}
 
The cross-sections for each electroweakino pair-production process are calculated at next-to-leading-order\footnote{For simplified model results published by ATLAS, \textsc{Resummino} \cite{Fuks:2013vua} is used to provide additional next-to-leading-logarithm corrections to the production cross-section. To reduce resource usage, \textsc{Resummino} has not been applied here which could cause small differences in cross-sections for similar processes.} using \textsc{Prospino} \cite{prospino} for each model that passes the initial constraints.
At this stage, a filter is applied to halt the processing of models that the ATLAS searches are very unlikely to have sensitivity to. This filter requires $m(\Ch{1})<1200\ \GeV$, as the analyses considered have no sensitivity to scenarios with a $\Ch{1}$ mass above this limit, and that the total cross-section for electroweak production of SUSY particles $\sigma_{\text{EWK}}>7 \times 10^{-5}\ \text{pb}$, as lower cross-sections would be expected to yield less than ten events in the full Run~2 data sample. In addition models with predicted stable or effectively stable \Ch{1} and \N{2} are also filtered, as the event generation cannot handle additional stable charginos and neutralinos. Models rejected by this filter are included in the final model sets used in Section~\ref{sec:result}, but are considered to be not excluded here, even if some might be excluded by dedicated long-lived particle searches (for example the ATLAS search for heavy long-lived charged particles with large ionisation losses has sensitivity to long-lived charginos~\cite{SUSY-2018-42}).
 
Events are then simulated using Monte Carlo (MC) techniques for each model that passes these filters. Events are simulated at leading-order using \MGNLO{2.9.5}~\cite{Alwall:2011uj} and \PYTHIA{8.306}~\cite{Sjostrand:2014zea}. Only electroweak production processes are generated, and coloured sparticles are not included in the subsequent decays. Assessing the detector effects on simulated events requires them to be processed either through a full simulation of the ATLAS detector~\cite{SOFT-2010-01} based on
\GEANT~\cite{Agostinelli:2002hh}, or a faster version of the simulation (\textsc{AtlFastII}), which relies on a parameterisation for the response of the calorimeters and on \GEANT for the other components of the detector, then reconstructed with the same algorithms as those used for the data. Producing such `detector-level' reconstructed samples is computationally expensive, so instead events are initially simulated at particle-level using the \textsc{SimpleAnalysis} framework~\cite{ATL-PHYS-PUB-2022-017}. This includes parameterised response functions that aim to emulate detector effects such as reconstruction and identification inefficiencies
and resolutions. These are typically parameterised as functions of the transverse momentum and rapidity for the different types of particles. \textsc{SimpleAnalysis} is used to obtain an approximate result that indicates if the model is likely to be excluded or not. Validation studies of each \textsc{SimpleAnalysis} implementation are performed to confirm that the yields obtained agree with the yields from fully reconstructed samples within a reasonable margin (usually around 20\%). All
electroweakino production processes are produced simultaneously in \textsc{MadGraph} and, to keep processing time manageable, up to a single additional jet from initial-state radiation is allowed in the \textsc{MadGraph} calculation and the 4-flavour merging scheme is used to combine them with \PYTHIA. The number of events generated is required to correspond to at least five times the Run~2 integrated luminosity, subject to an upper limit of 250\,000 events. For high cross-section models exceeding
this limit, additional filtered samples are produced at particle level to improve the statistical precision of the evaluation.
 
ATLAS searches for SUSY typically define a set of statistically independent event selections. `Signal regions' (SR) are event selections where a statistically significant excess of events would be observed over the SM prediction if the signal were present. These are supplemented by auxiliary `control regions' (CR) that are used to produce semi-data-driven estimates of SM backgrounds. A simultaneous profile likelihood fit is used to constrain the MC yields with the observed data in the CRs. The CRs are designed to be both orthogonal and similar to the SRs, whilst also having little signal contamination.   When performing the statistical analysis  the likelihood is constructed as a product of Poisson probability density functions, describing the observed number of events in each CR/SR, and Gaussian distributions that describe the nuisance parameters associated with each of the systematic uncertainties. Poisson distributions are used for MC statistical uncertainties, and systematic uncertainties that are correlated between different samples can be accounted for by using the same nuisance parameter. Exclusion limits are set on SUSY model parameters using the CL$_{\text{s}}$ prescription~\cite{Read:2002hq}. The SRs and CRs in different ATLAS searches are not statistically independent and cannot be combined easily. Therefore, in this paper, only the search with the best expected exclusion limit for a given model is used to decide if a model is excluded or not.
 
To emulate this calculation using the particle-level signal samples and determine whether a model is excluded, a profile-likelihood fit using the procedure outlined in Ref.~\cite{ATL-PHYS-PUB-2019-029} is performed using the particle-level signal yields along with a detector-level estimate of the background yields and observed data for each analysis.\footnote{Most Run~2 ATLAS SUSY analyses have published a `likelihood'~\cite{ATL-PHYS-PUB-2019-029} on the HEPData platform~\cite{hepdata}, but in the cases where this was not available this was obtained internally.} A hypothesis test is performed using the \textsc{pyHF} framework~\cite{pyhf,pyhf_joss} to produce an observed and expected CL$_{\text{s}}$ value for each analysis. Statistical and systematic uncertainties in the signal yields are not included in this calculation, but all background systematic and statistical uncertainties are retained. For some analyses signal contamination in the CRs is also neglected when performing the particle-level evaluation, however it is always considered when calculating the detector-level CL$_{\text{s}}$ values discussed next.
 
Using particle-level signal yields and not applying the complete set of signal uncertainties means that the CL$_{\text{s}}$ values calculated using this procedure are only approximate.
To determine the final exclusion, the particle-level CL$_s$ values are used to categorise the models as follows.
Models deemed to be `likely excluded' have an expected CL$_{\text{s}}<0.001$ for at least one of the considered electroweak searches. Those deemed `likely not excluded' have an expected CL$_{\text{s}}>0.1$ for every analysis.
These boundaries are chosen by checking that the particle-level classifications for a subset of models are consistent with their full detector-level evaluation.
Finally, those identified as `ambiguous' have an expected CL$_{\text{s}}$ that satisfies 0.001$\leq$ CL$_{\text{s}}$ $\leq$ 0.1 for at least one analysis and CL$_{\text{s}}\geq 0.001$ for every analysis.  For the first two categories these particle-level evaluations are considered final and used for the results presented in Section~\ref{sec:result}. For the models deemed `ambiguous' detector-level MC samples are produced using the \GEANT simulation with parameterised simulation of the calorimeters. Whilst computationally expensive, the categorisation process outlined in this section ensures this only has to be done for 5\%--10\% of models.
The full ATLAS analyses are then applied to the resulting MC samples using RECAST -- a framework for re-using existing analyses to interpret new physics models.  The RECAST of an analysis contains all of the original analysis steps in Docker containers~\cite{merkel2014docker}, to allow a full re-execution of the event selection, calculation of systematic uncertainties and statistical analysis for a new signal sample. Finally, the CL$_{\text{s}}$ output by RECAST is used to categorise ambiguous models as excluded (CL$_{\text{s}} < 0.05$) or not excluded (CL$_{\text{s}} > 0.05$) based on the full detector-level analysis.


\subsection{Scan configurations}
\label{sec:scanconfig}

\begin{table}[t]
\caption{Parameter ranges that are randomly sampled for the `EWKino' scan.}
\label{tab:ewkino_ranges}
\centering
\begin{tabular}{ l r r l } \toprule
Parameter & Min & Max & Note \\ \midrule
$M_{\tilde{L}_1}$ (=$M_{\tilde{L}_2}$) & 10 \TeV & 10 \TeV & Left-handed slepton (first two gens.) mass \\
$M_{\tilde{e}_1}$ (=$M_{\tilde{e}_2}$) & 10 \TeV & 10 \TeV & Right-handed slepton (first two gens.) mass \\
$M_{\tilde{L}_3}$                      & 10 \TeV & 10 \TeV & Left-handed stau doublet mass \\
$M_{\tilde{e}_3}$                      & 10 \TeV & 10 \TeV & Right-handed stau mass \\ \midrule
$M_{\tilde{Q}_1}$ (=$M_{\tilde{Q}_2}$) & 10 \TeV & 10 \TeV & Left-handed squark (first two gens.) mass \\
$M_{\tilde{u}_1}$ (=$M_{\tilde{u}_2}$) & 10 \TeV & 10 \TeV & Right-handed up-type squark (first two gens.) mass \\
$M_{\tilde{d}_1}$ (=$M_{\tilde{d}_2}$) & 10 \TeV & 10 \TeV & Right-handed down-type squark (first two gens.) mass \\
$M_{\tilde{Q}_3}$                      & 2 \TeV  & 5 \TeV & Left-handed squark (third gen.) mass \\
$M_{\tilde{u}_3}$                      & 2 \TeV  & 5 \TeV & Right-handed top squark mass \\
$M_{\tilde{d}_3}$                      & 2 \TeV  & 5 \TeV & Right-handed bottom squark mass \\ \midrule
$M_1$                                & $-2$ \TeV  & 2 \TeV & Bino mass parameter \\
$M_2$                                & $-2$ T\eV  & 2 \TeV & Wino mass parameter \\
$\mu$                                  & $-2$ \TeV  & 2 \TeV & Bilinear Higgs boson mass parameter \\
$M_3$                                  & 1 \TeV  & 5 \TeV & Gluino mass parameter \\ \midrule
$A_t$                                & $-8$ \TeV  & 8 \TeV & Trilinear top coupling \\
$A_b$                                & $-2$ \TeV  & 2 \TeV & Trilinear bottom coupling \\
$A_{\tau}$                           & $-2$ \TeV  & 2 \TeV & Trilinear $\tau$-lepton coupling \\
$M_A$                                  & 0 \TeV  & 5 \TeV & Pseudoscalar Higgs boson mass \\
$\tan\beta$                            & 1 & 60 & Ratio of the Higgs vacuum expectation values \\ \bottomrule
\end{tabular}
\end{table}
 
Two pMSSM scans are processed with different parameter ranges and sampling strategies. In both scans, the sleptons and all squarks
and gluinos are `decoupled', i.e., mass parameters are set sufficiently high as to not influence the production or decays of the electroweakinos. The first scan focuses on general electroweakino production and is denoted the `EWKino scan'. The parameter ranges for the EWKino scan are listed in Table \ref{tab:ewkino_ranges}.
The EWKino scan comprises 12\,280 models that survive all of the constraints and filters when starting from an initial set of 20\,000 models randomly sampled (with a flat prior) from these ranges.
 
The magnitudes of the $M_1$, $M_2$ and $\mu$ parameters, which control the bino, wino and higgsino masses respectively, are varied between 0 and 2 \TeV.
The ATLAS Run 2 sensitivity to electroweakinos does not extend beyond masses of 1 \TeV, so these ranges capture a range of relevant senarios including those where the heavier \Ch{2}, \N{3} and \N{4} (not present in simplified models) are both below and above this limit.
Whilst the third generation squark masses are scanned over and those parameters affect for example the Higgs boson mass, they have no significant impact on the phenomenology or analysis exclusion in the electroweak sector and are not included in the event generation. This is also the case for the gluino mass parameter $M_{3}$.

\begin{figure}[t]
\centering
\includegraphics[width=0.47\textwidth]{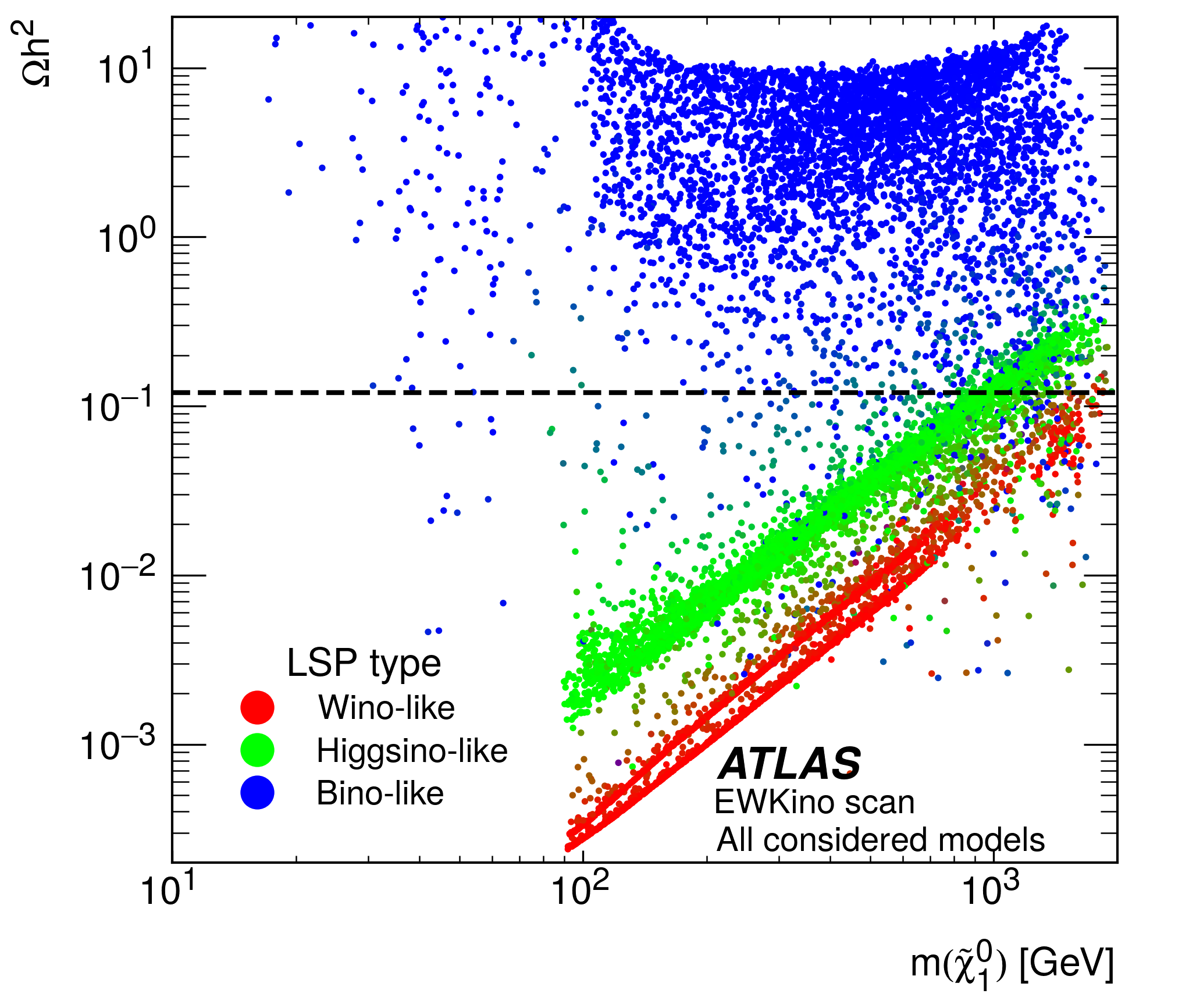}\label{DM:LSP_type}
\caption{Scatter plot of models from the EWKino scan in the $\Omega h^2$ versus $m(\N{1})$ plane, coloured with RGB value set to the (wino, bino, higgsino) fraction of the LSP\@. The experimentally measured value of the relic density~\cite{Planck:2018vyg}, $\Omega h^2 = 0.120\pm0.001$, is indicated by a horizontal dashed line.}
\label{fig:EWKinoscanLSPtype}
\end{figure}

Figure~\ref{fig:EWKinoscanLSPtype} shows a scatter plot of the models from the EWKino scan in the $\Omega h^2$ versus $m(\N{1})$ plane, coloured by the dominant component (wino, bino or higgsino) of the LSP\@.
The experimentally measured value of the relic density~\cite{Planck:2018vyg}, $\Omega h^2 = 0.12$, is indicated by a horizontal dashed line.
Here the LSP is defined as `bino-like' if the bino fraction is greater than the wino/higgsino fraction, and similar for `wino-like' and `higgsino-like'.
With decoupled sleptons (that limit options for co-annihilation), models with a bino-like LSP typically overestimate the DM relic density unless there are additional annihilation mechanisms available.
A higgsino- or wino-like LSP, in general, provides a DM relic density prediction below the measured value, unless the LSP mass is around 1~\TeV\ (higgsino) or 3~\TeV\ (wino). This is because a higgsino- or wino-like LSP has enhanced co-annihilation with the chargino and/or second neutralino that by construction are close in mass to a wino- or higgsino-like LSP.
 
Applying the DM relic density constraint $\Omega h^2 \leq 0.12$ (where the inequality allows the neutralino LSP to be a subdominant component of DM) would thus remove almost all models in the EWKino scan with a bino-like LSP\@. To avoid over-saturating the relic density additional annihilation mechanisms are required for bino-like LSP models. These typically lead to models with more compressed mass splittings between the LSP and the $\N{2},\Ch{1}$ as well as concentrating models in the `$Z$/$h$/$A$/$H$ funnel' regions. In the funnel regions the LSP mass is approximately half the mass of the $Z$, Higgs boson, heavy neutral scalar Higgs boson $H$, or pseudoscalar Higgs boson $A$, such that self-annihilation into a $Z$/$h$/$A$/$H$ is enhanced, resulting in a DM relic density below the experimental value. The ranges involved in the EWKino scan do not generate enough models in the funnel regions to demonstrate the sensitivity of the ATLAS searches. To compensate for this and to produce a sizeable sample of bino-LSP models with $\Omega h^2 \leq 0.12$, a second scan is performed that oversamples such models.  The ranges of this scan are the same as those in Table \ref{tab:ewkino_ranges} except that the bino mass parameter range is tightened to $|M_{1}| < 500$ \GeV\ in order to focus on low-mass bino models. 437,500 models are randomly sampled from these ranges with a flat prior, and only those with a bino-like LSP and $\Omega h^2 \leq 0.12$ are kept. These constraints, along with the additional filters, reduce the number of models to 8\,897. This scan is referred to as the `BinoDM' scan. In the EWKino scan the relic density constraint is not applied in the initial selection of models. Table \ref{tab:scans} summarises the bino mass parameter ranges and the number of models passing each step of the workflow, for the two scans.

\begin{table}
\caption{Set-ups for the two pMSSM scans performed and number of models passing each step and constraint. Whilst listed as part of the selections applied to produce the model set, the DM relic density is not considered in the model generation for the EWKino scan, where it is instead only used as part of the DM constraints discussed in Table~\ref{tab:externalconstraints} of Section~\ref{sec:framework}.
}
\label{tab:scans}
\centering
\begin{tabular}{ | c | r r | } \hline
\textbf{Scan name} & \textbf{EWKino} & \textbf{BinoDM} \\
\hline
\hline
$|M_1|$ range & 0 -- 2 \TeV & 0 -- 500 \GeV \\
LSP type & Neutralino & Bino-like neutralino \\
\hline
\hline
\textbf{Number of models generated:} & & \\
Sampled &  20\,000 & 437\,500 \\
Successful generation & 16\,667 & 370\,017 \\
Correct LSP type & 15\,321 & 286\,267 \\
Satisfy DM relic density constraint $\Omega h^2\leq 0.12$ & N/A & 11\,122 \\
Satisfy LEP chargino mass constraint & 13\,969 & 10\,174 \\
$120~\GeV < \mathrm{m}(h) < 130~\GeV$ &  12\,280 & 8\,897 \\
Satisfy non-DM external constraints & 7\,956 & 5\,752 \\
Satisfy all external constraints & 2\,460 & 1\,769 \\
\hline
\end{tabular}
\end{table}
 
\begin{figure}[t]
\centering
\subfloat[][EWKino scan]{ \includegraphics[width=0.43\linewidth]{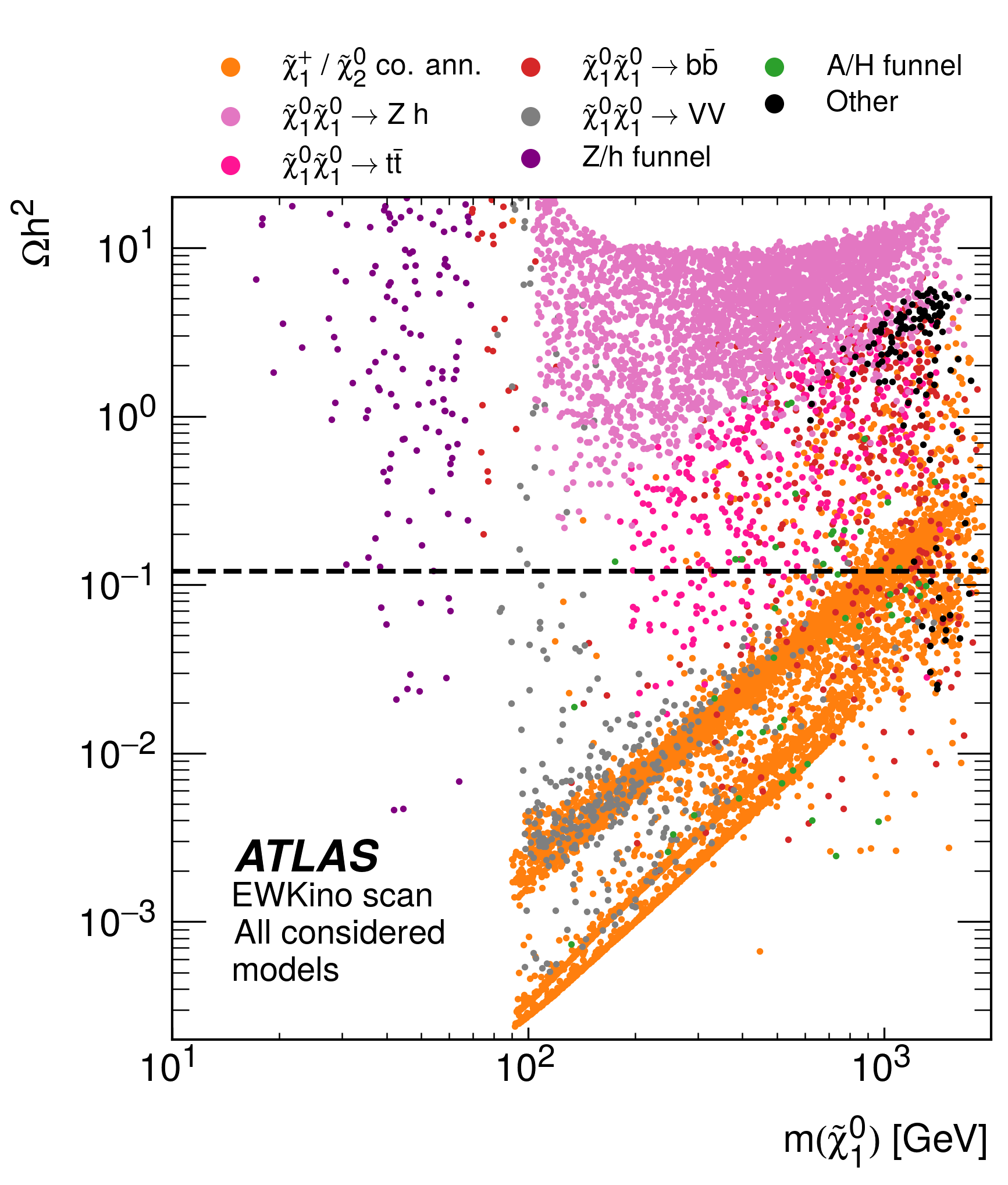} }
\subfloat[][BinoDM scan]{ \includegraphics[width=0.43\linewidth]{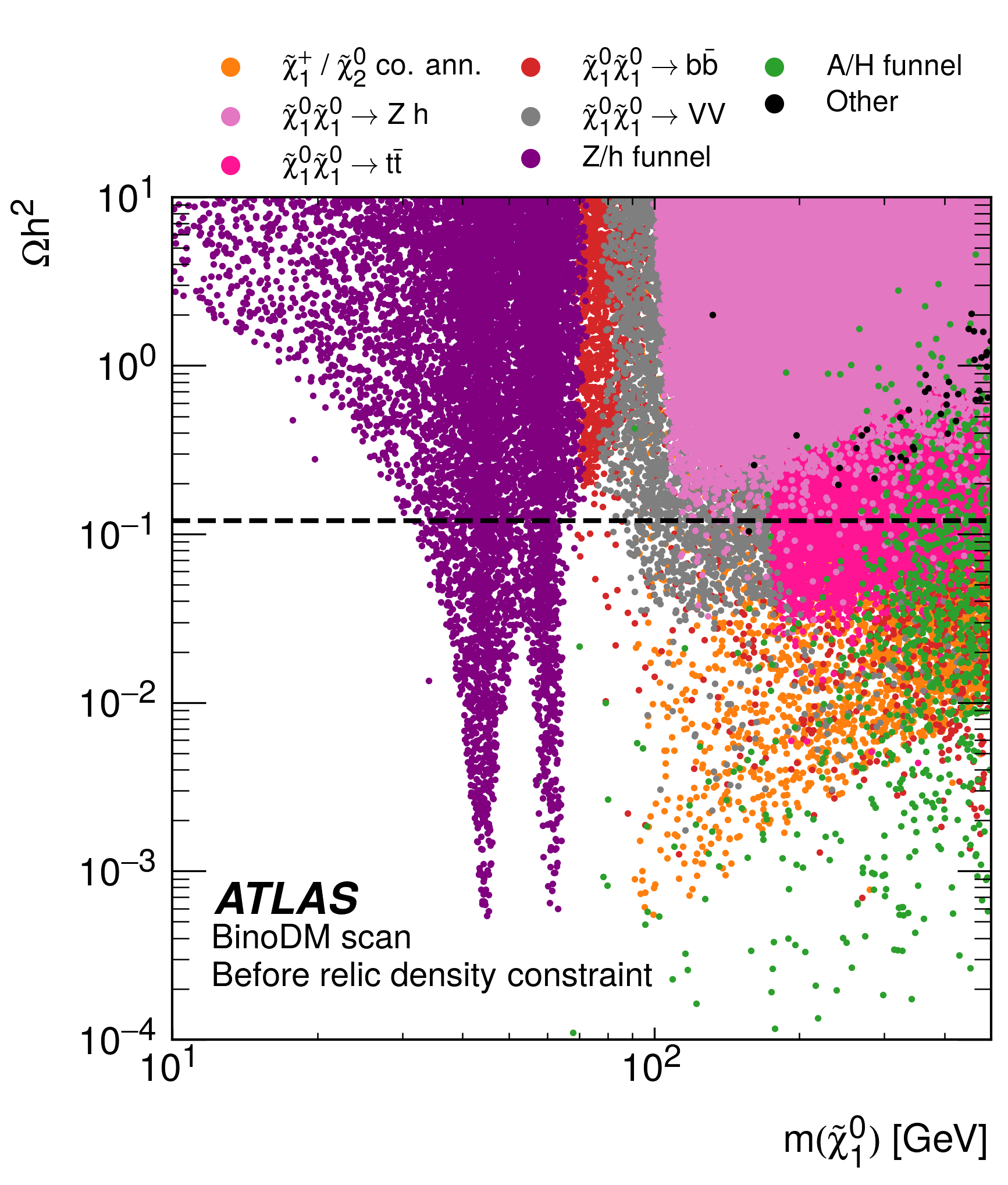}  }
 
\caption{ Scatter plot of models selected from (a) the EWKino and (b) the BinoDM scan (before the relic density constraint is applied) in the $\Omega h^2$ versus $m(\N{1})$ plane, coloured by the dominant annihilation mechanism. No additional external constraints are applied.}
\label{fig:annihilationplots}
\end{figure}

Figure~\ref{fig:annihilationplots} shows models from the EWKino and BinoDM scans plotted in the $\Omega h^2$ versus $m(\N{1})$ plane before the DM relic density constraint is applied, coloured by the dominant annihilation mechanism. Several regions of interest can be observed. The first are the `$Z$/$h$ funnel' regions in purple, which are particularly important for the BinoDM scan.
In order to couple to the Higgs boson and $Z$ bosons such that these annihilation mechanisms are allowed, the LSP must have a higgsino component.
These `funnel regions' are of particular interest as they can satisfy the DM relic density constraint and also have an LSP mass that overlaps with the region of sensitivity for many of the Run~2 ATLAS SUSY searches.
Similarly, the $A$ and $H$ funnel regions  are shown in green. For $m(\N{1})>100\ \GeV$, in the  `bulk region', there are a variety of other (co-)annihilation mechanisms contributing.
When $m(\N{1})>173\,\GeV$ (i.e. the LSP is similar to or greater in mass than the top quark), the $\N{1}\N{1}\rightarrow t\bar{t}$ self-annihilation process is allowed, coloured in pink.
The orange points also show models where the LSP is close in mass to a wino- or higgsino-like \Ch{1} and \N{2}, causing enhanced co-annihilation.
In general, in the BinoDM scan the DM relic density constraint favours an LSP with a bino/higgsino mix. A pure-bino LSP is disfavoured unless its mass is very close to the $Z$/$h$ pole, or the \Ch{1}/\N{2} are very nearby in mass.
 
\subsection{Summary of ATLAS searches and measurements considered}
\label{sec:searches}

This sub-section describes the eight ATLAS searches and the additional constraints from ATLAS Higgs boson measurements that are included in the results in Section~\ref{sec:result}.
The ATLAS Run~2 searches for electroweak SUSY took a `signature-driven' approach, meaning that multiple final states can have sensitivity to the same SUSY production mode due to differing decays of the gauge bosons in the decay chain.
The simplified models used to optimise and interpret the searches typically assumed that the electroweakinos are pure bino, wino or higgsino states in terms of cross-section evaluation, with the dominant SUSY production mode being determined by the next-lightest SUSY particle (NLSP), which then decays into a particular SM final state and the LSP with a fixed branching fraction.
In Run~1 the main simplified model presented was the `bino-wino' scenario where the LSP is a pure bino with the next lightest SUSY particles being mass degenerate wino-like $\N{2}$ and $\chinoonepm$.
The dominant production modes are then associated  $\chinoonepm\ninotwo$ production or $\chinoonep\chinoonem$ pair production.
The larger data samples in Run~2 have enabled multiple searches targeting this scenario with complementary sensitivity in different areas of the LSP versus NLSP mass plane.
Assuming boson-mediated decays, the $\chinoonepm$ decays via a $W$ boson either leptonically or hadronically, whilst the $\ninotwo$ can decay via either a $Z$ boson, which decays leptonically or hadronically, or via a Higgs boson, which can then decay according to its SM branching fractions.
The Run~1 electroweak SUSY searches focused on leptonic final states for these models as they provide a clean final state for triggering and lower SM backgrounds. The larger data samples in Run~2 have allowed more challenging final states to be probed.
 
Table~\ref{table:ewk_analyses} summarises the Run~2 electroweak SUSY searches considered in this analysis.
The `2L0J' search~\cite{SUSY-2018-32} targeted  $\chinoonep\chinoonem$ with $W$ boson mediated decays into final states with exactly two opposite sign electrons or muons, low hadronic activity and missing transverse momentum.
The `3L' search~\cite{SUSY-2019-09} targeted associated $\chinoonepm\ninotwo$ production with $WZ$ mediated decays where both the $Z$ and $W$ boson decay leptonically. Scenarios where the decays proceed via on-shell and off-shell $Z/W$ bosons were targeted with separate selections.
In addition to these fully leptonic searches, channels probing hadronic decays of the $W$ or $Z$ were considered.
This incudes the `FullHad' search~\cite{SUSY-2018-41} which considered both $\chinoonep\chinoonem$ and $\chinoonepm\ninotwo$ production with  boson-mediated decays (including $\ninotwo$ decays into both $Z$ and $h$) into final states involving hadronic jets, and the `2L2J' search~\cite{SUSY-2018-05} which targeted $\chinoonepm\ninotwo$ where the $\ninotwo$ decays leptonically via a $Z$ and the $\chinoonepm$ decays hadronically. An additional dedicated search, denoted `1Lbb'~\cite{SUSY-2019-08},
specifically targeted the scenario where the $\ninotwo$ decays via an on-shell Higgs boson into two $b$-tagged jets, and the $\chinoonepm$ decays leptonically.

Unlike the other searches cosidered, the `4L' search~\cite{SUSY-2018-02} targeted general gauge-mediated (GGM)~\cite{Meade:2008wd} scenarios with a gravitino LSP\@. However it does provide sensitivity to the pMSSM models considered in this paper that produce four leptons through long decay chains following production of the heavier electroweakinos.
 
For the bino-wino scenario without additional radiation, the amount of missing transverse momentum in the final state is largely determined by the mass splitting between the LSP and the  $\chinoonepm, \ninotwo$. Many of the analyses referred to above included multiple search regions targeting different mass splittings, however a dedicated `compressed'~\cite{SUSY-2018-16} search was also performed that targeted models with smaller mass splittings in final states with `soft' low-transverse momentum leptons, and high missing transverse momentum generated by the SUSY production system recoiling against initial-state radiation. This search was also the first analysis to target SUSY scenarios with a pure higgsino LSP\@.  Finally, the only search in this paper that was optimised for pure wino LSP scenarios (while also targeting models with a higgsino LSP) is the `disappearing-track' search~\cite{SUSY-2018-19}. This targeted anomaly-mediated SUSY breaking (AMSB)~\cite{Giudice:1998xp,Randall:1998uk} models that naturally give a pure wino LSP as well as providing interpretations for compressed higgsino scenarios. This search targeted long-lived charginos using a distinct signature of a short track that `disappears' in the ATLAS tracking detector associated with missing transverse momentum.
 
\begin{table}[t]
\caption{
List of ATLAS electroweak SUSY analyses considered in this analysis, and the relevant original simplified models targeted by them.
}
\label{table:ewk_analyses}
\begin{tabular}{ l >{\raggedright\arraybackslash}p{0.75\textwidth} }
\toprule
Analysis & Relevant simplified models targeted \\
\midrule
\midrule
FullHad~\cite{SUSY-2018-41} &   Wino $\chinoonepm\ninotwo$ via $WZ$, Wino $\chinoonepm\ninotwo$ via $Wh$, Wino $\chinoonep\chinoonem$ via $WW$ \\
1Lbb~\cite{SUSY-2019-08} & Wino $\chinoonepm\ninotwo$ via $Wh$ \\
2L0J~\cite{SUSY-2018-32} & Wino $\chinoonep\chinoonem$ via $WW$, slepton pairs \\
2L2J~\cite{SUSY-2018-05} & Wino $\chinoonepm\ninotwo$ via $WZ$ \\
3L~\cite{SUSY-2019-09} & Wino $\chinoonepm\ninotwo$ via $WZ$, Wino $\chinoonepm\ninotwo$ via $Wh$, higgsino $\chinoonepm\ninotwo\ninoone$ \\
4L~\cite{SUSY-2018-02} & Higgsino GGM \\
Compressed~\cite{SUSY-2018-16} & Wino $\chinoonepm\ninotwo$ via $WZ$, higgsino $\chinoonepm\ninotwo\ninoone$ \\
Disappearing-track~\cite{SUSY-2018-19} & Wino $\chinoonep\chinoonem$ and $\chinoonepm\ninoone$ \\
 
\bottomrule
\end{tabular}
\end{table}
 
When performing the workflow described in the previous sub-section, several modifications are made for specific searches:
\begin{itemize}
\item In the cases where the time to run the full analysis likelihood with particle-level signal yields is too computationally expensive, the `simplified likehood' procedure described in Ref.~\cite{ATL-PHYS-PUB-2021-038} is used instead. This makes several simplifications including combining all background components into a single sample and reducing all nuisance parameters in the full likelihood to a single constrained parameter. This significantly increases the speed of the statistical fit, at the price of some accuracy. This approach is used for the compressed and 3L off-shell analyses.
\item For the disappearing-track analysis the upper limits on production cross-sections are used to determine exclusion.
This is justified in this case since the disappearing-track analysis acceptance is largely determined by the $\Ch{1}$ mass and lifetime, and not by other model parameters such as $\tan\beta$ or $\mu$. Only direct $\Ch{1}$ and \N{1} production is considered, i.e., production through decay of heavier states is ignored, leading to slightly conservative limits.
A linear interpolation between the available upper limits is used to determine the appropriate limit for each pMSSM model.
To ensure a reliable result, only models with chargino mass and lifetime within the ranges for which limits are available are considered.
If the total production cross-section for a model, $\sigma(\Ch{1}\N{1}) + \sigma(\ChP{1}\ChM{1})$, is greater than the corresponding limit then the model is considered excluded.
\end{itemize}
 
In addition to the ATLAS SUSY searches shown in Table~\ref{table:ewk_analyses}, two additional constraints related to measurements of the Higgs boson are applied when assessing the ATLAS Run~2 constraints on the pMSSM models. The first is the most recent combined upper limit on the branching ratio of the Higgs boson into invisible particles  of $\mathcal{B}$($h\rightarrow\textup{inv}$)$<0.107$ ~\cite{ATLAS:2023tkt}. Models where Higgs boson decays into SUSY particles would increase this value above this limit are considered excluded. The second constraint applied is the ATLAS combined constraint on the CP-odd Higgs boson mass in the MSSM from Higgs boson cross-sections and branching fractions~\cite{ATLAS:2019nkf}, which is $m(A)>480\,\GeV$.
This $m(A)$ bound is approximate, particularly at high tan$\beta$ where additional SUSY corrections will contribute in the pMSSM\@. However, these corrections are suppressed by $\mu/M_{\mathrm{SUSY}}$ where $M_{\mathrm{SUSY}}$ is the SUSY scale. In the scans presented here, $\mu$ is much smaller than $M_{\mathrm{SUSY}}$ such that these corrections are expected to be small.


\FloatBarrier
\clearpage
 
\section{Results}
\label{sec:result}

This section summarises the constraints from the ATLAS searches for electroweak SUSY on the pMSSM models selected by the scans described in Section~\ref{sec:scanconfig}. To determine whether a model is excluded, the analysis with the best (i.e. smallest) expected CL$_{\text{s}}$ of the analyses in Table~\ref{table:ewk_analyses} is used.  If the observed CL$_{\text{s}}$ for that analysis is less than 0.05, the model is considered to be excluded. Statistical combinations of analyses are not
performed in this study so these results provide a conservative estimate of the ATLAS constraints.
As outlined in the Section~\ref{sec:searches}, models with $\mathcal{B}(h\rightarrow\textup{inv})\geq0.107$ or $m(A)\leq480\,\GeV$ are also considered to be excuded by ATLAS.
 
Throughout this section, results are presented as projections of the pMSSM models in one or two dimensions. For the one-dimensional plots the number of models satisfying different model selection criteria is presented as a function of a single pMSSM parameter or observable, along with the fraction of `all considered models' satisfying that selection. Throughout this section `all considered models' is the total number of models generated in each scan that satisfy the LEP chargino and LHC Higgs boson mass constraints
discussed in Section~\ref{sec:scanconfig}  (constituting 12\,280 and 8\,897 models for the EWKino and BinoDM scans respectively as shown in Table~\ref{tab:scans}). The BinoDM scan also includes the relic density requirement. The two-dimensional plots are presented  in terms of the fraction of models excluded by the ATLAS Run~2 constraints described in Section~\ref{sec:searches} in a given bin. When calculating fractional exclusion plots three sets of model selections are considered: plots
labelled `all considered models' correspond to the selection described above, plots labelled `all non-DM constraints' include only the subset of these models that satisfy the flavour and electroweak precision measurements in Table~\ref{tab:externalconstraints} and those labelled `all external constraints' include the further subset that also satisfy the DM constraints in Table~\ref{tab:externalconstraints}. For the BinoDM scan the relic density requirement is applied in the initial model selection stage, meaning that the only difference between the models selected with `all non-DM constraints' and `all external constraints' is the application of the direct detection constraints on the spin-dependent and spin-independent cross-sections.

\subsection{Constraints on the LSP mass}
\label{results:1D}
 
\begin{figure}
\centering
\subfloat[][EWKino scan]{\includegraphics[width=0.45\textwidth]{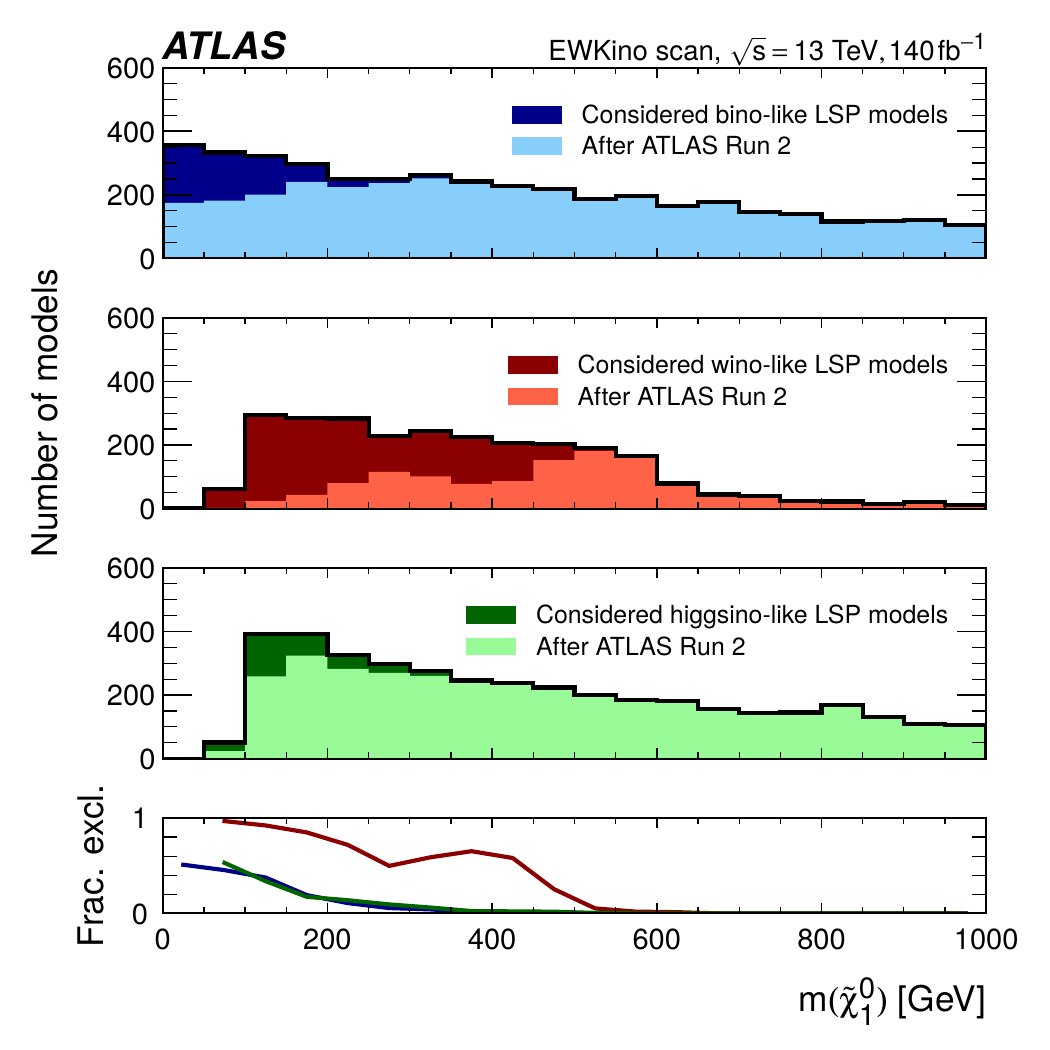}\label{1D:N1_EWKino_split}}
\qquad
\subfloat[][EWKino scan]{\includegraphics[width=0.45\textwidth]{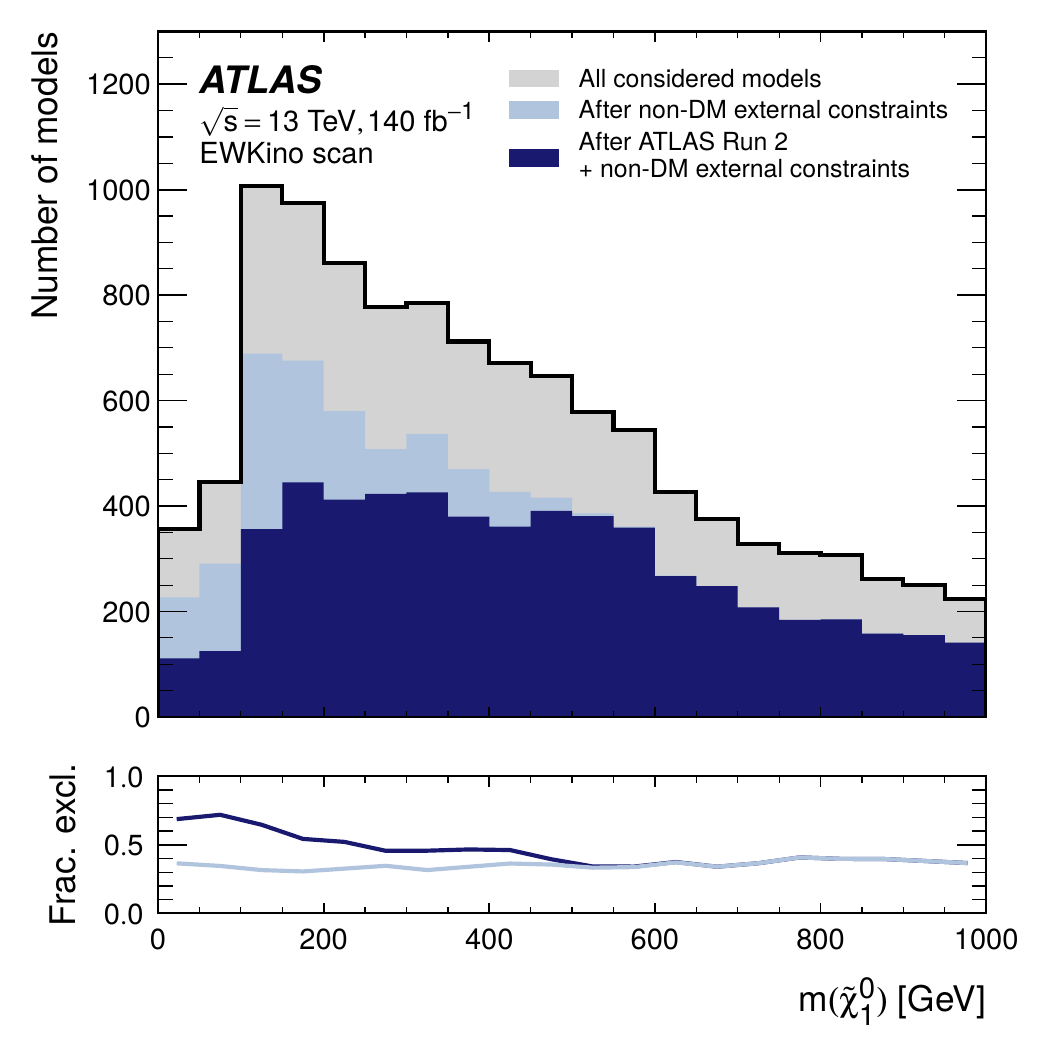}\label{1D:N1_EWKino}}
\qquad
\subfloat[][BinoDM scan]{\includegraphics[width=0.45\textwidth]{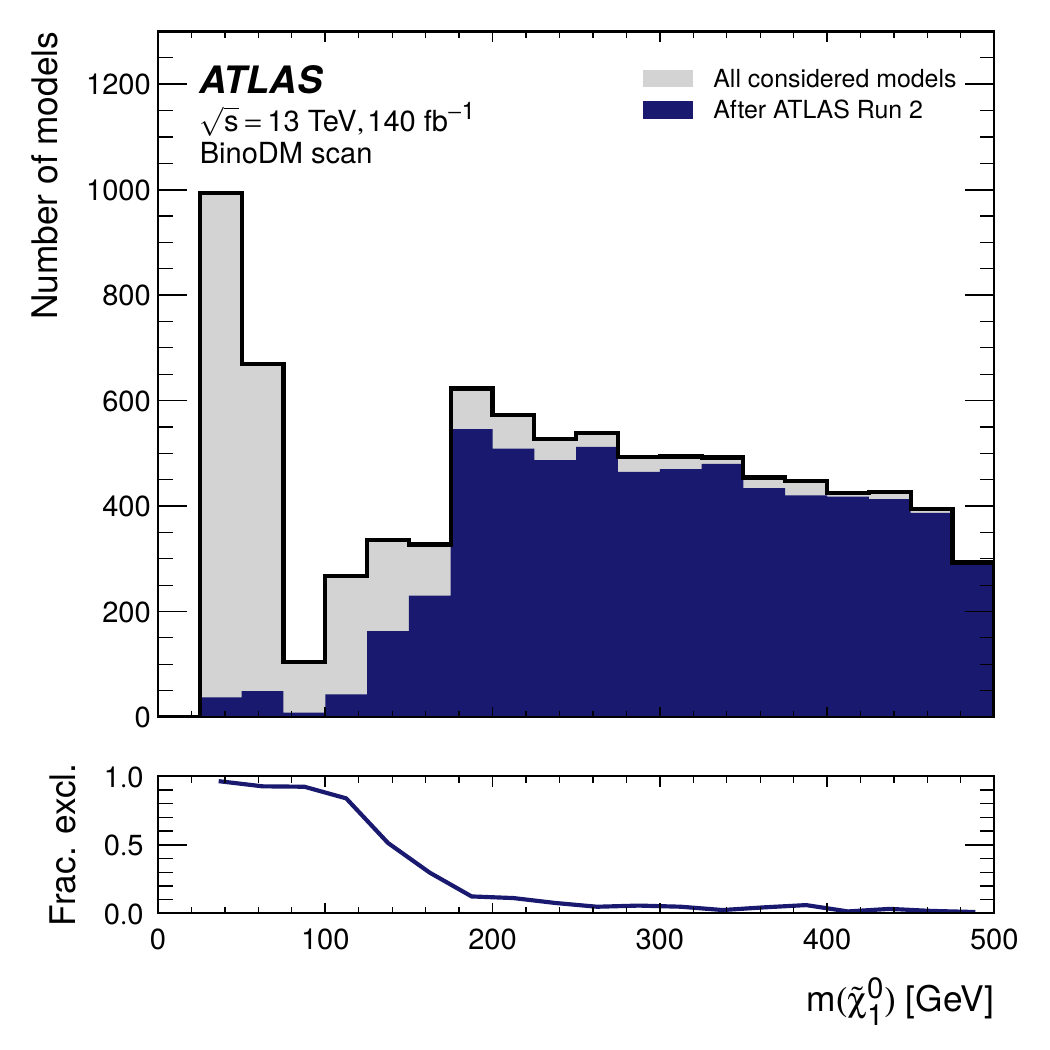}\label{1D:N1_Bino_split}}
\qquad
\subfloat[][BinoDM scan]{\includegraphics[width=0.45\textwidth]{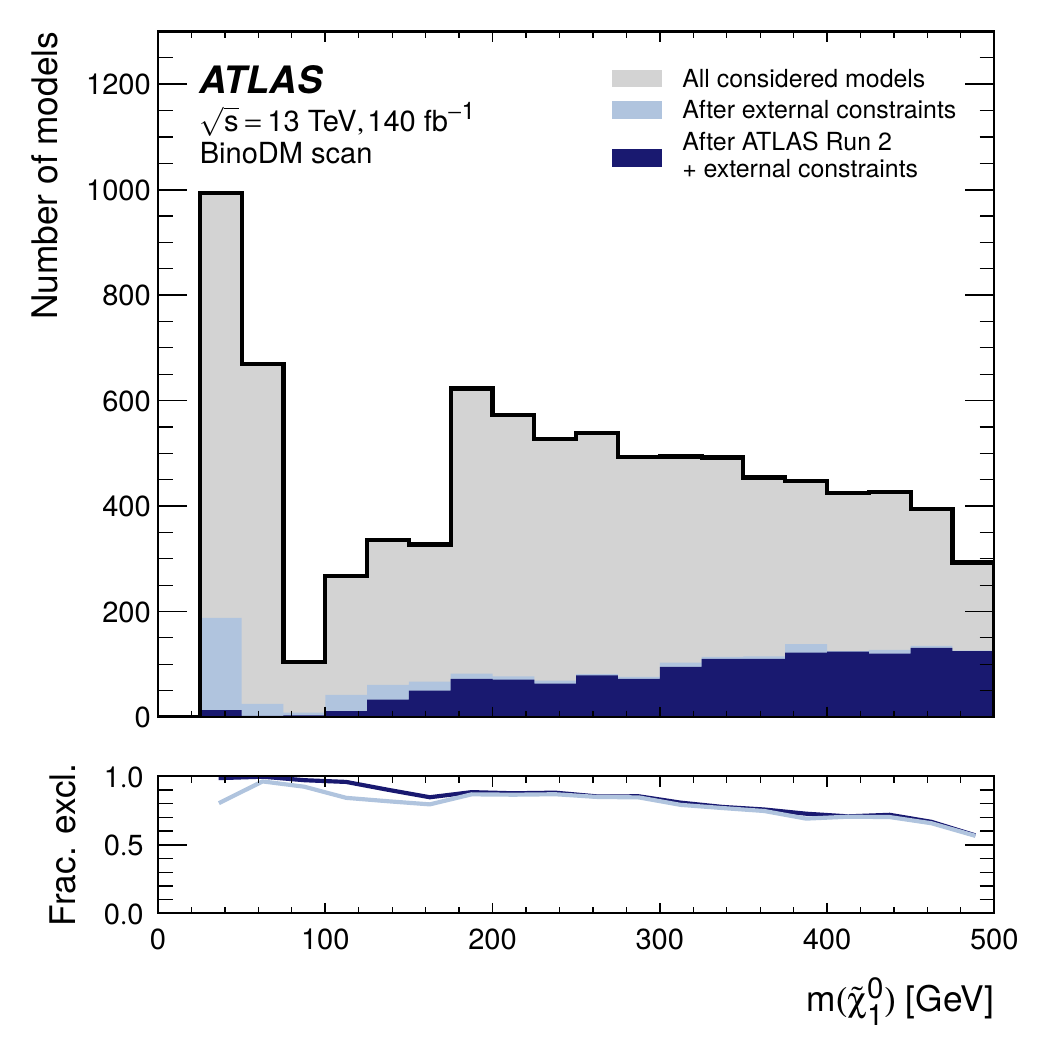}\label{1D:N1_Bino}}
\caption{
Distributions of the LSP mass.
\protect\subref{1D:N1_EWKino_split} The distribution of all considered EWKino scan models before and after the ATLAS Run~2 constraints are applied, split by the dominant component (bino, wino or higgsino) of the LSP.
\protect\subref{1D:N1_EWKino} The distribution of all considered EWKino scan models, the remaining models after non-DM external constraints are applied and the remaining models after non-DM external constraints and the ATLAS Run~2 constraints are applied.
\protect\subref{1D:N1_Bino_split} The distribution of all considered BinoDM scan models before and after the ATLAS Run~2 constraints are applied.
\protect\subref{1D:N1_Bino} The distribution of all considered BinoDM scan models, the remaining models  after all the external constraints are applied and the remaining models after all external constraints and the ATLAS Run~2 constraints are applied.
The lower panels show the fraction of models excluded in each case.
}
\label{1D:N1}
\end{figure}

Figure~\ref{1D:N1} shows one-dimensional distributions of the LSP mass for models selected by the EWKino and BinoDM scans, with the fraction of models excluded in each bin being indicated by the lower panels.
Figure~\ref{1D:N1_EWKino_split} shows all of the EWKino scan models before and after applying the ATLAS Run~2 constraints, separated by LSP type.
Figure~\ref{1D:N1_EWKino} shows the distribution of all EWKino scan models, the distribution after non-DM external constraints are applied and finally the distribution after non-DM external constraints and the ATLAS Run~2 constraints are applied.
The impact of the ATLAS constraints on the EWKino scan is most significant at lower LSP mass.
Figure~\ref{1D:N1_EWKino_split} shows that almost all of the models with an LSP mass below 100~\GeV\ are bino-like LSP models, due to the LEP chargino constraint, and around 50\% of these models are excluded by ATLAS\@.
Though not shown here, these low mass bino-LSP models are also disfavoured by the DM relic density constraint -- the complementarity between the ATLAS results and the DM external constraints is discussed separately in Section~\ref{results:DMcomplementarity}.
For an LSP mass less than 400~\GeV, the ATLAS Run~2 searches exclude more than 50\% of the wino LSP models in each bin, which is driven by the disappearing-track analysis.
The ATLAS Run~2 exclusion fractions for both bino- and higgsino-like LSPs in the EWKino scan is less than about 20\% for LSP masses above 200~\GeV.
 
Figures~\ref{1D:N1_Bino_split} and \ref{1D:N1_Bino} show the one-dimensional distribution of the LSP mass for the models in the BinoDM scan.
The distribution is shown considering only the ATLAS Run~2 constraints and when also applying the external constraints. A bino-like LSP with a mass below 100~\GeV\ is almost entirely excluded in the BinoDM scan by the ATLAS constraints, particularly when also considering external constraints. This is discussed further in the next section.

\subsection{Constraints on electroweakino masses and branching ratios}
\label{results:2D}
 
\begin{figure}[h]
\centering
\subfloat[][All considered models]{\includegraphics[width=0.47\textwidth]{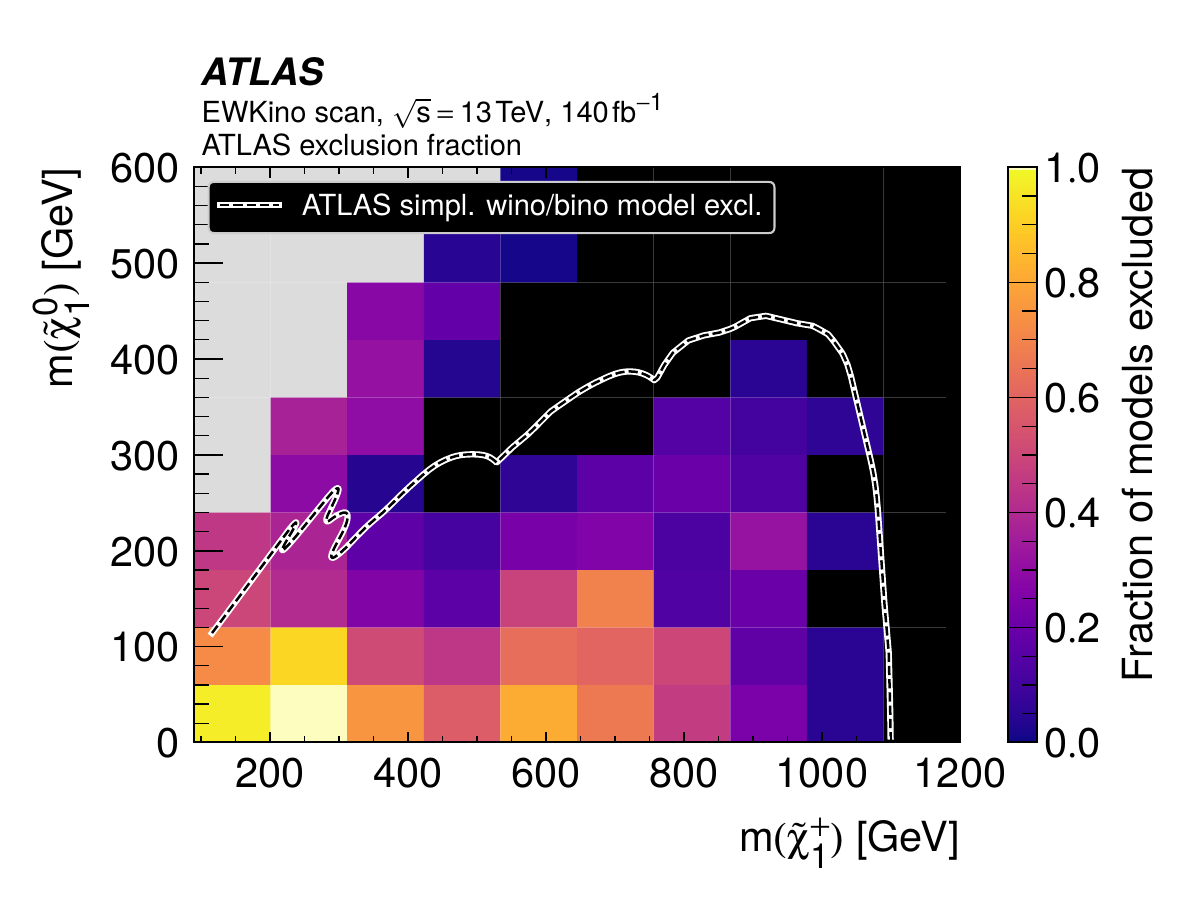}\label{excl:C1_N1_EWK_ATLAS}}
\qquad
\subfloat[][All considered models]{\includegraphics[width=0.47\textwidth]{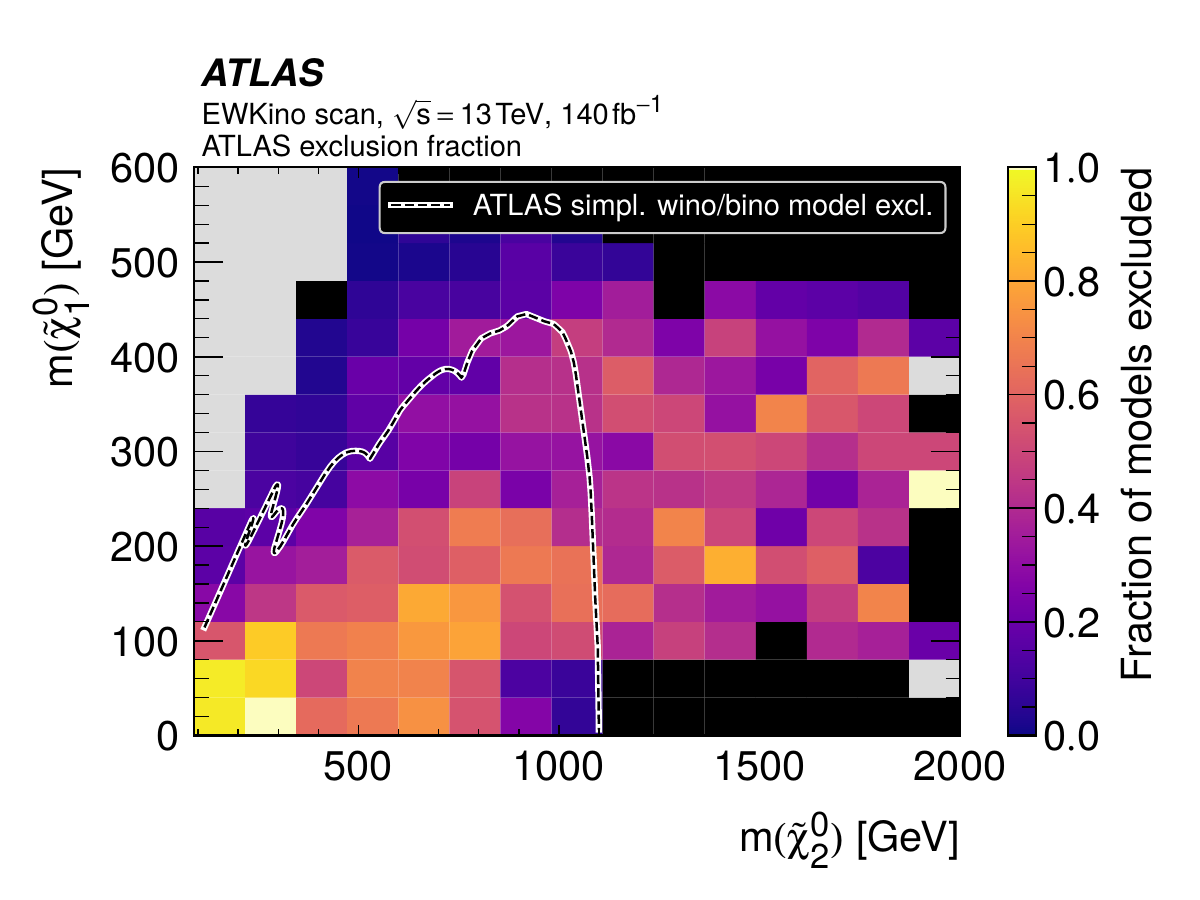}\label{excl:N2_N1_EWK_ATLAS}}
\qquad
\subfloat[][Models satisfying non-DM external constraints]{\includegraphics[width=0.47\textwidth]{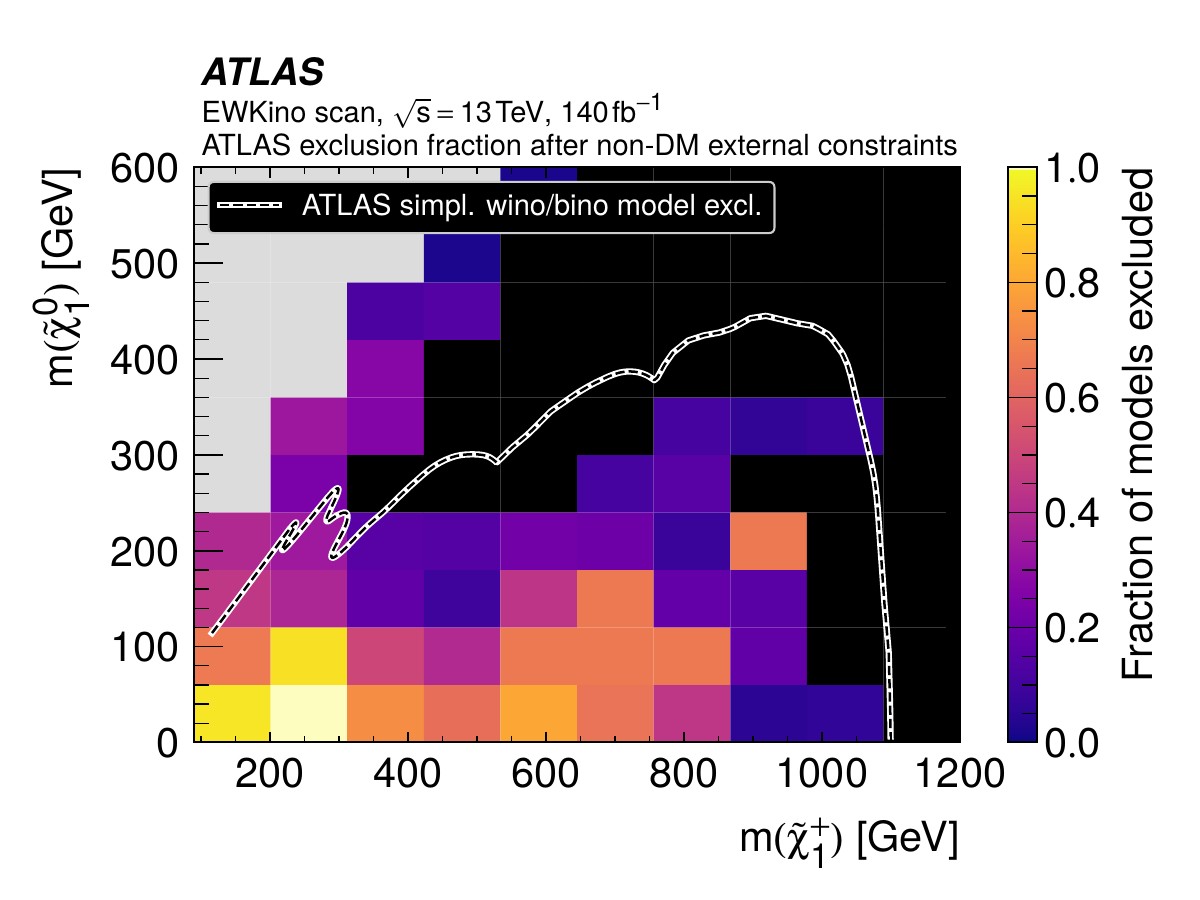}\label{excl:C1_N1_EWK_ATLASextnoDM}}
\qquad
\subfloat[][Models satisfying non-DM external constraints]{\includegraphics[width=0.47\textwidth]{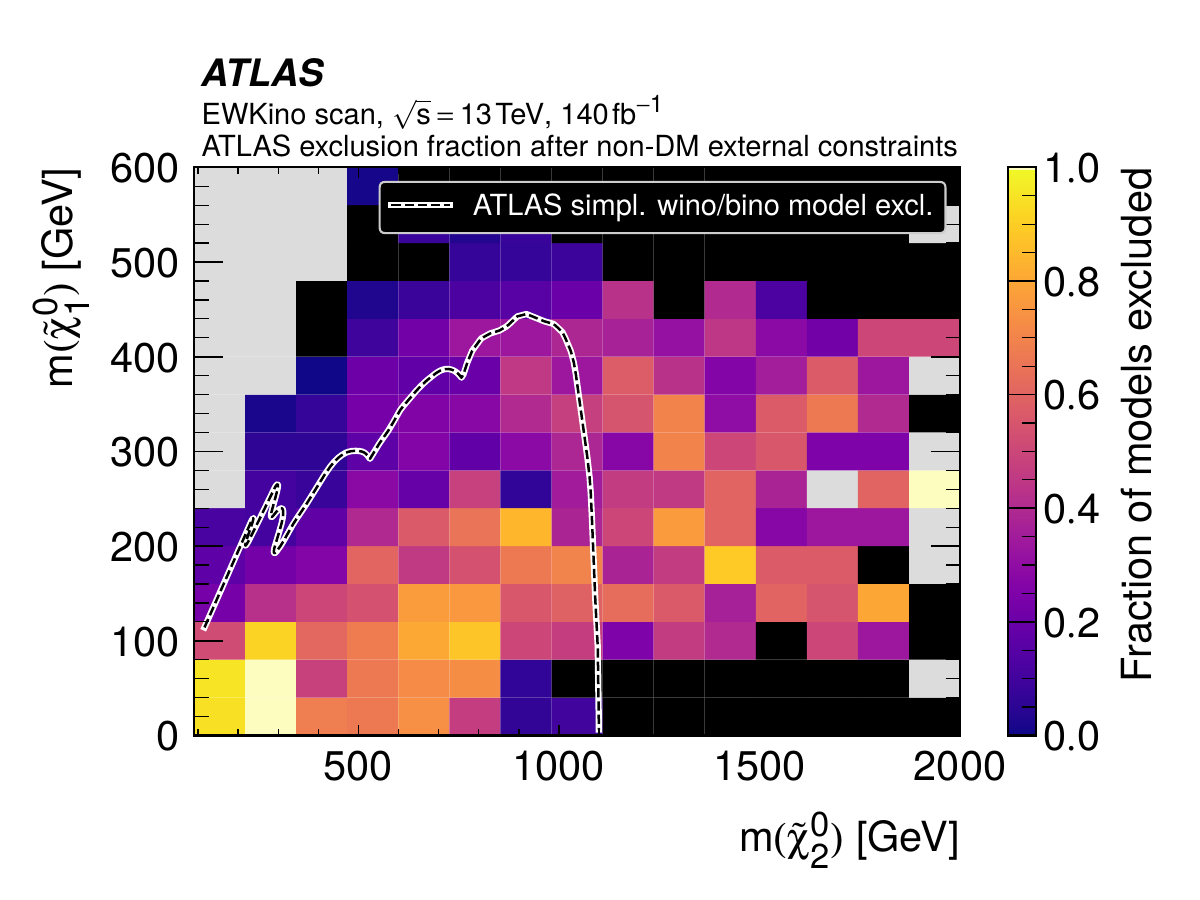}\label{excl:N2_N1_EWK_ATLASextnoDM}}
\qquad
\subfloat[][Models satisfying all external constraints]{\includegraphics[width=0.47\textwidth]{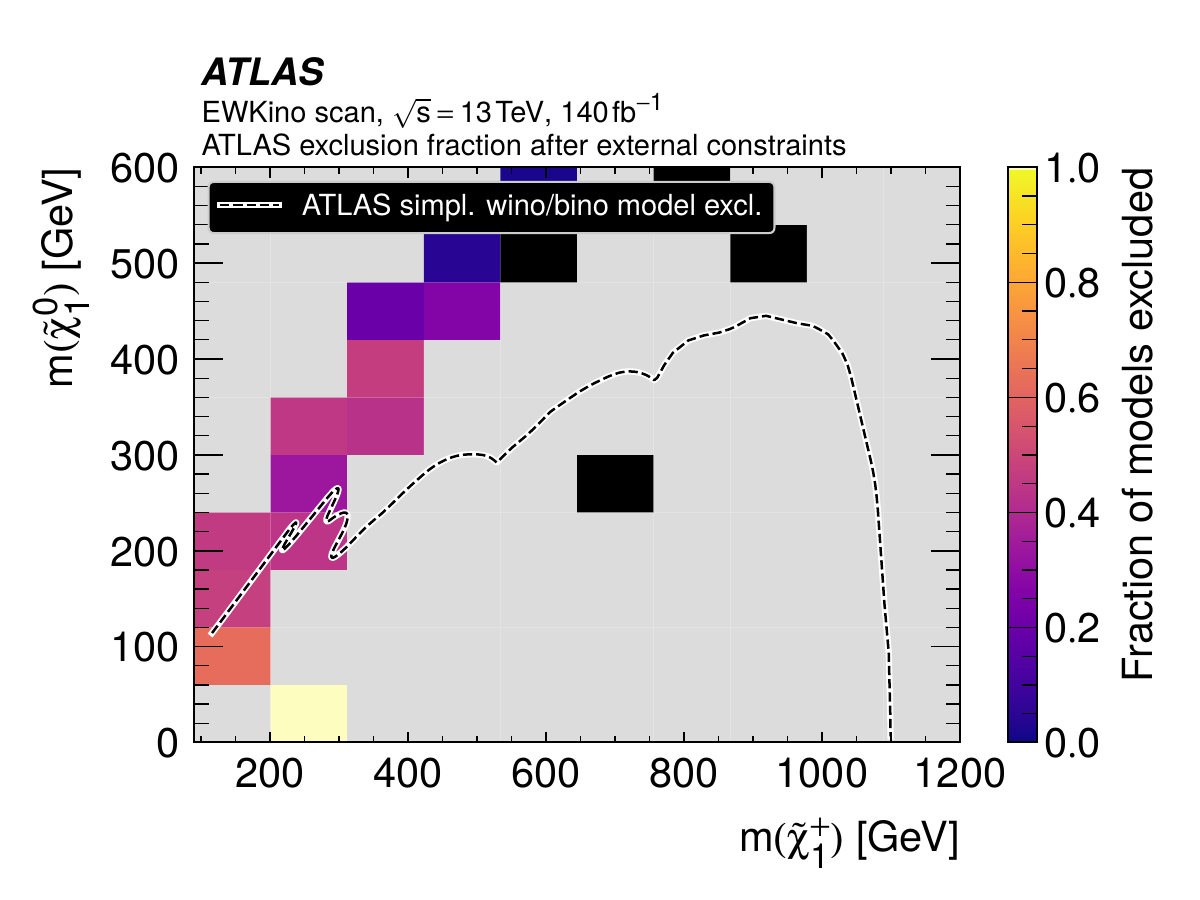}\label{excl:C1_N1_EWK_ATLASext}}
\qquad
\subfloat[][Models satisfying all external constraints]{\includegraphics[width=0.47\textwidth]{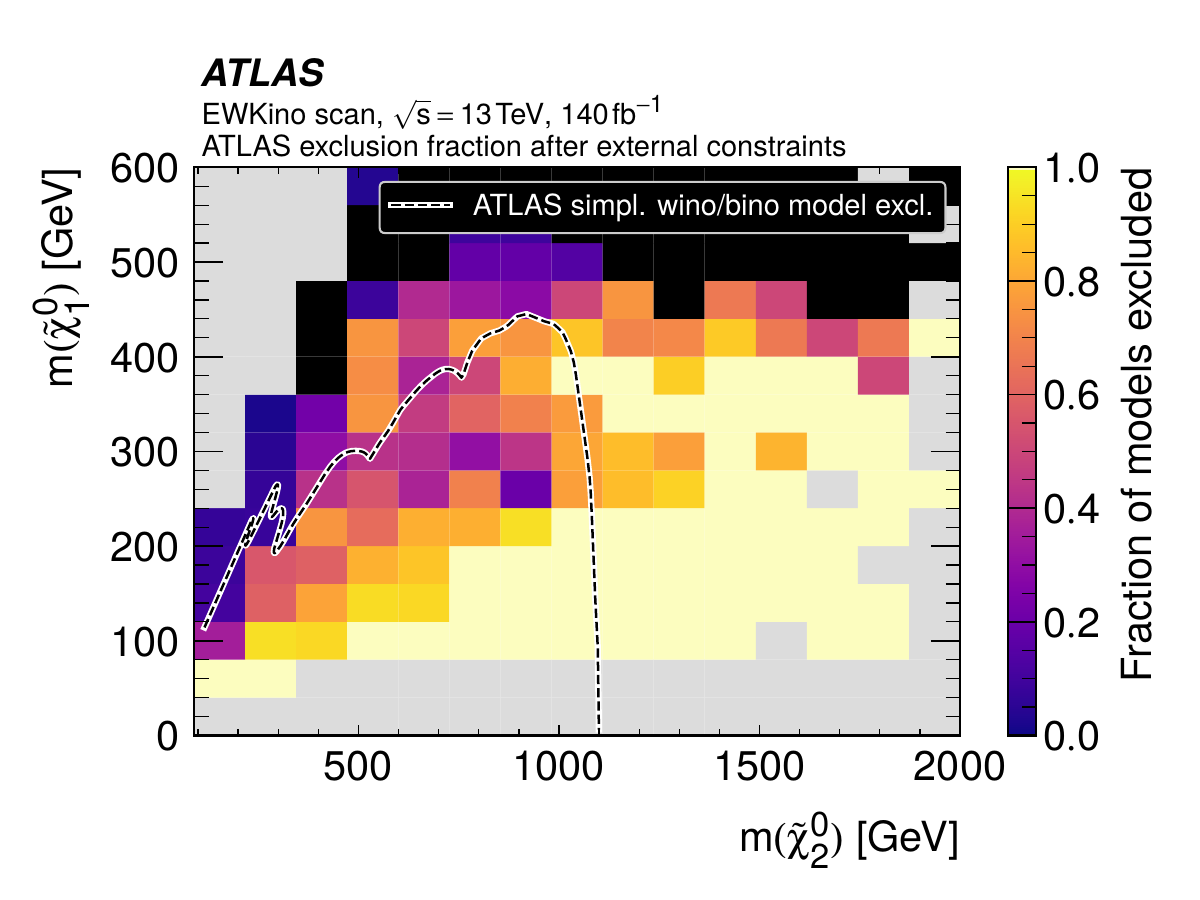}\label{excl:N2_N1_EWK_ATLASext}}
\caption{The fraction of EWKino scan models excluded by ATLAS Run~2 results. The first column shows the $m(\Ch{1})$--$m(\N{1})$ plane and the second column the $m(\N{2})$--$m(\N{1})$ plane. The first row includes all considered models, the second includes models that satisfy the non-DM external constraints and the third row models that satisfy all external constraints. The overlaid dashed line shows the envelope of the 3L~\cite{SUSY-2019-09}, 2L2J~\cite{SUSY-2018-05} and FullHad~\cite{SUSY-2018-41} exclusion of a wino \Ch{1}/\N{2} simplified model with $\mathcal{B}(\ninotwo\rightarrow Z\ninoone)=100\%$ and $\mathcal{B}(\Ch{1}\rightarrow W^\pm\ninoone)=100\%$. Bins in grey have no models to consider, while for bins in cream (black) all models are (not) excluded.}
\label{excl:EWK}
\end{figure}

For the EWKino scan, Figure \ref{excl:EWK} shows the fraction of models excluded by the ATLAS Run~2 electroweak searches in the $m(\Ch{1})$--$m(\N{1})$ and $m(\N{2})$--$m(\N{1})$ planes. The plots are overlaid with a contour indicating the exclusion for relevant simplified models. Assuming wino-like $\N{2}\Ch{1}$ and $\Ch{1}\Ch{1}$ production with a bino-like LSP, $\mathcal{B}(\N{2}\rightarrow Z\N{1})=100\%$ and $\mathcal{B}(\Ch{1}\rightarrow W^{\pm}\N{1})=100\%$, this contour is calculated as the envelope of the exclusion contours of the 3L~\cite{SUSY-2019-09}, 2L2J~\cite{SUSY-2018-05} and FullHad~\cite{SUSY-2018-41} analyses, which have complementary sensitivity for different sparticle masses.
Before considering the external constraints, most of the excluded models are inside the simplified model contours, but only at very low sparticle masses does the exclusion approach 100\%. For the $m(\Ch{1})$--$m(\N{1})$ plane, most of the models inside the contour are removed by the external constraints, primarily due to the relic density requirement suppressing most models with a bino LSP, with most of the remaining models lying along the diagonal line with small mass splittings between the lightest chargino and the LSP\@. The ATLAS Run~2 searches exclude at least 50\% of the models, even outside the simplified model contour, up to a chargino mass around 400~\GeV. For the $m(\N{2})$--$m(\N{1})$ plane, the region away from the compressed region still contains many models after the external constraints are applied, and there is a large region outside the simplified model contour with $m(\N{2})>1\,100$~\GeV\ where the ATLAS Run~2 results exclude all of the models. These are driven by models with a wino-like $\N{1}$ and $\Ch{1}$ and a bino- or higgsino-like $\N{2}$ with a large mass splitting, where the disappearing-track analysis provides sensitivity to the compressed $\N{1}$ and $\Ch{1}$.
 
\begin{figure}[h]
\centering
\subfloat[][All considered models]{\includegraphics[width=0.47\textwidth]{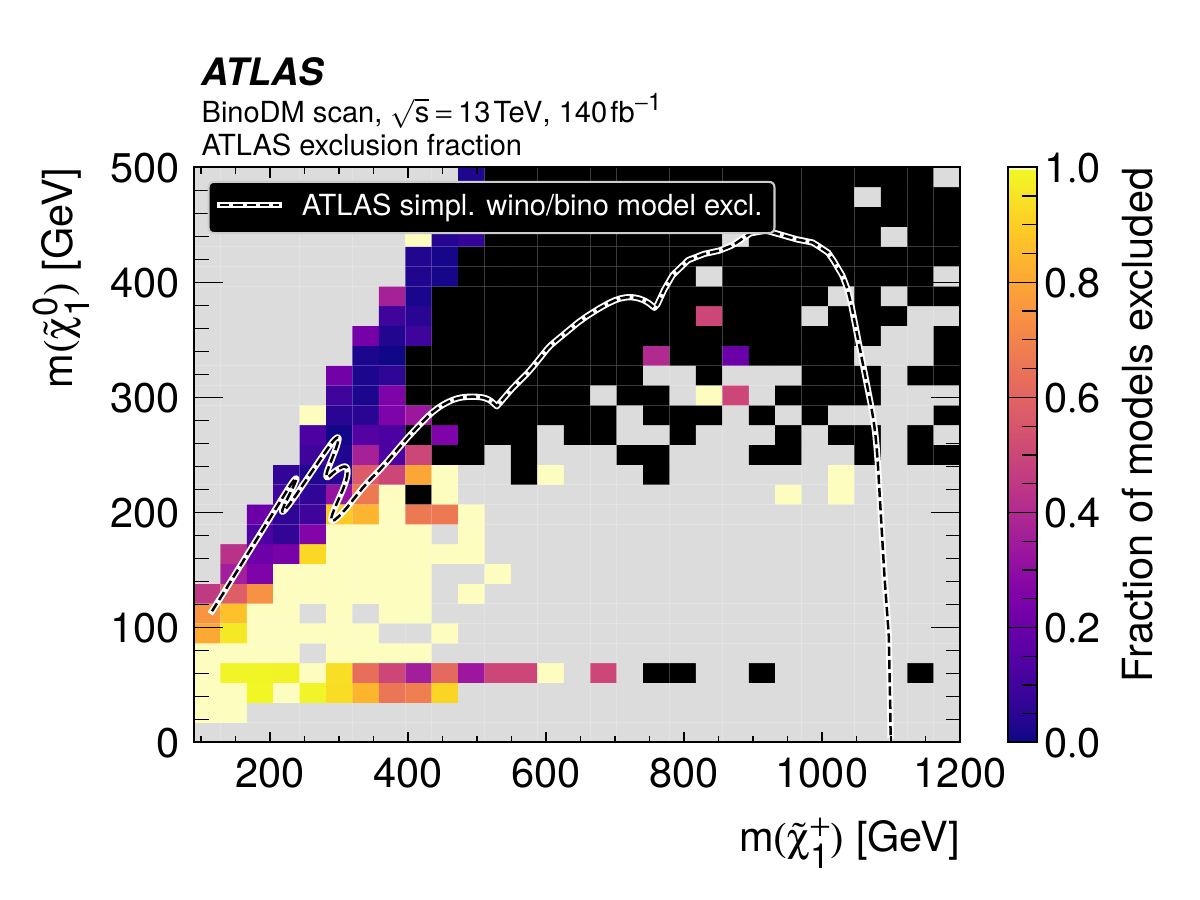}\label{excl:C1_N1_Bino_ATLAS}}
\qquad
\subfloat[][All considered models]{\includegraphics[width=0.47\textwidth]{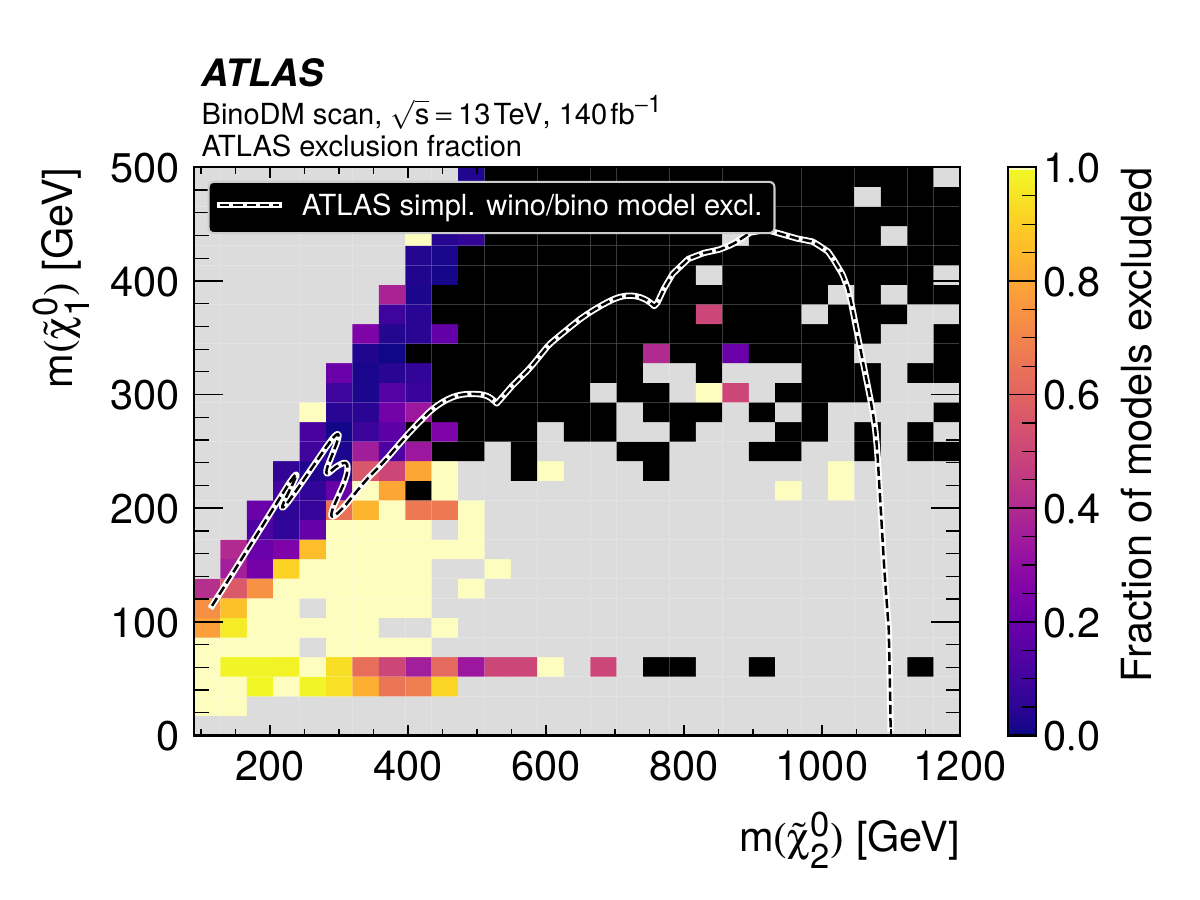}\label{excl:N2_N1_Bino_ATLAS}}
\qquad
\subfloat[][Models satisfying non-DM external constraints]{\includegraphics[width=0.47\textwidth]{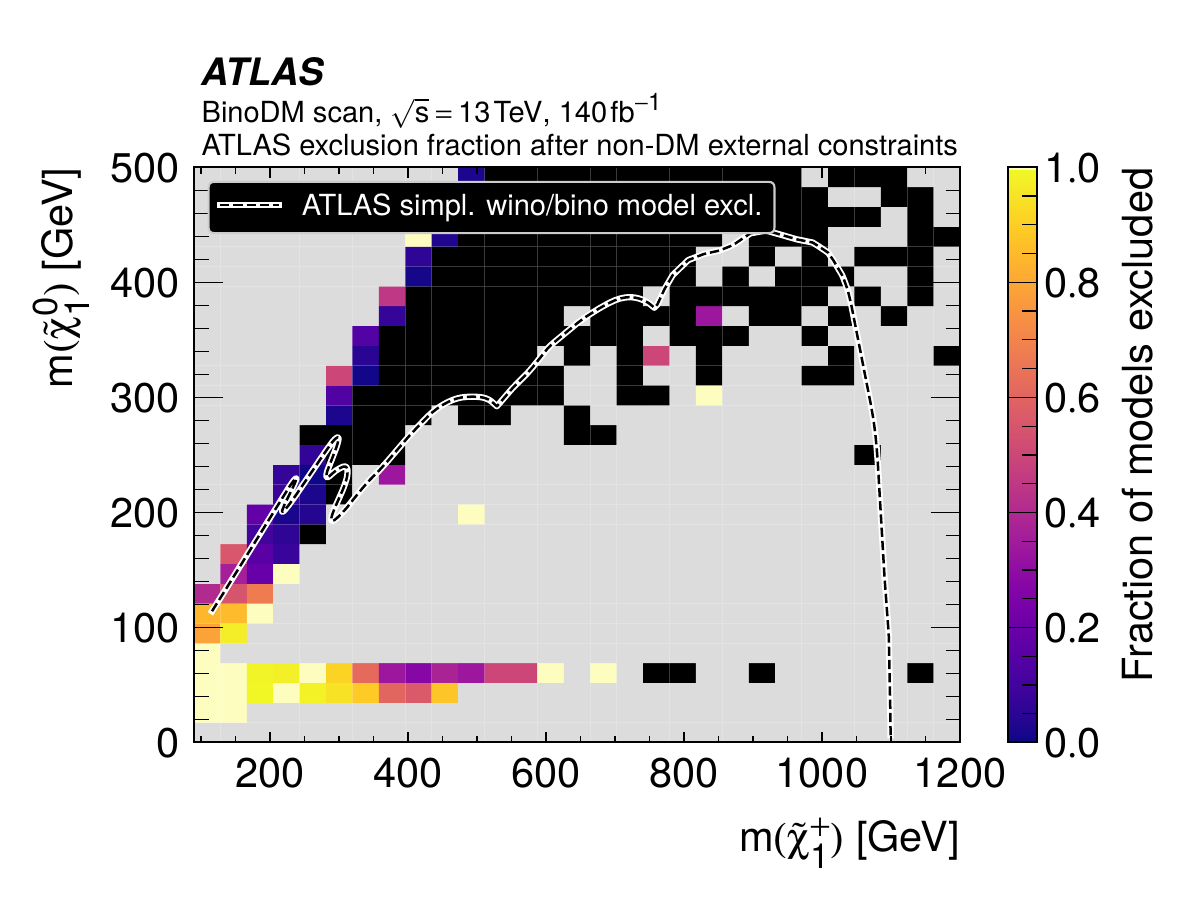}\label{excl:C1_N1_Bino_ATLASextnoDM}}
\qquad
\subfloat[][Models satisfying non-DM external constraints]{\includegraphics[width=0.47\textwidth]{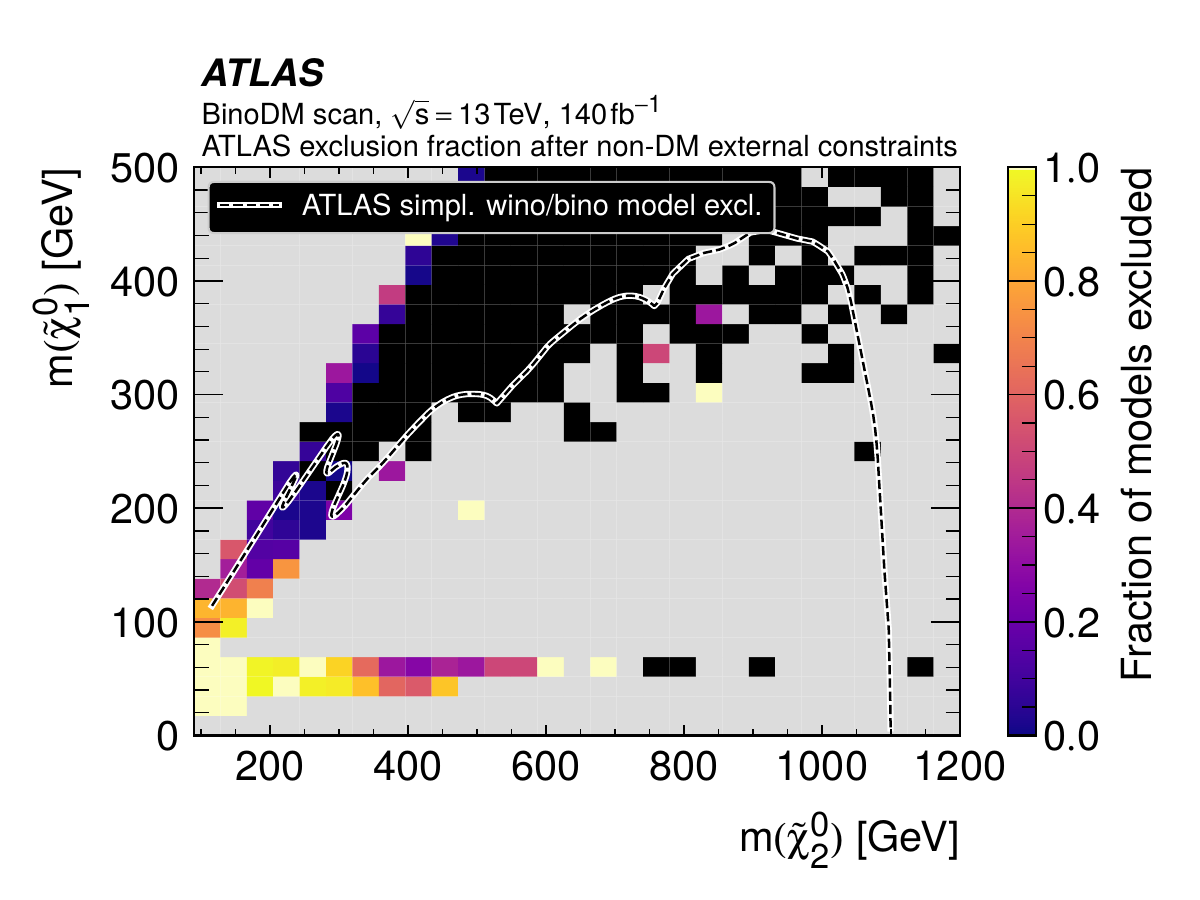}\label{excl:N2_N1_Bino_ATLASextnoDM}}
\qquad
\subfloat[][Models satisfying all external constraints]{\includegraphics[width=0.47\textwidth]{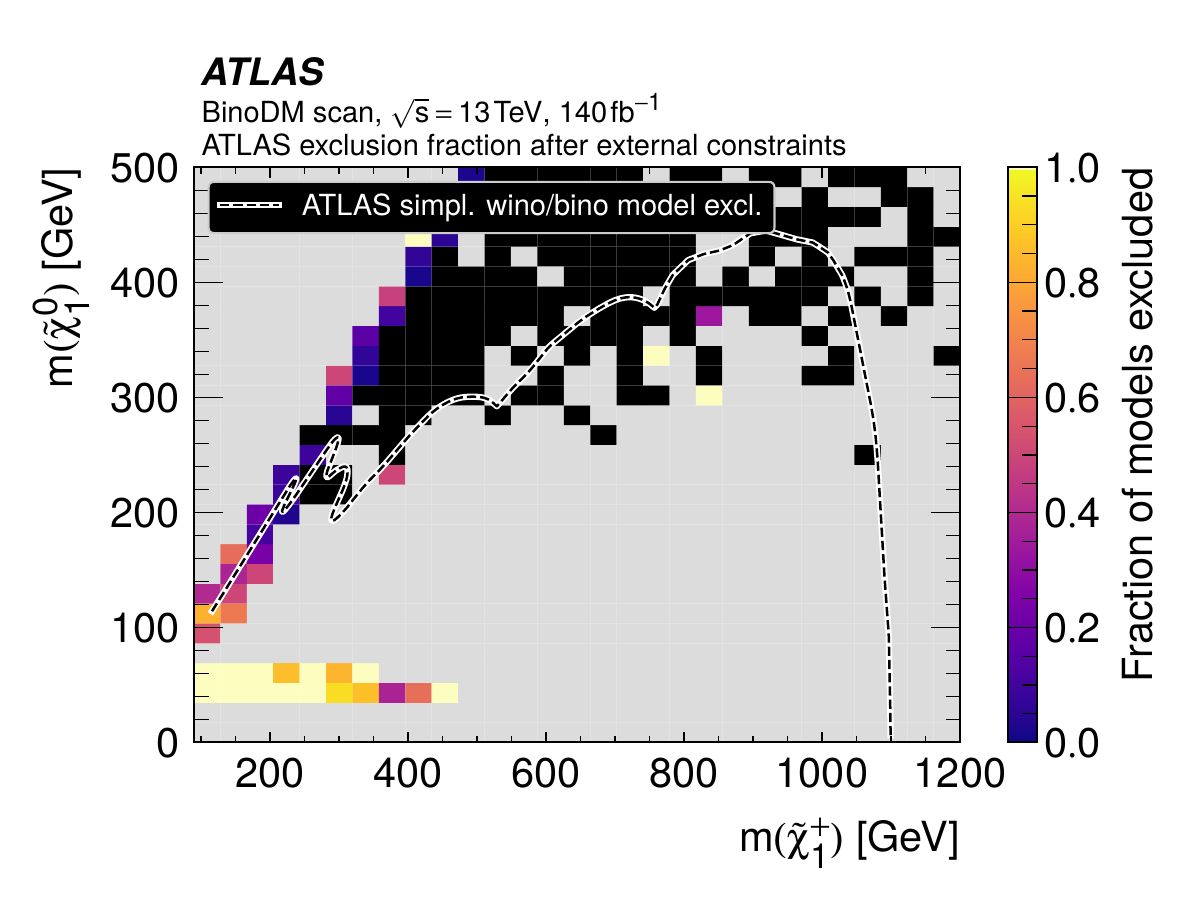}\label{excl:C1_N1_Bino_ATLASext}}
\qquad
\subfloat[][Models satisfying all external constraints]{\includegraphics[width=0.47\textwidth]{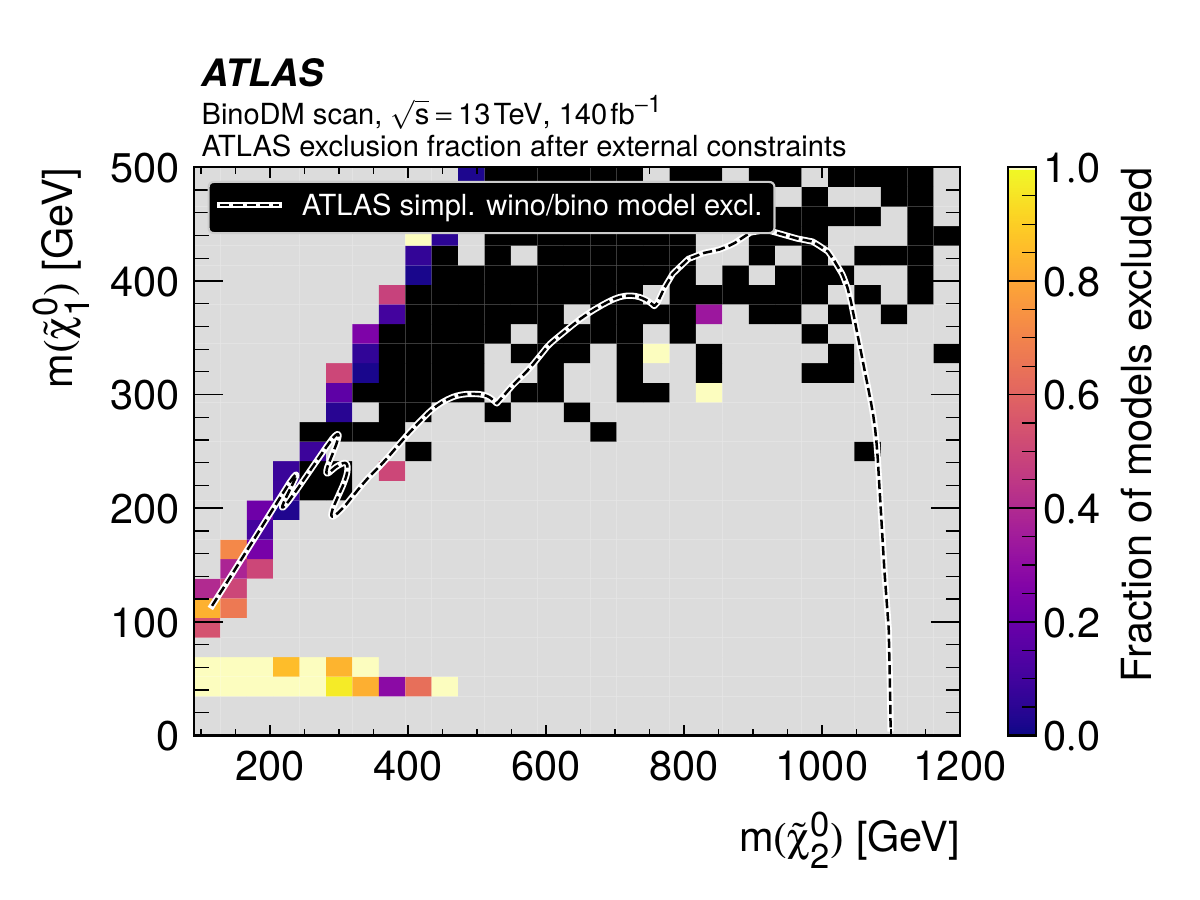}\label{excl:N2_N1_Bino_ATLASext}}
\caption{Fraction of BinoDM scan models excluded by the ATLAS Run~2 results. The first column shows the $m(\Ch{1})$--$m(\N{1})$ plane and the second column the $m(\N{2})$--$m(\N{1})$ plane. The first row includes all considered models, the second includes models that satisfy the non-DM external constraints and the third row models that satisfy all external constraints. The overlaid dashed line shows the envelope of the 3L~\cite{SUSY-2019-09}, 2L2J~\cite{SUSY-2018-05} and FullHad~\cite{SUSY-2018-41} exclusion of a wino \Ch{1}/\N{2} simplified model with $\mathcal{B}(\ninotwo\rightarrow Z\ninoone)=100\%$ and $\mathcal{B}(\Ch{1}\rightarrow W^\pm\ninoone)=100\%$. Bins in grey have no models to consider, while for bins in cream (black) all models are (not) excluded.}
\label{excl:Bino}
\end{figure}

The same plots are shown for the BinoDM scan in Figure~\ref{excl:Bino}. As the relic density requirement is applied to all models in this scan, there are fewer models inside the simplified model contour.
The $Z$/$h$ funnel regions are visible in all plots and strongly constrained by the ATLAS results.
A region of 100\% exclusion is observed between the funnel and compressed regions in Figures~\ref{excl:C1_N1_EWK_ATLAS} and \ref{excl:N2_N1_EWK_ATLAS}.
These are mostly models in the $A$-funnel region that are excluded by the ATLAS constraint $m(A)>480$~\GeV.
This region is absent in Figures~\ref{excl:C1_N1_EWK_ATLASextnoDM}-\subref{excl:N2_N1_EWK_ATLASext} due to the $\mathcal{B}$($b\rightarrow s\gamma$) external constraint that is impacted by loop contributions involving $A$.
Similar to the EWKino scan, there are models with compressed spectra, i.e., $m(\N{2})$ or $m(\Ch{1})\approx m(\N{1})$ where the ATLAS results exclude a significant fraction of the models. This illustrates the complementarity of the ATLAS searches and external constraints. To demonstrate the progress of the ATLAS Run~2 search programme relative to Run~1, Figures~\ref{excl:C1_N1_EWK_ATLASext} and~\ref{excl:C1_N1_Bino_ATLASext} can be compared with Figure~2(a) of the Run~1
electroweak reinterpretation paper~\cite{SUSY-2015-12} that reported fractional exclusion in the $m(\Ch{1})$--$m(\N{1})$ plane by the ATLAS searches for models selected using a profile-likelihood scan using external constraints. The distribution of models seen here is similar to those in the Run~1 analysis, however the ATLAS exclusion in the funnel region extends to higher chargino masses (the region with $\approx100\%$ exclusion is O(100)~\GeV\ higher using the Run~2 constraints) and there is now sensitivity to the compressed region at lower masses due to the compressed and disappearing-track analyses.
 
\begin{figure}[h]
\centering
\subfloat[][EWKino scan ($\Delta m(\Ch{1},\N{1})<5$ \GeV)]{\includegraphics[width=0.47\textwidth]{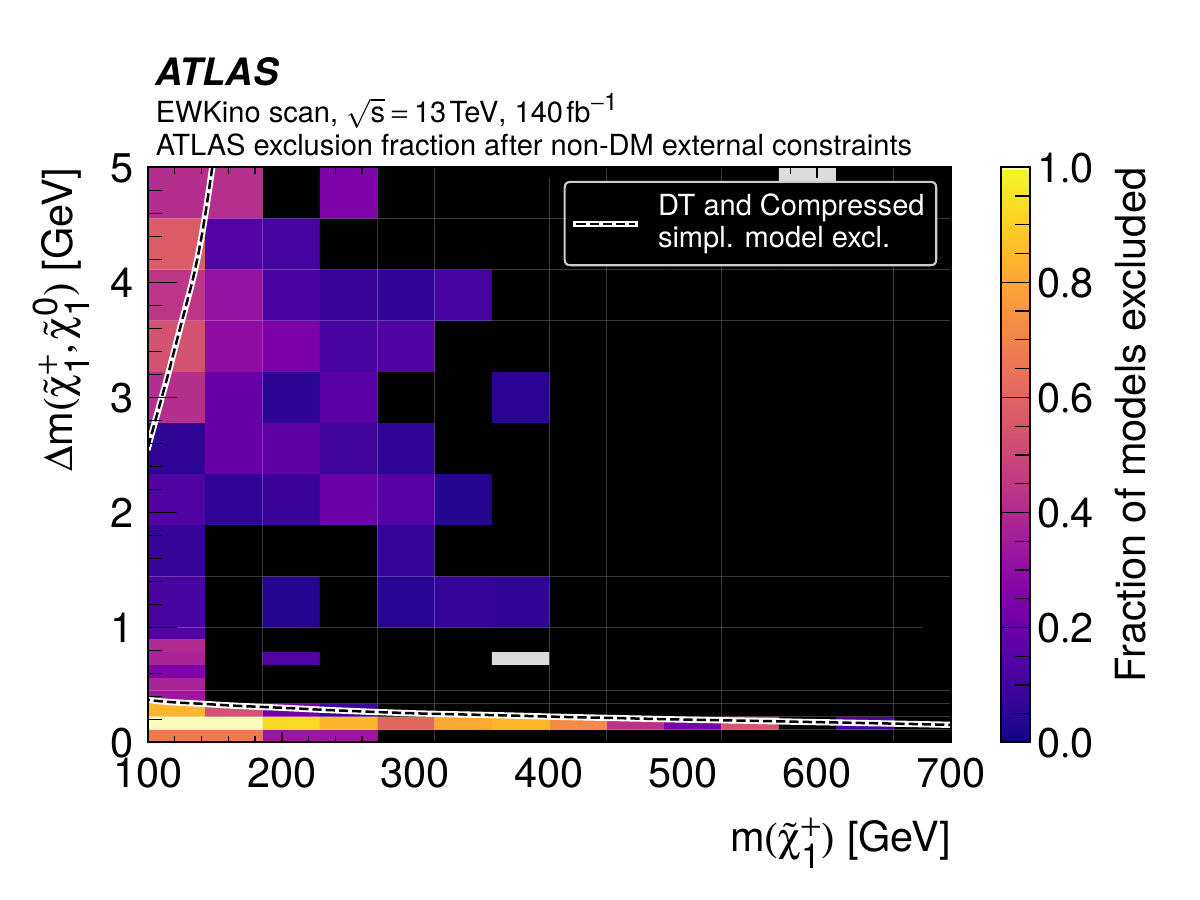}\label{excl:dMC1N1_EWK_ATLASext}}
\qquad
\subfloat[][BinoDM scan ($\Delta m(\Ch{1},\N{1})<20$ \GeV)]{\includegraphics[width=0.47\textwidth]{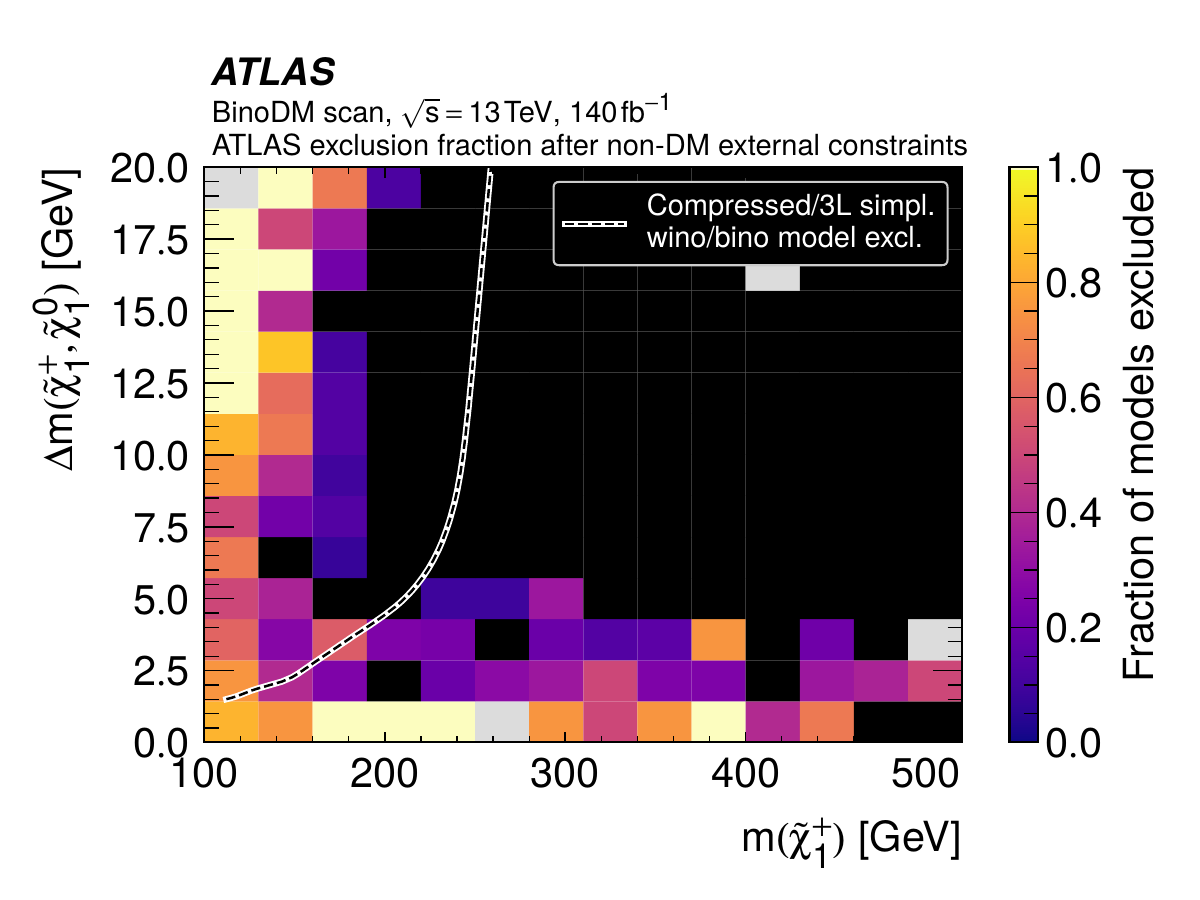}\label{excl:dMC1N1_Bino_ATLASext}}
\qquad
\caption{Fraction of EWKino (left) and BinoDM (right) scan models excluded in the $m(\Ch{1})$--$\Delta m(\Ch{1},\N{1})$ plane. Only models satisfying the non-DM external constraints are included. \protect\subref{excl:dMC1N1_EWK_ATLASext} is overlaid with the disappearing-track~\cite{SUSY-2018-19} and compressed~\cite{SUSY-2018-16} exclusion of a higgsino simplified model. \protect\subref{excl:dMC1N1_Bino_ATLASext} is overlaid with the envelope of the compressed~\cite{SUSY-2018-16} and 3L~\cite{SUSY-2019-09} exclusion of a wino/bino $\Ch{1}\N{2}\rightarrow W Z \N{1}\N{1}$ simplified model. Bins in grey have no models to consider, while for bins in cream (black) all models are (not) excluded.}
\label{excl:deltaM}
\end{figure}

\begin{figure}[h]
\centering
\subfloat[][EWKino scan]{\includegraphics[width=0.47\textwidth]{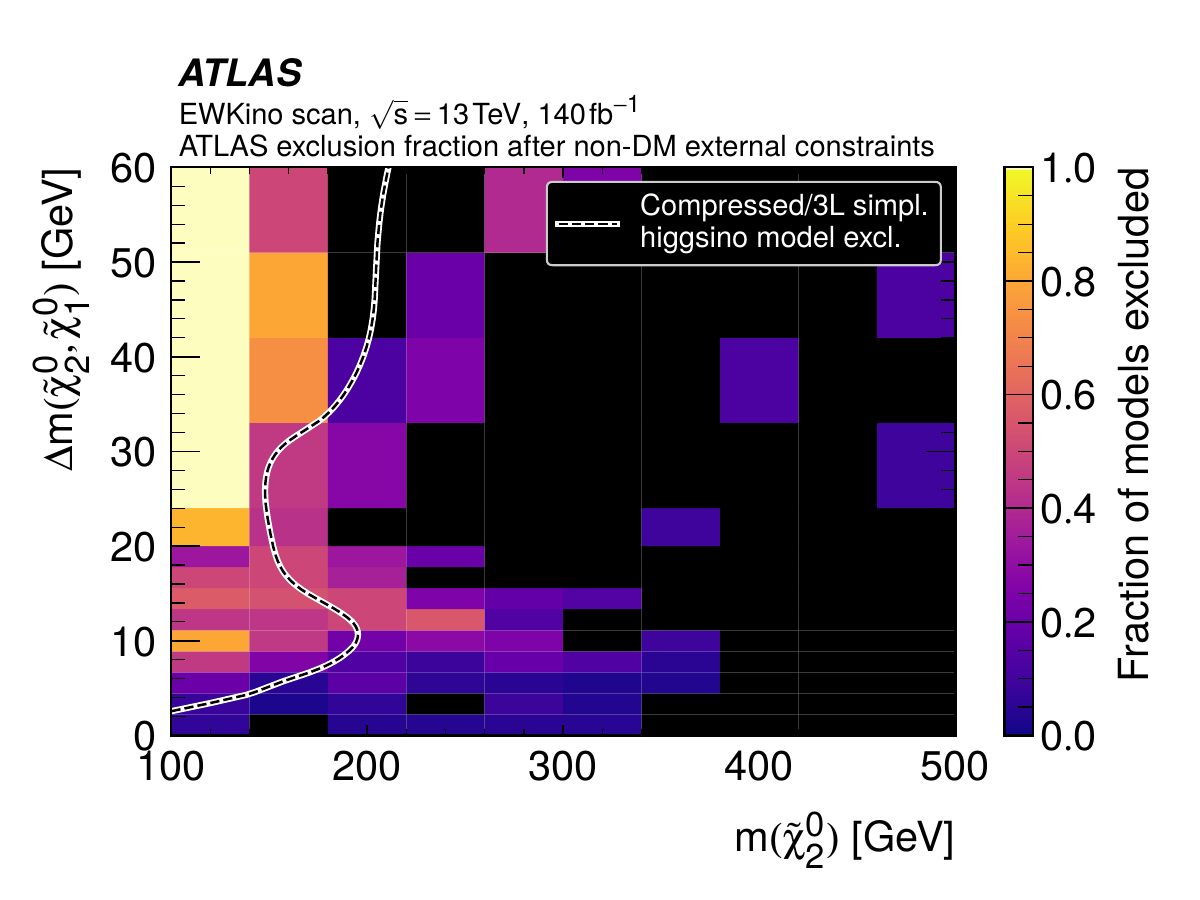}\label{excl:dMN2N1_EWK_ATLASext}}
\qquad
\subfloat[][BinoDM scan]{\includegraphics[width=0.47\textwidth]{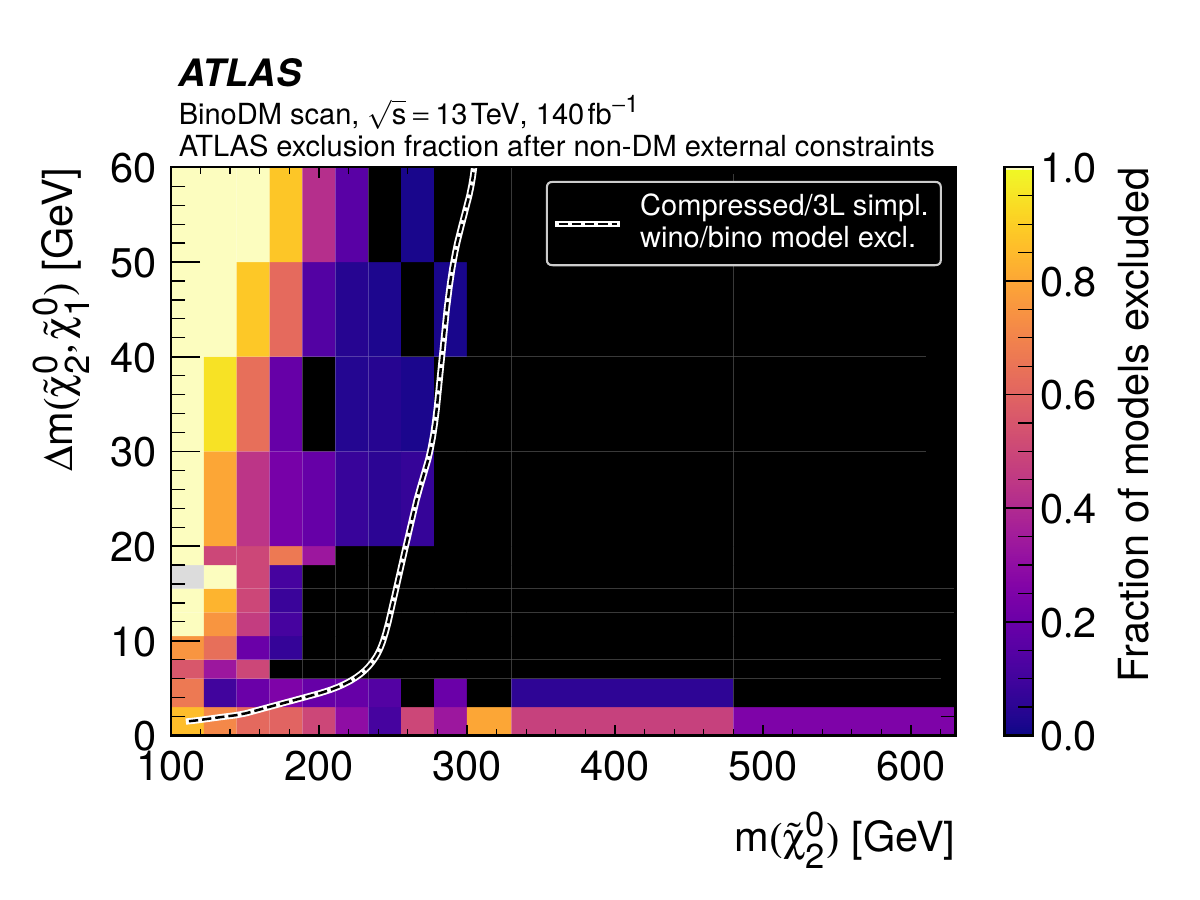}\label{excl:dMN2N1_Bino_ATLASext}}
\caption{Fraction of EWKino (left) and BinoDM (right) scan models excluded in the $m(\N{2})$--$\Delta m(\N{2},\N{1})$  plane. Only models satisfying the non-DM external constraints are included. The overlaid dashed lines show the envelope of the compressed~\cite{SUSY-2018-16} and 3L~\cite{SUSY-2019-09} exclusion of relevant higgsino or wino/bino $\Ch{1}\N{2}\rightarrow W Z \N{1}\N{1}$ simplified models. Bins in grey have no models to consider, while for bins in cream (black) all models are (not) excluded.}
\label{excl:N2_dM}
\end{figure}

Figures~\ref{excl:EWK} and~\ref{excl:Bino} illustrate that once external constraints are applied, many of the surviving models have compressed mass spectra.
Figure~\ref{excl:deltaM} shows the fractions of models excluded in the $\Delta m(\Ch{1},\N{1})=m(\Ch{1})-m(\N{1})$ versus $m(\Ch{1})$ plane once non-DM external constraints are applied.
For both scans the ATLAS results have sensitivity to low chargino masses.
For the EWKino scan (left plot), the highest exclusion for low chargino masses occurs for $\Delta m(\Ch{1},\N{1}) \approx$ 0.1--0.2~\GeV, which is consistent with the O(160 \MeV) mass splittings expected for pure wino scenarios that the disappearing-track analysis is optimised for~\cite{SUSY-2018-19}.
For the BinoDM scan (right plot), at chargino masses less than 400~\GeV\ the lowest $\Delta m(\Ch{1},\N{1})$ bins show exclusion fractions greater than 0.8. This can be attributed both to the disappearing-track analysis and to production mechanisms involving higher mass electroweakinos such as $\Ch{2}$ and $\N{4}$. A similar effect is seen in Figure~\ref{excl:N2_dM} that shows the fractions of models excluded for each scan in the $m(\N{2})$--$\Delta m(\N{2},\N{1})$ plane. In general, many bins within the simplified model contours in all the two-dimensional plots show less than 100\% exclusion.
This is mostly due to the pMSSM models having smaller branching fractions and cross-sections than the simplified models.
 
\begin{figure}[h!]
\centering
\subfloat[][EWKino scan]{\includegraphics[width=0.47\textwidth]{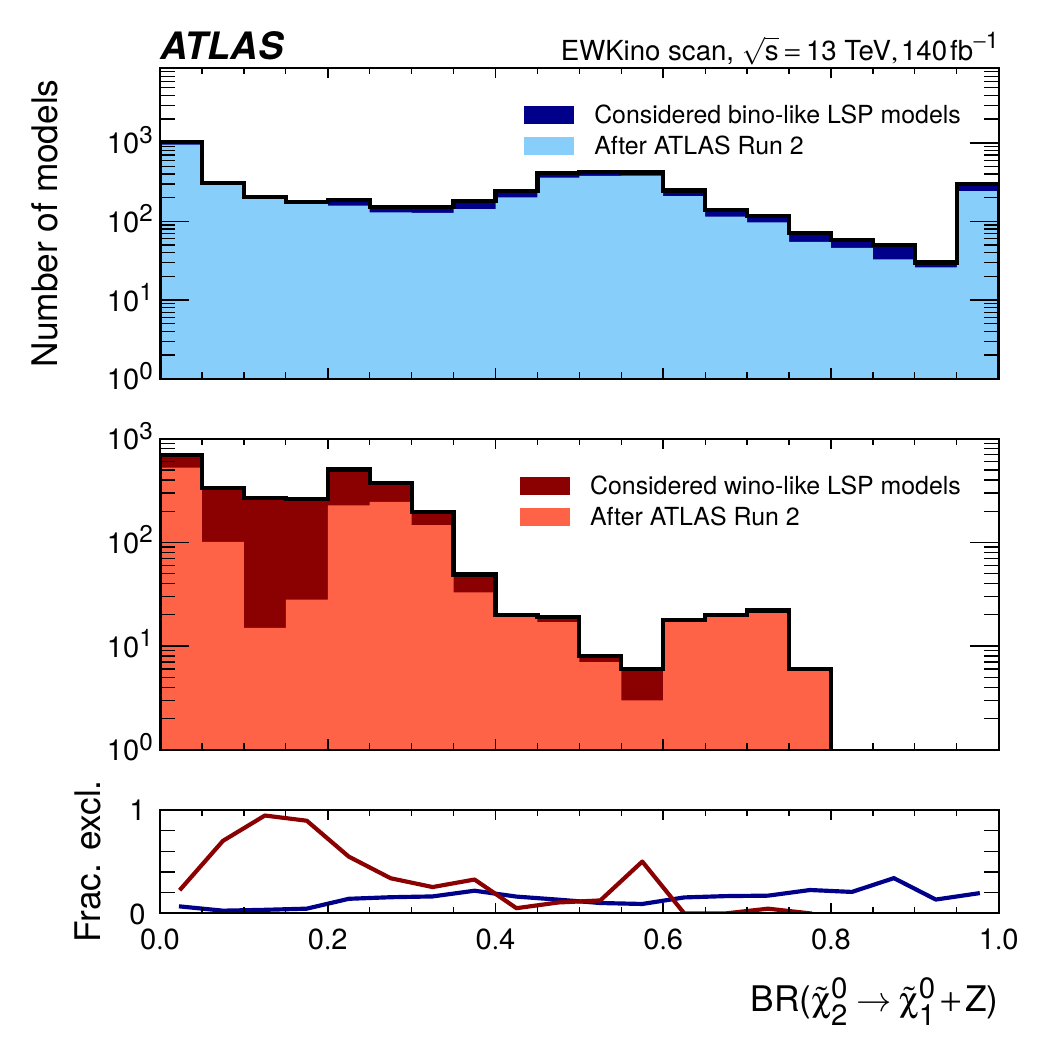}\label{1D:BRZ_EWKino_ext}}
\qquad
\subfloat[][BinoDM scan]{\includegraphics[width=0.47\textwidth]{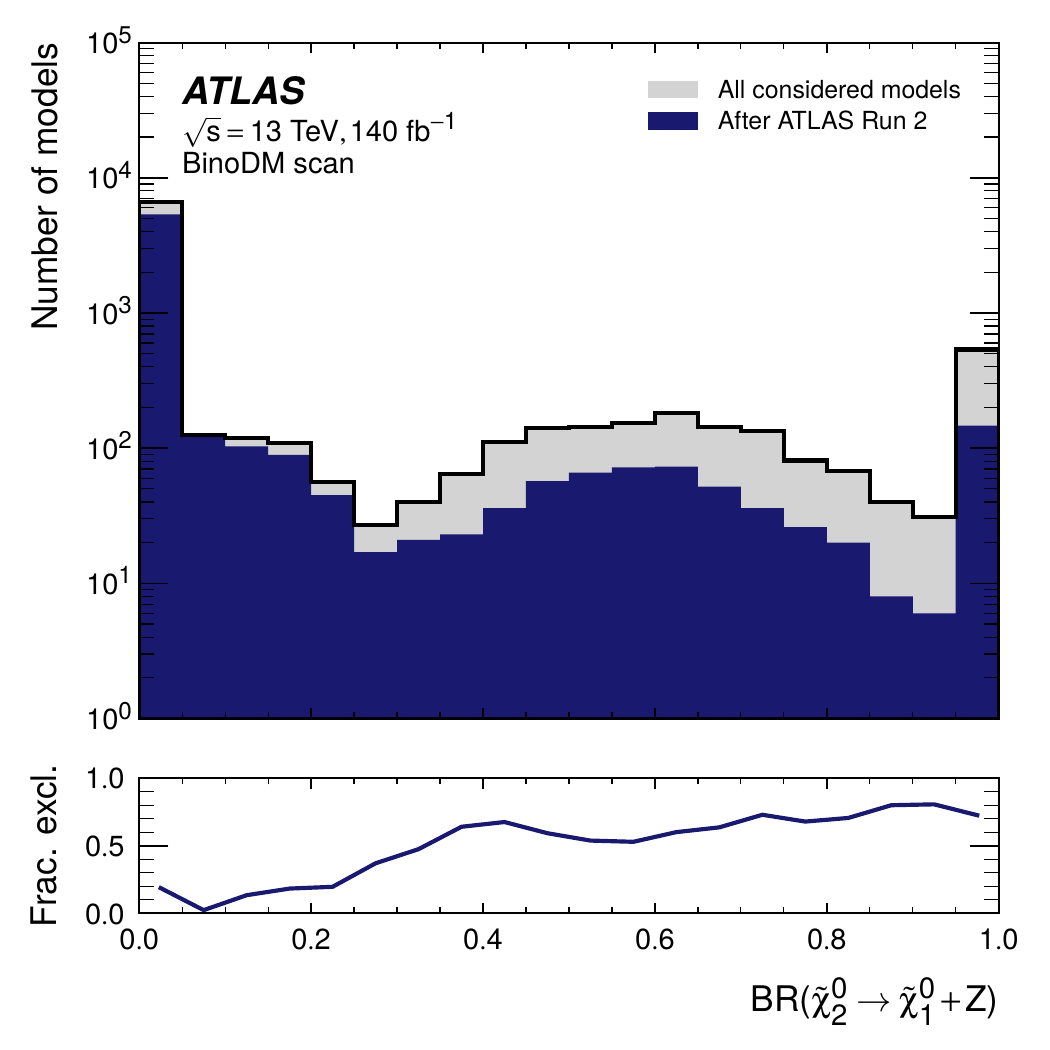}\label{1D:BRZ_Bino_ext}}
\qquad
\subfloat[][EWKino scan]{\includegraphics[width=0.47\textwidth]{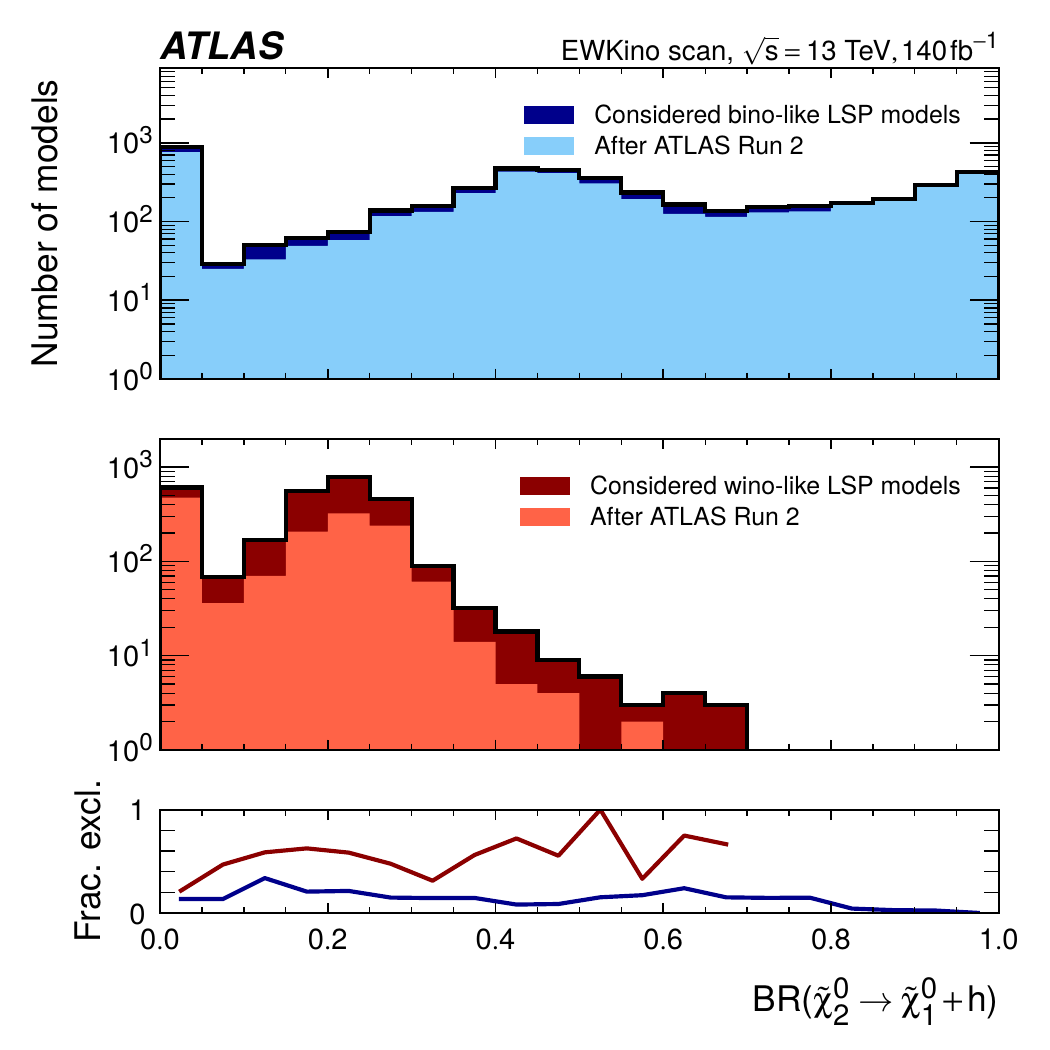}\label{1D:BRh_EWKino_ext}}
\qquad
\subfloat[][BinoDM scan]{\includegraphics[width=0.47\textwidth]{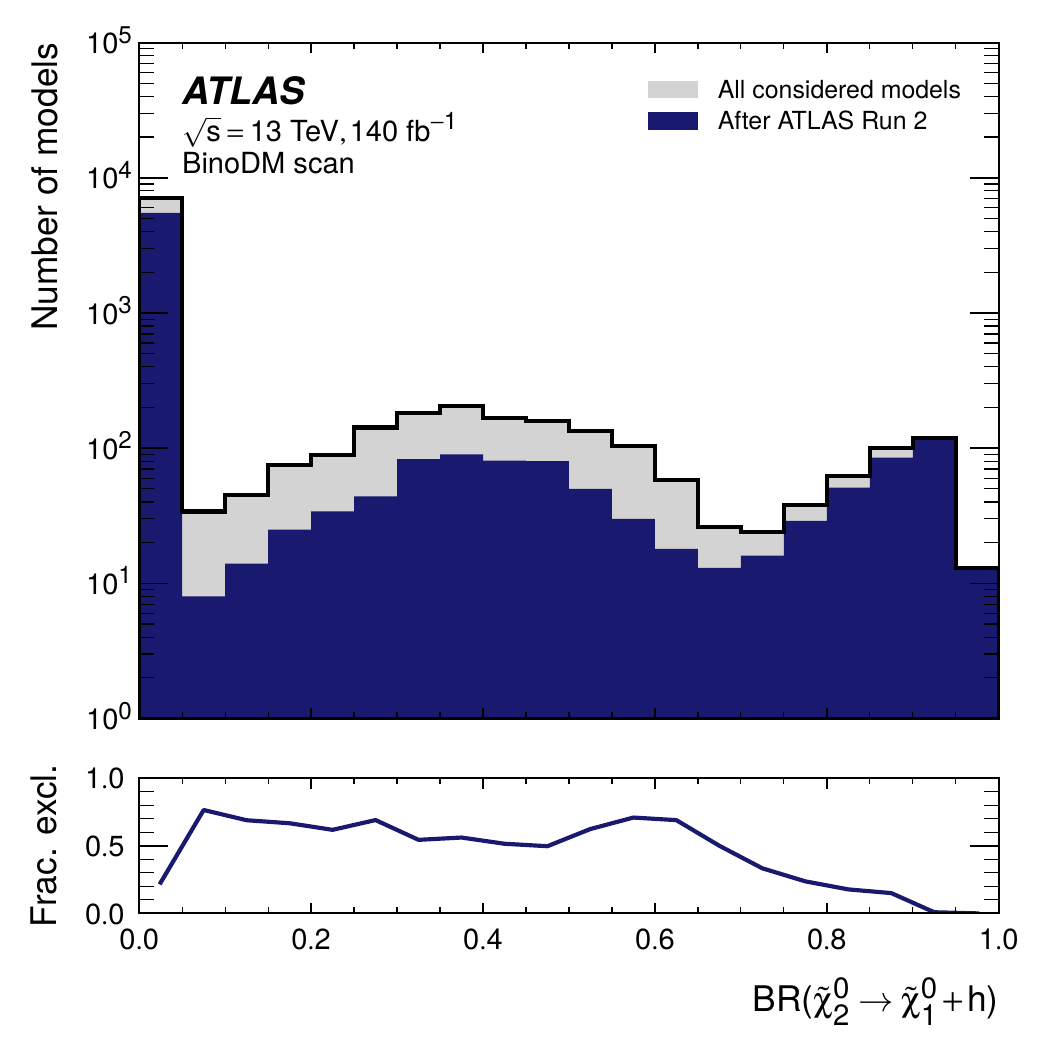}\label{1D:BRh_Bino_ext}}
\caption{Distributions of (first row) $\mathcal{B}$($\N{2}\rightarrow\N{1} Z$) and (second row) $\mathcal{B}$($\N{2}\rightarrow\N{1} h$). The first column shows models in the EWKino scan split by LSP type and the second column models in the BinoDM scan. The distributions show all considered models before and after the ATLAS Run~2 constraints are applied. The lower panels show the fraction of models excluded in each case.}\label{1D:BRs}
\end{figure}

Simplified models used in SUSY searches typically assume fixed branching ratios into particular final states (often 100\%), however in general these vary as a function of the pMSSM parameters. Figure~\ref{1D:BRs} shows one-dimensional distributions of $\mathcal{B}(\N{2}\rightarrow\N{1} Z)$ and $\mathcal{B}(\N{2}\rightarrow\N{1} h)$ for both scans before and after applying the ATLAS Run~2 constraints (no external constraints are considerered here). These quantities are branching fractions for two-body decays via on-shell $Z$ and $h$ bosons, therefore the large peaks at zero in these distributions contain models with $\Delta m(\N{2},\N{1})<m(Z/h)$ such that the three-body off-shell decay modes dominate.
For the EWKino scan, Figures~\ref{1D:BRZ_EWKino_ext} and \ref{1D:BRh_EWKino_ext} are split by LSP type as the distribution and ATLAS sensitivity is distinct for each. These on-shell decay modes are not relevant to models with a higgsino-like LSP, so these are not shown.
For models with a bino-like LSP in both the EWKino and BinoDM scan, the highest exclusion by the ATLAS results occurs for large values of $\mathcal{B}(\N{2}\rightarrow\N{1} Z)$, as expected due to the large number of searches optimised for simplified models with $\mathcal{B}(\N{2}\rightarrow\N{1} Z)=100\%$ whilst only the 1Lbb search directly targets the $\N{2}\rightarrow\N{1} h$ decay for low mass $\N{2}$.
For models with a wino-like LSP, the largest exclusion occurs when  $\mathcal{B}(\N{2}\rightarrow\N{1} Z)\approx15\%$ and larger $\mathcal{B}(\N{2}\rightarrow\N{1} h)$.
These branching ratios are impacted by the mass splittings between the \N{2} and \N{1}, which also affects the lifetime of the \Ch{1}. This peak in exclusion corresponds to models with a chargino lifetime of around $0.1$ns that aligns with the region of sensitivity of the disappearing-track analysis.

\FloatBarrier
 
\subsection{Dark matter phenomenology}
\label{results:DMcomplementarity}
 
This section assesses the impact of the searches on models with particular DM annihilation phenomenology and  investigates the complementarity between collider and non-collider searches for DM by considering the relic density $\Omega h^2$ and DM--nucleon scattering cross-sections. Most direct DM detection experiments target DM scattering on nucleons. Currently, the most stringent limit on the spin-independent WIMP--nucleon scattering cross-section comes from the LUX-ZEPLIN (LZ) experiment~\cite{LUX-ZEPLIN:2022xrq} whilst the most stringent constraint on the spin-dependent WIMP--proton scattering cross-section comes from the PICO-60 experiment~\cite{PICO:2019vsc}. The impact of these limits on the models considered here is weakened by the scaling of the cross-sections by $(\Omega h^2 / 0.12)$ to account for lower than measured DM relic density as  explained in Section~\ref{sec:framework}.
 
\begin{figure}
\centering
 
\subfloat[][Models satisfying all non-DM constraints]{\includegraphics[width=0.45\textwidth]{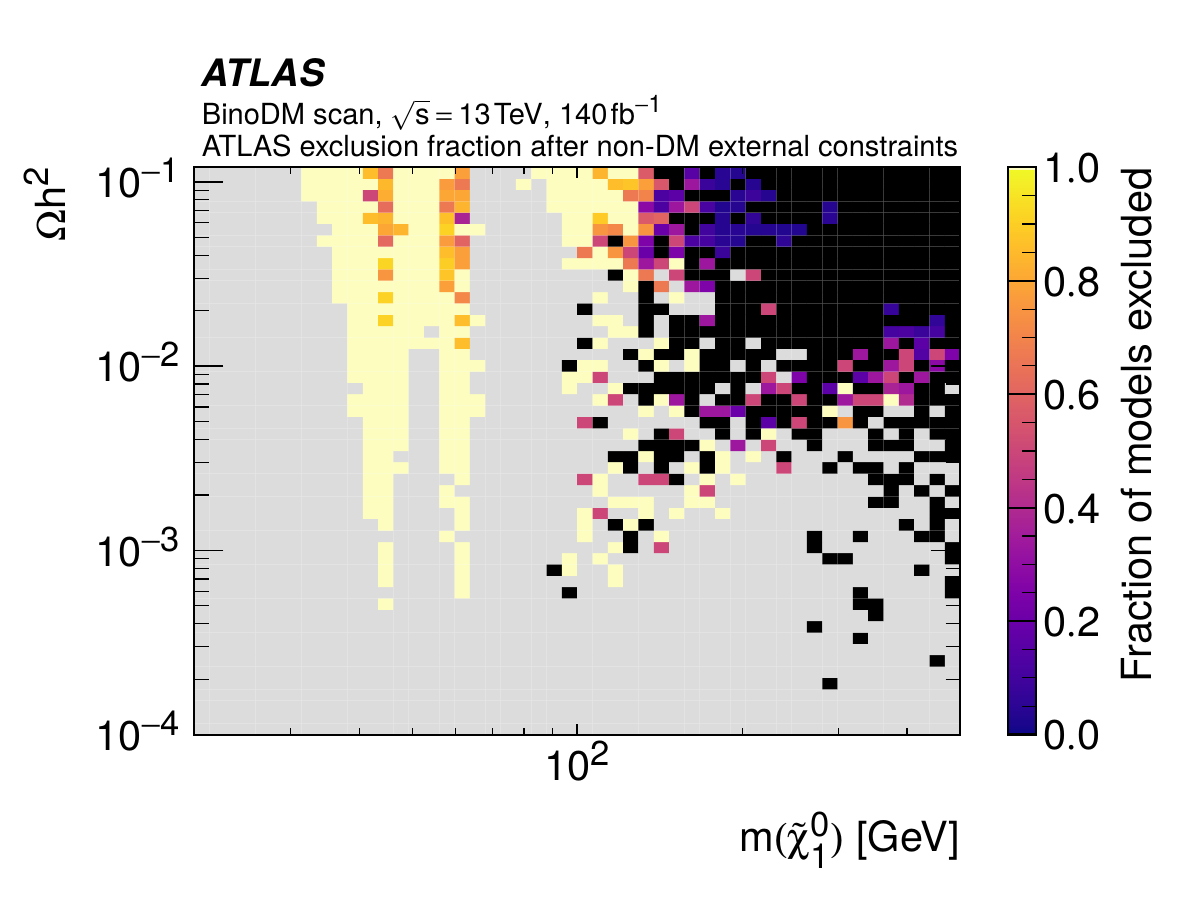}\label{Bino_DM_ATLASextNoDm}}
\qquad
\subfloat[][Models satisfying all external constraints]{\includegraphics[width=0.45\textwidth]{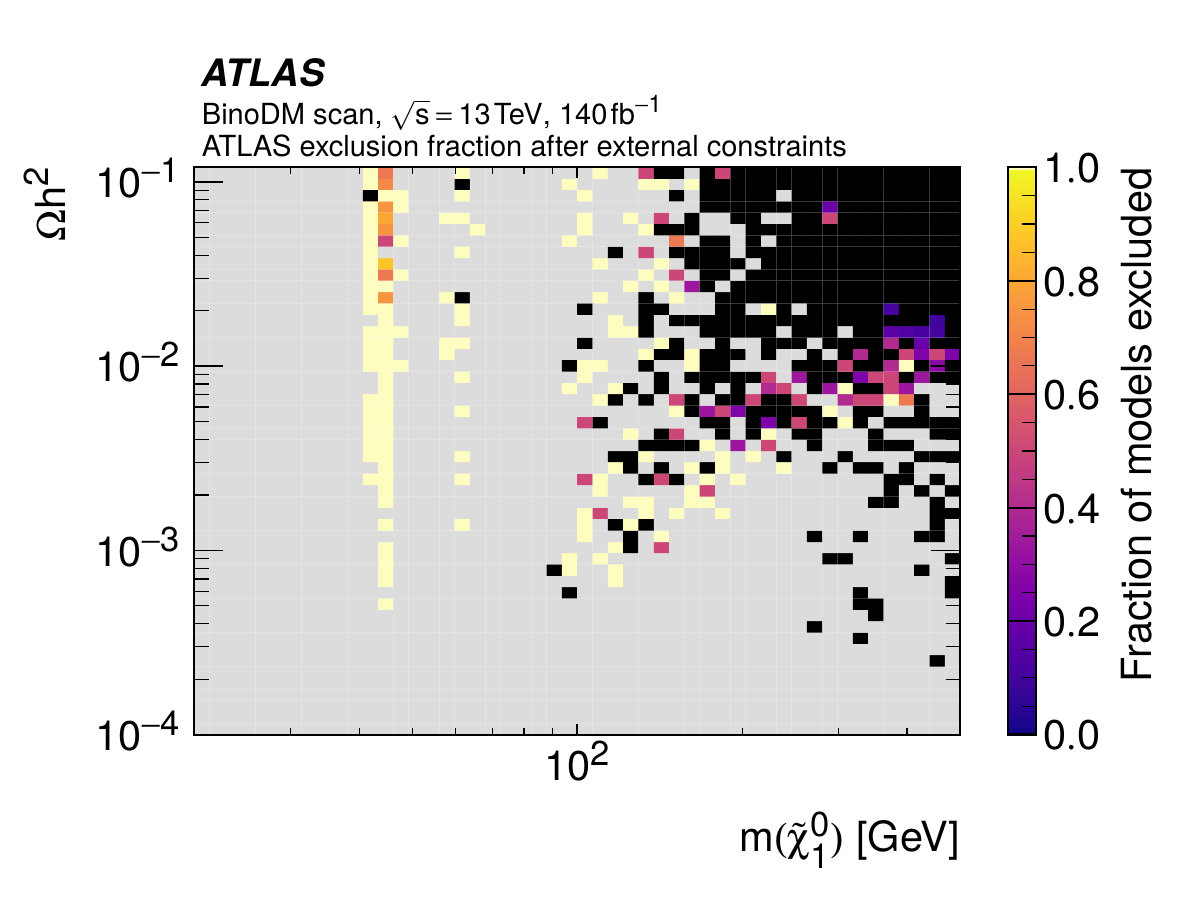}\label{Bino_DM_ATLASext}}
\qquad
\caption{
Exclusion of models from the BinoDM scan in the $m(\N{1})$--$\Omega h^2$ plane with (a) all non-DM constraints applied and (b) all external constraints applied. Bins in grey have no models to consider, while for bins in cream (black) all models are (not) excluded.}
\label{excl:DMrelic}
\end{figure}

Figure \ref{excl:DMrelic} shows the fraction of BinoDM models excluded by the ATLAS Run~2 results in the $m(\N{1})$--$\Omega h^2$ plane for all considered models satisfying the non-DM external constraints and for models satisfying all external constraints. The results from the direct detection searches strongly constrain the funnel regions, and once the ATLAS Run~2 results are applied only a few models remain viable in the funnel regions. The ATLAS constraints at higher LSP masses are weaker.
 
\begin{figure}
\centering
 
\subfloat[][]{\includegraphics[width=0.45\textwidth]{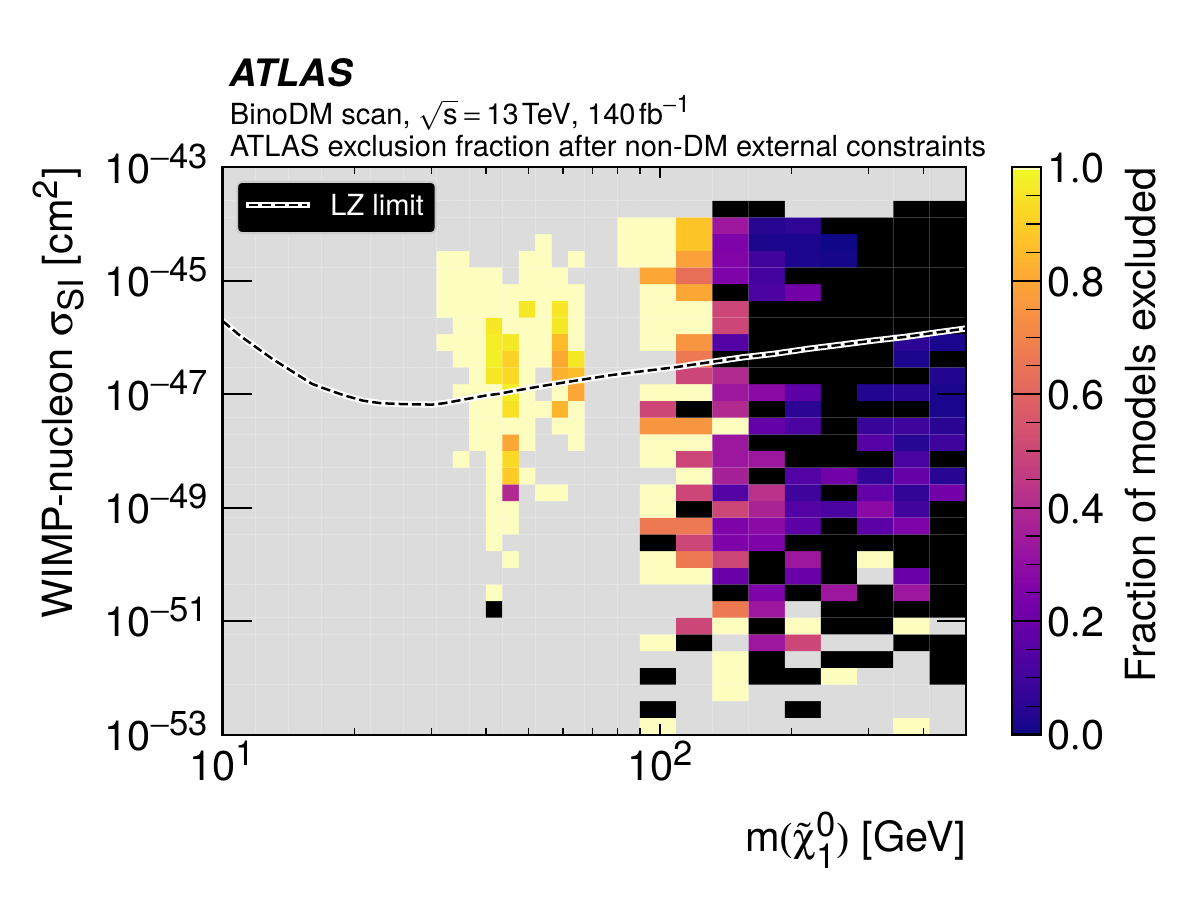}\label{Bino_DDSI_ATLASext}}
\qquad
\subfloat[][]{\includegraphics[width=0.45\textwidth]{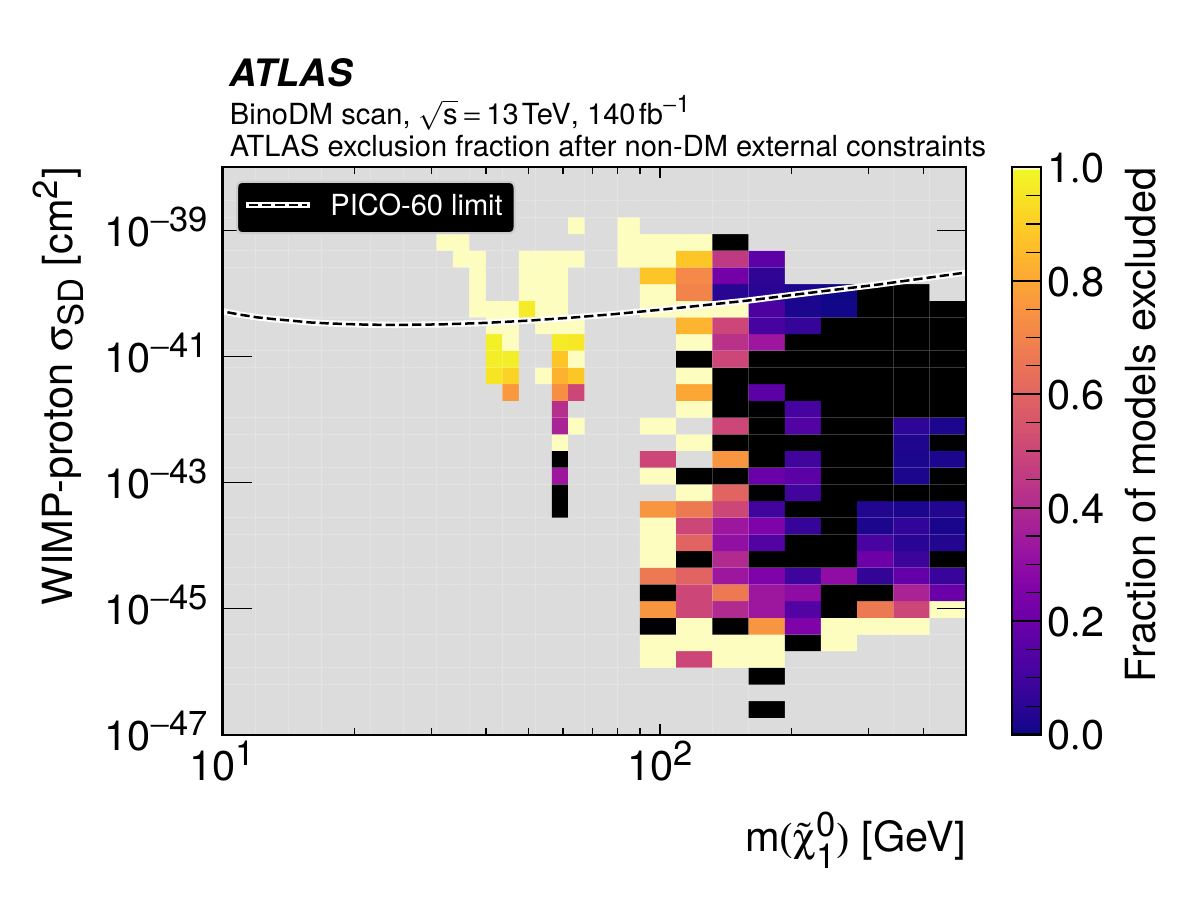}\label{Bino_DDSD_ATLASext}}
 
\caption{
The fraction of models excluded by the ATLAS Run~2 searches from the BinoDM scan in (a) the WIMP--nucleon spin-independent scattering cross-section $\sigma_{\textup{SI}}$ versus $m(\N{1})$ plane and (b) the WIMP--proton spin-dependent scattering cross-section $\sigma_{\textup{SD}}$ versus $m(\N{1})$ plane. The pMSSM model cross-sections are scaled by $\Omega h^2 / 0.12$ to provide a fair comparison with the LZ~\cite{LUX-ZEPLIN:2022xrq} and PICO-60~\cite{PICO:2019vsc} upper limit contours shown as dashed lines on the two plots, respectively. Only models that satisfy the non-DM external constraints are included. Bins in grey have no models to consider, while for bins in cream (black) all models are (not) excluded.}
\label{excl:DMdd}
\end{figure}

\begin{figure}[h!]
\centering
\subfloat[][All considered models]{\includegraphics[width=0.47\textwidth]{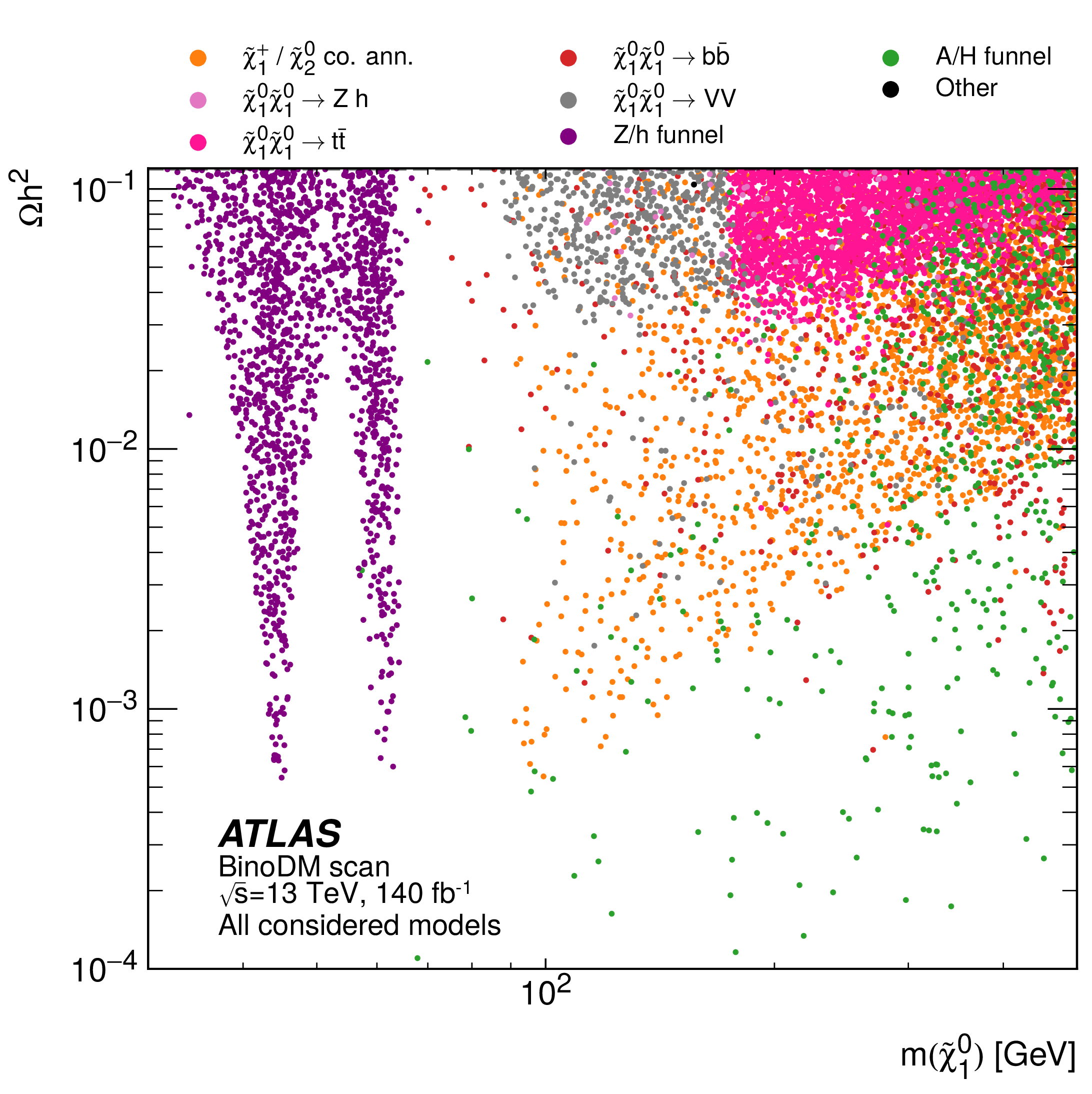}\label{excl:DMall}}
\qquad
\subfloat[][Models not excluded by ATLAS]{\includegraphics[width=0.47\textwidth]{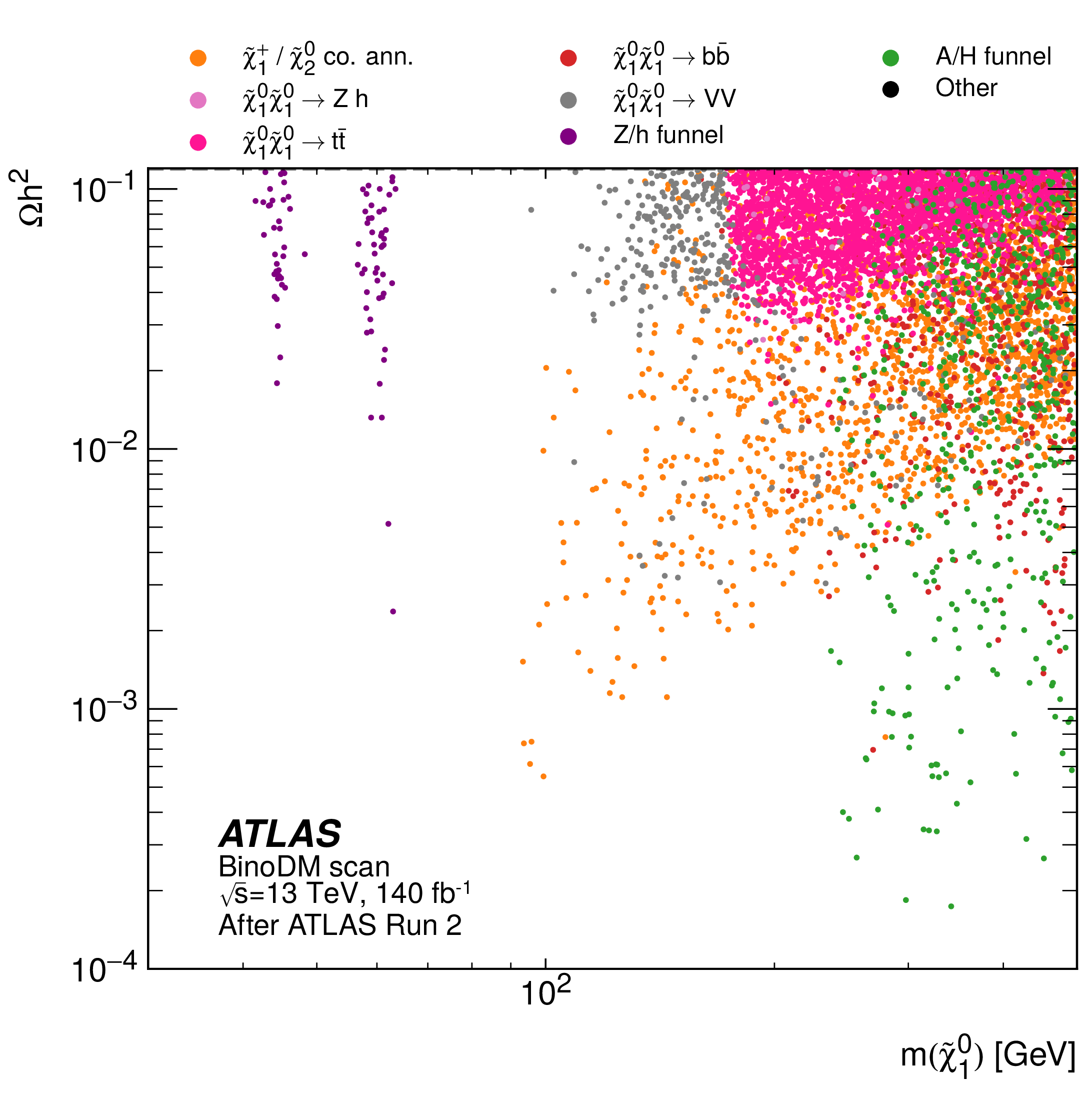}\label{excl:DMpostATLAS}}
\qquad
\subfloat[][Models not excluded by external constraints]{\includegraphics[width=0.47\textwidth]{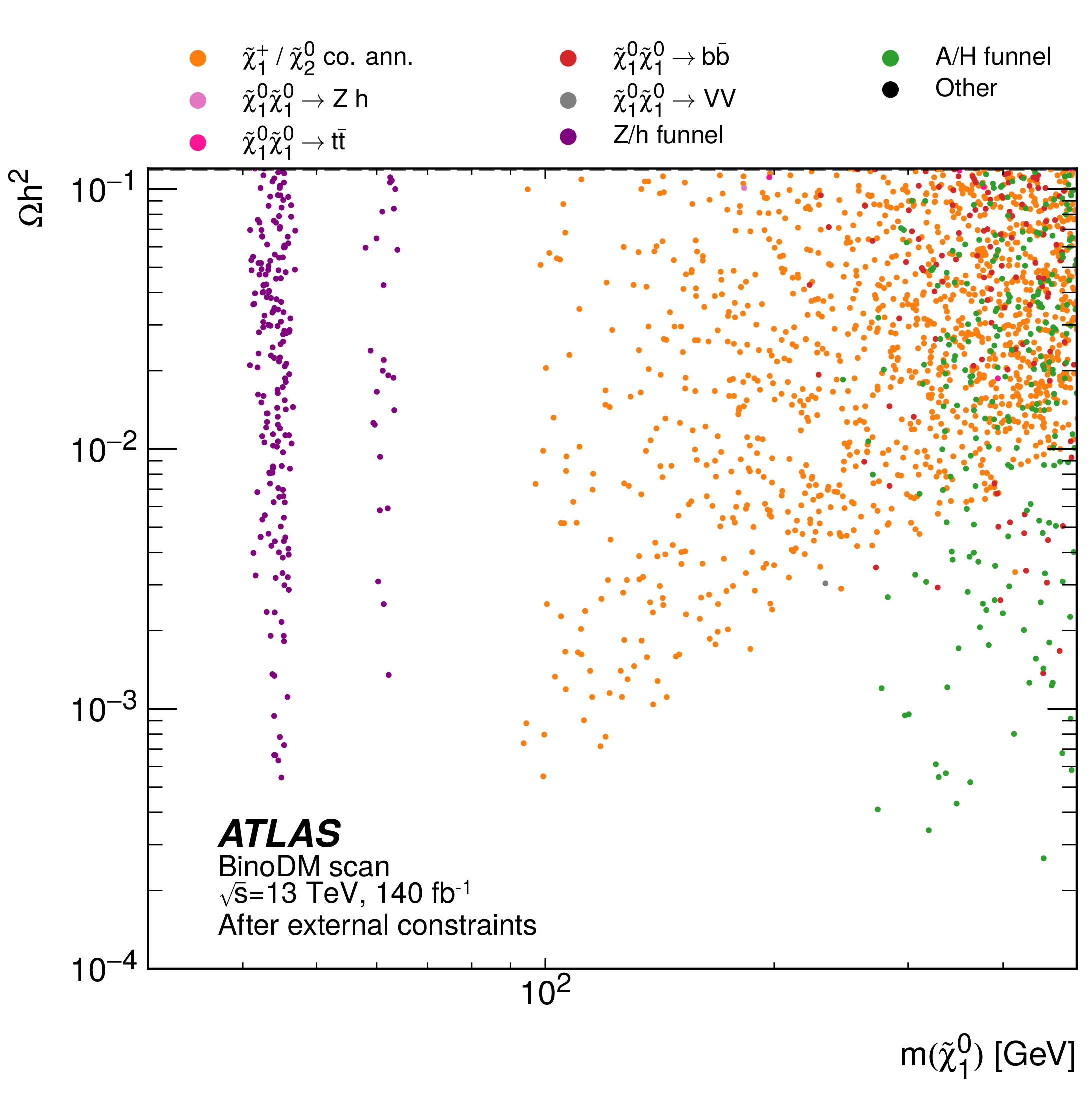}\label{excl:DMpostExt}}
\qquad
\subfloat[][Models not excluded by ATLAS or external constraints]{\includegraphics[width=0.47\textwidth]{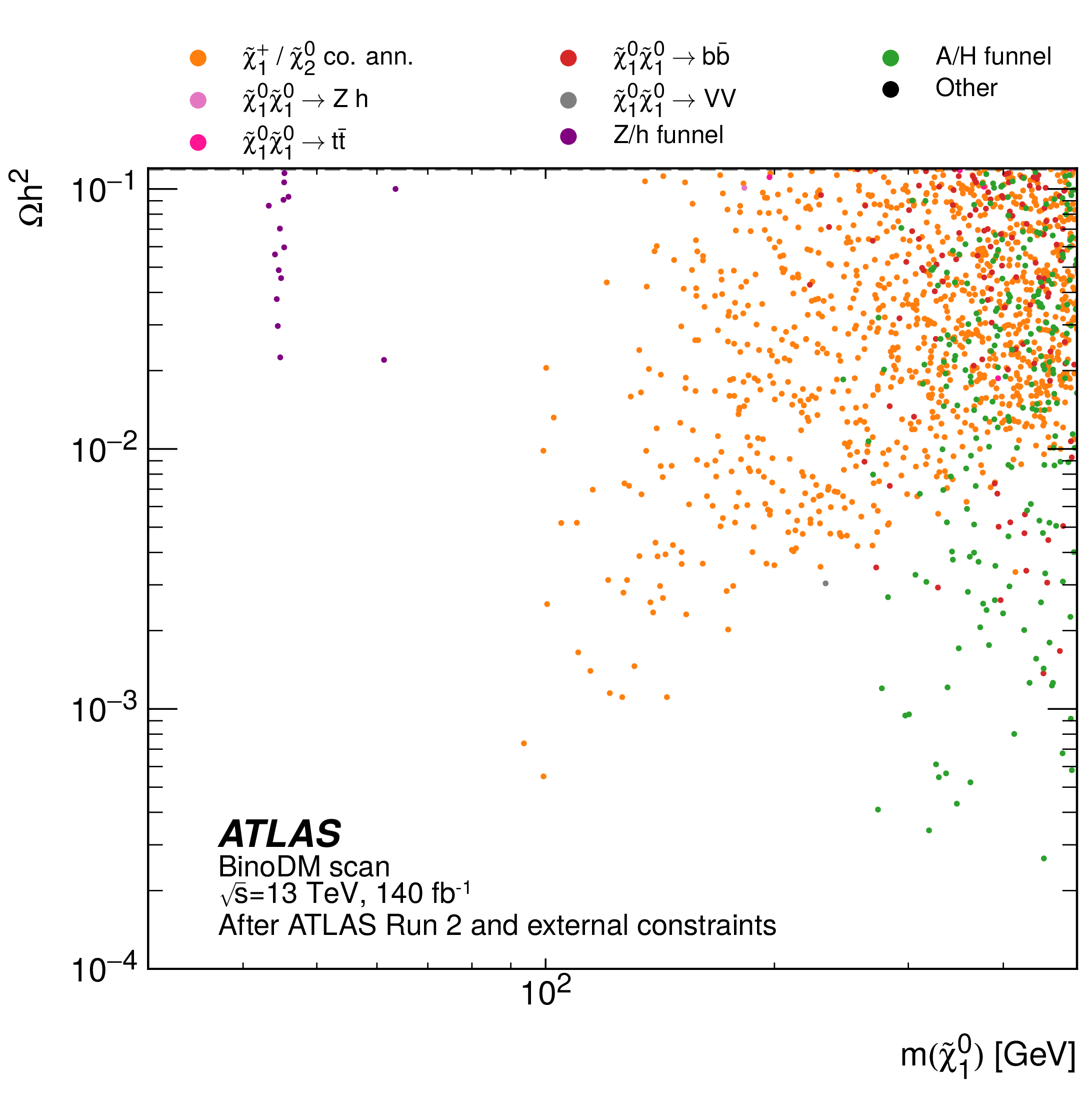}\label{excl:DMleft}}
\caption{
Scatter plots of models from the BinoDM scan in the  $m(\N{1})$--$\Omega h^2$ plane, coloured by the dominant annihilation mechanism for the corresponding model. The plot is shown for all considered models, after ATLAS Run~2 constraints, after external constraints, and after both ATLAS Run~2 and external constraints.}
\label{excl:DM_scatter}
\end{figure}

To further illustrate the complementarity between the collider and non-collider results for a bino-like LSP, Figure~\ref{excl:DMdd} shows the fraction of models in the BinoDM scan (that satisfy all non-DM external constraints) that are excluded by the ATLAS Run~2 results as function of LSP mass and WIMP--nucleon spin-independent/dependent scattering cross-sections.
The ATLAS searches are observed to have a high sensitivity in regions where the direct-detection searches are insensitive and vice-versa, demonstrating the complementarity. Additionally, Figure~\ref{excl:DM_scatter} shows scatter plots of $\Omega h^2$  versus $m(\N{1})$ with different constraints applied to the model set and each model point coloured by the dominant LSP- annihilation mechanism.
The $Z$/$h$ funnel regions are almost entirely excluded when considered both ATLAS Run 2 and external constraints, as are most models up to $m(\N{1}) \approx 200$~\GeV. Most models with \Ch{1}/\N{2} co-annihilation as the dominant mode are still viable.
 
\begin{figure}[h!]
\centering
\subfloat[][]{\includegraphics[width=0.45\textwidth]{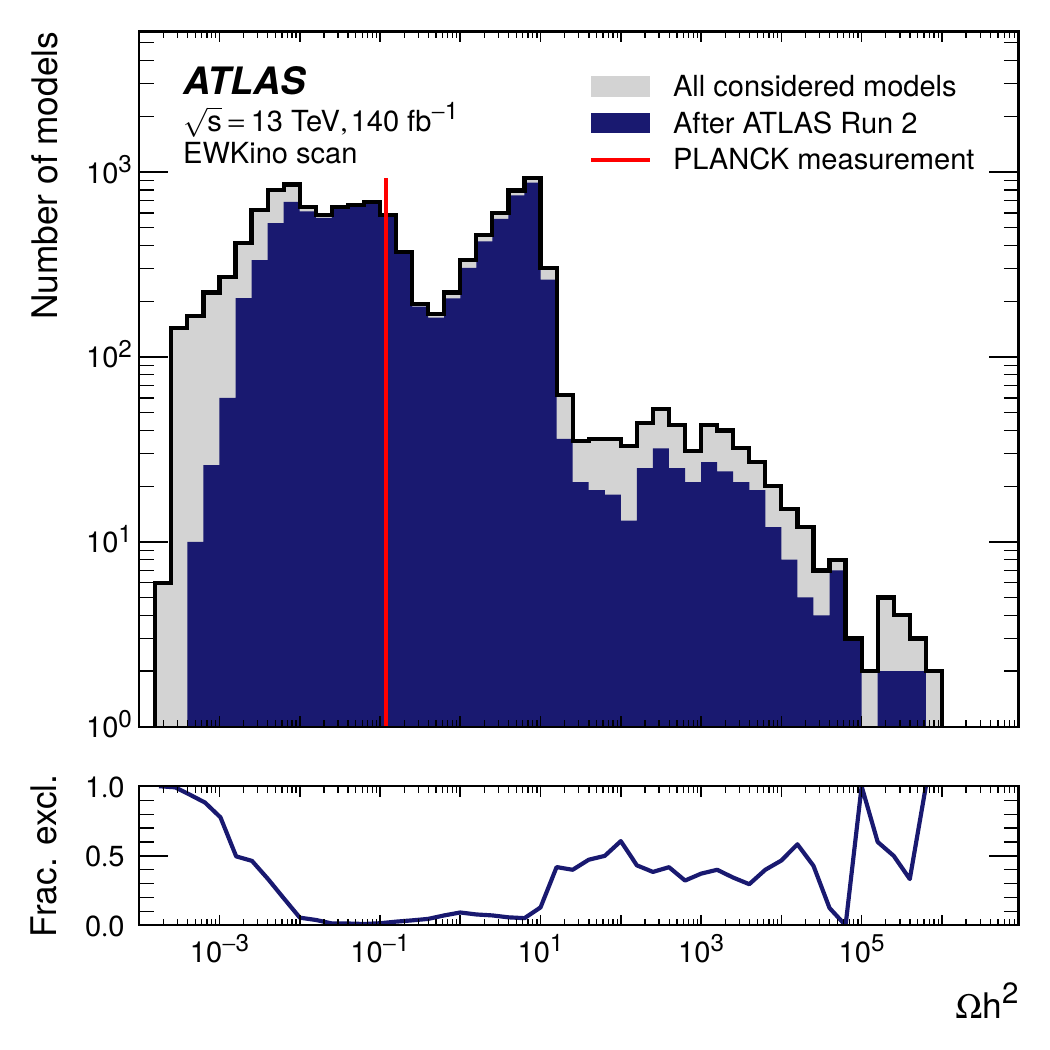}}
\qquad \\
\subfloat[][]{\includegraphics[width=0.45\textwidth]{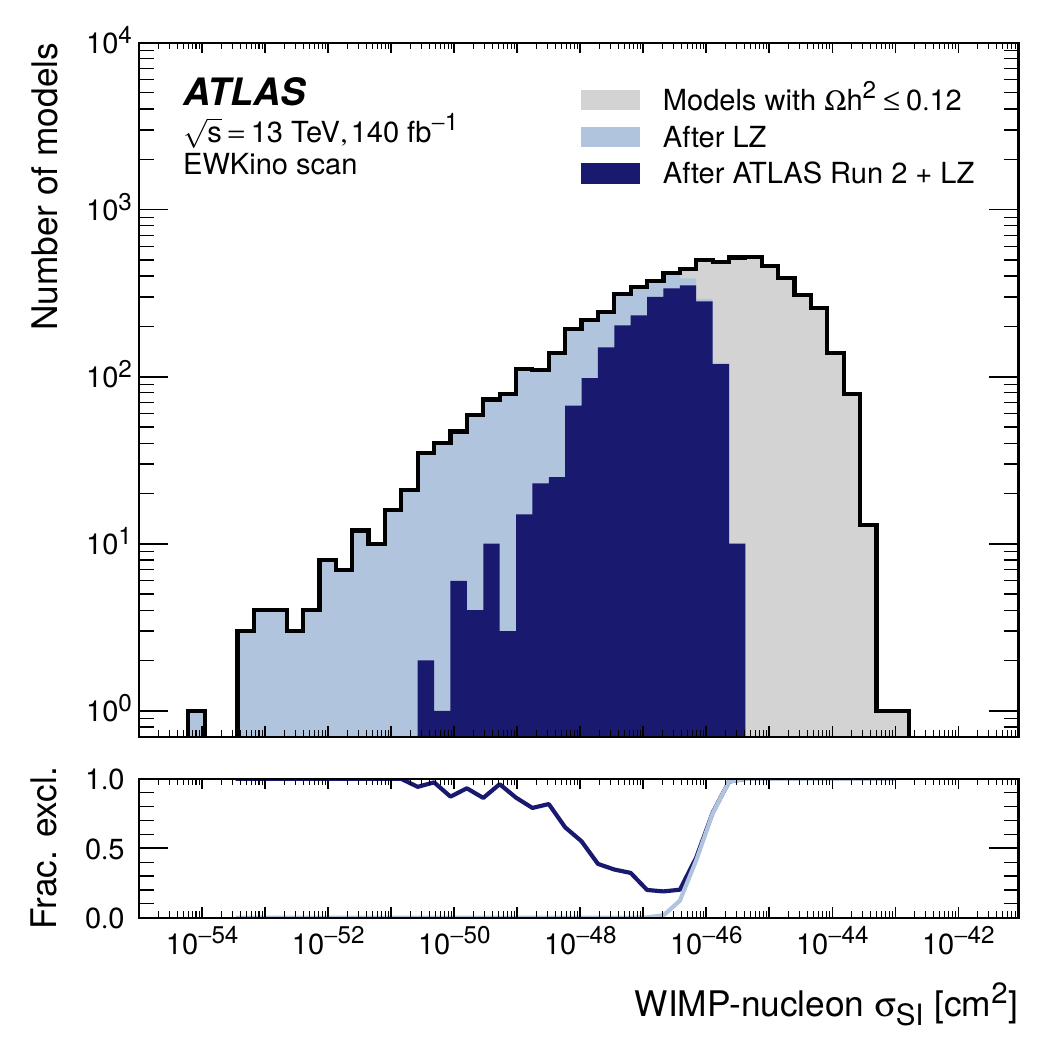}}
\qquad
\subfloat[][]{\includegraphics[width=0.45\textwidth]{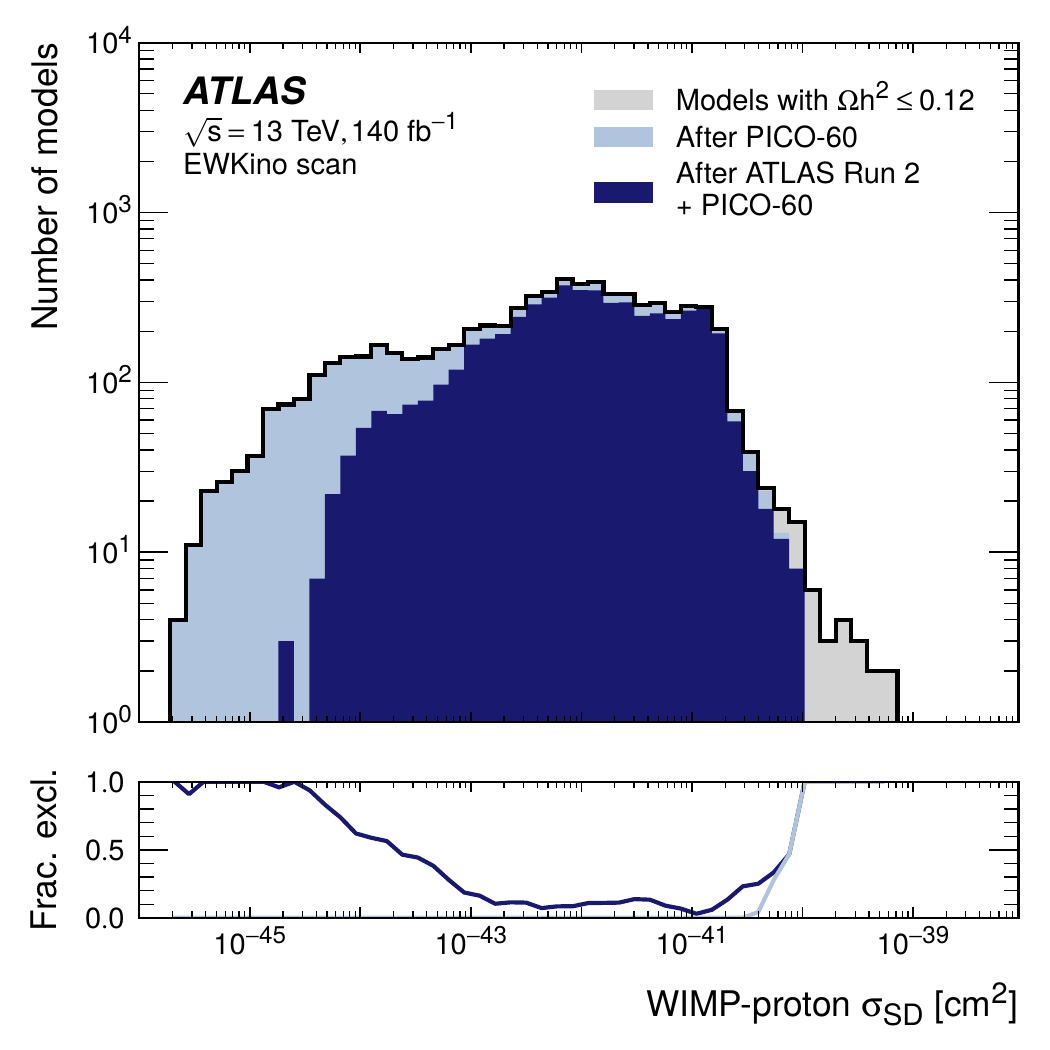}}
\caption{One-dimensional distributions of (a) the DM relic density, (b) the WIMP--nucleon spin-independent scattering cross-section and (c) the WIMP--proton spin-dependent scattering cross-section, for the EWKino scan. For (a) the distribution is shown before and after applying the ATLAS Run~2 constraints (with no additional constraints applied) with the measured relic density marked by the vertical line. The distributions in (b) and (c) are presented for all considered models that satisfy the DM relic density constraint, after applying the relevant direct detection constraint (LZ for the spin-independent cross-section and PICO-60 for the spin-dependent cross-section), and after applying both the relevant direct detection constraint and the ATLAS Run~2 constraints (with no additional external constraints applied). The pMSSM model cross-sections are scaled by $\Omega h^2 / 0.12$ to provide a fair comparison with the direct detection experiment limits.}
\label{fig:EWKinoDDcrosssec}
\end{figure}

Finally, Figure~\ref{fig:EWKinoDDcrosssec} illustrates the complementarity between the DM constraints and the ATLAS Run~2 constraints for the EWKino scan.
One dimensional distributions of the DM relic density, the WIMP--nucleon spin-independent scattering cross-section and the WIMP-proton spin-dependent scattering cross-section are presented for models selected by the EWKino scan.
For the DM relic density the distributions are presented before and after applying the ATLAS Run~2 constraints and for the scattering cross-sections one-dimensional distributions are presented after applying the relevant direct detection constraint (LZ for the spin-independent cross-section and PICO-60 for the spin-dependent cross-section), and after applying both the direct detection constraint and the ATLAS Run~2 constraints (with no additional external constraints applied).
This again demonstrates the complementarity between direct detection experiments and ATLAS searches, as ATLAS provides good exclusion of low cross-section scenarios that lie well below the LZ and PICO-60 limits.
 
\FloatBarrier
\subsection{Complementarity and overlap between ATLAS Run~2 results and external constraints}
\label{results:complementarity}
 
\begin{figure}[h!]
\centering
\subfloat[][EWKino scan: overlaps between searches.]{\includegraphics[width=0.47\textwidth]{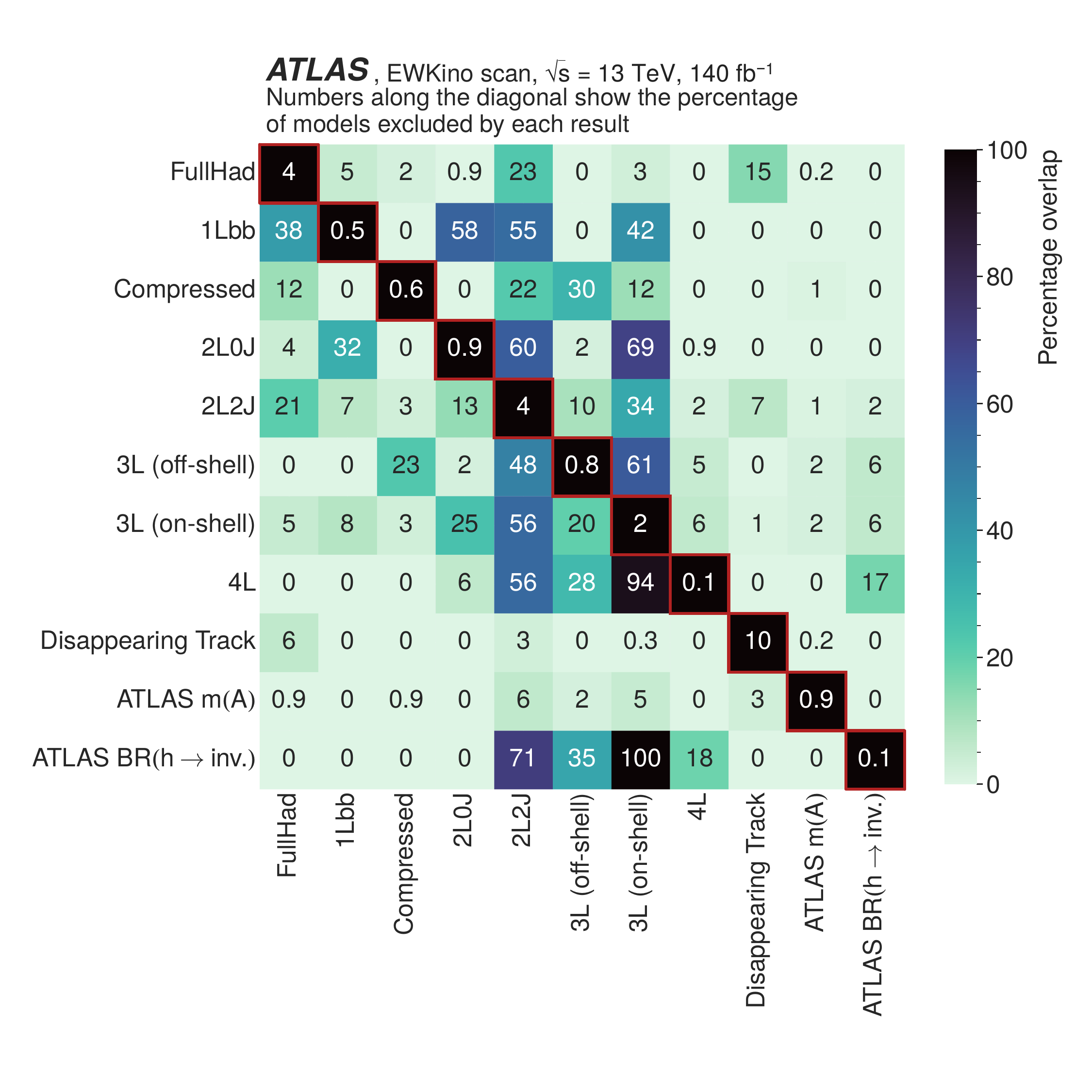}\label{search_overlap_ewk}}
\qquad
\subfloat[][BinoDM scan: overlaps between searches.]{\includegraphics[width=0.47\textwidth]{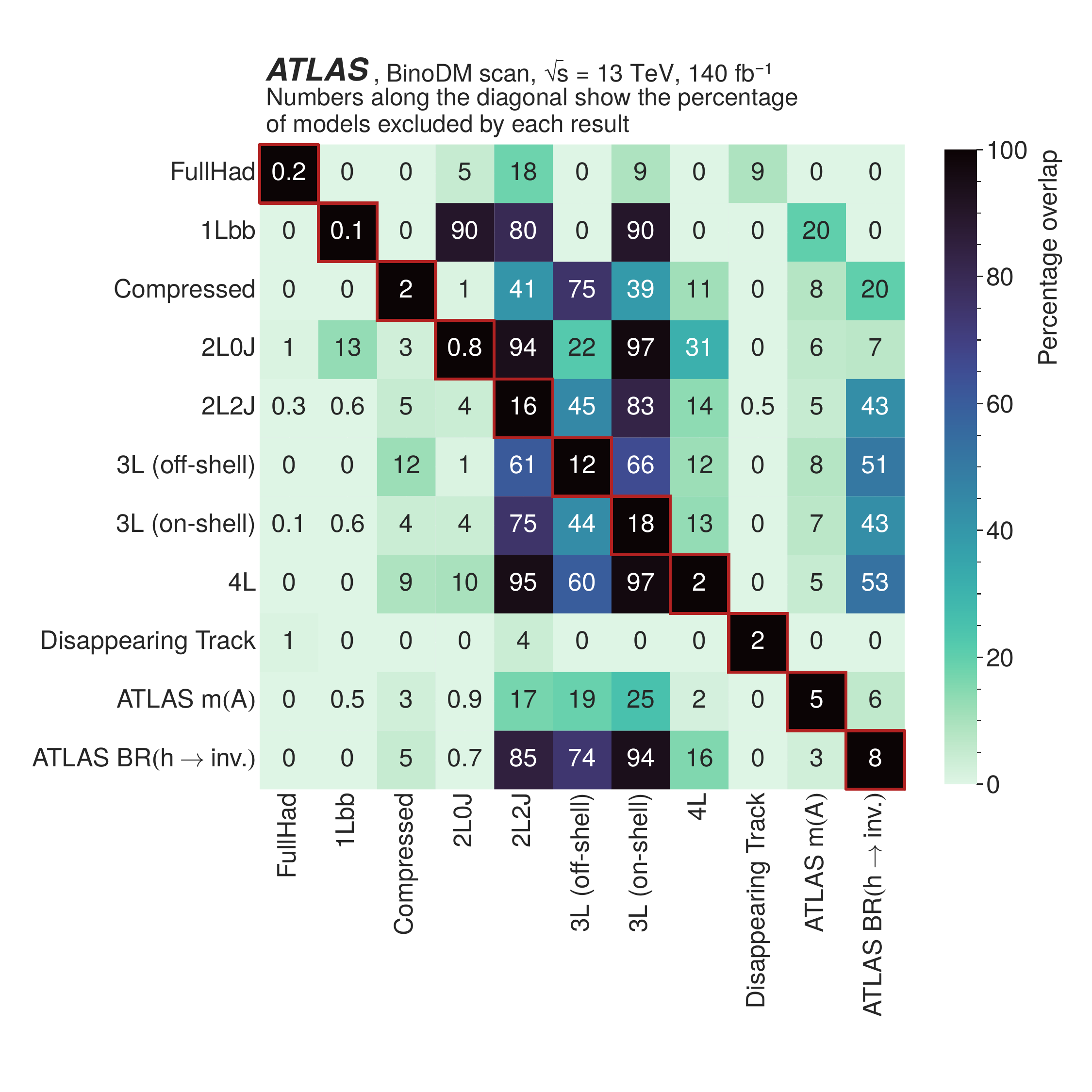}\label{search_overlap_bino}}
\qquad
\subfloat[][EWKino scan: overlaps between external constraints.]{\includegraphics[width=0.47\textwidth]{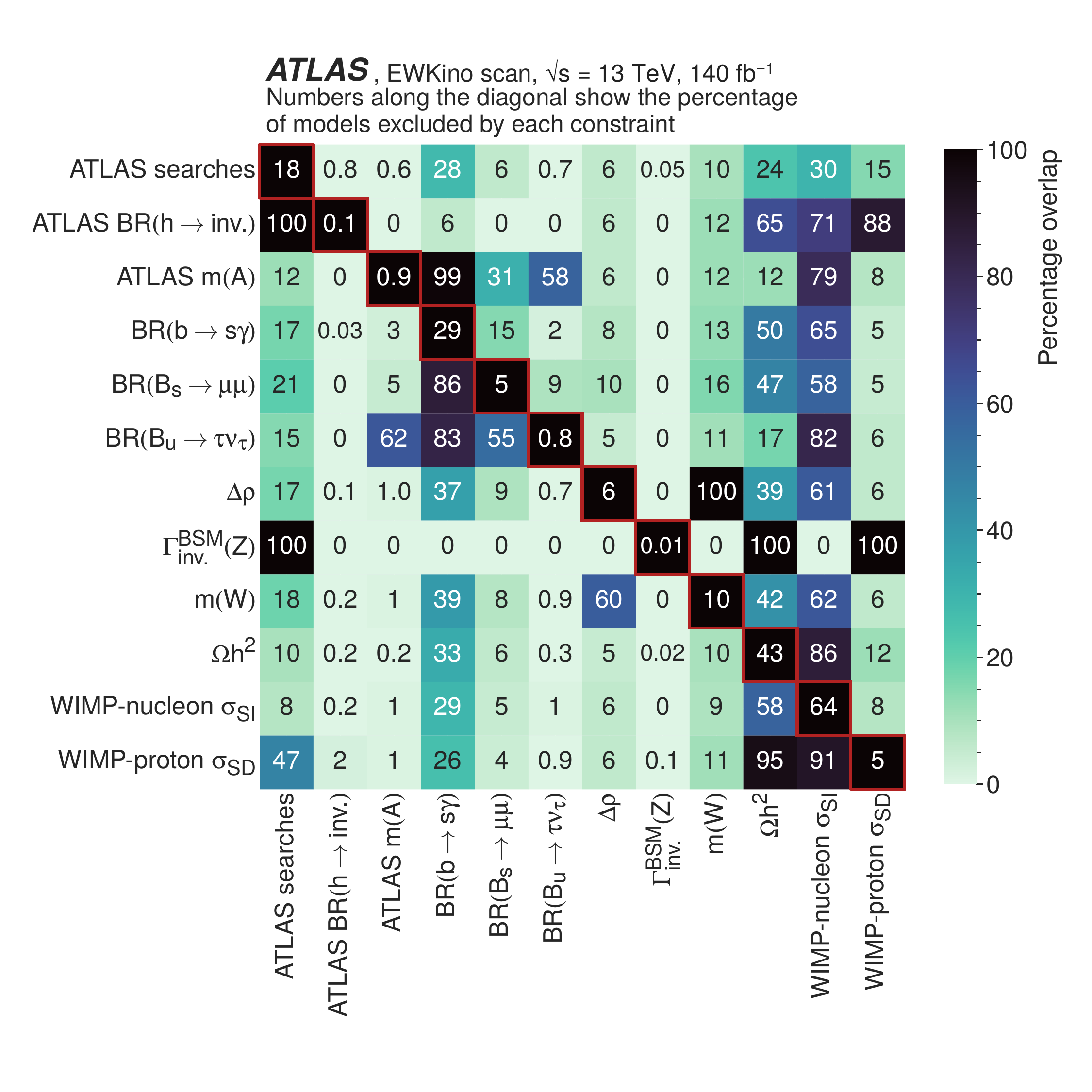}\label{constraint_overlap_ewk}}
\qquad
\subfloat[][BinoDM scan: overlaps between external constraints. ]{\includegraphics[width=0.47\textwidth]{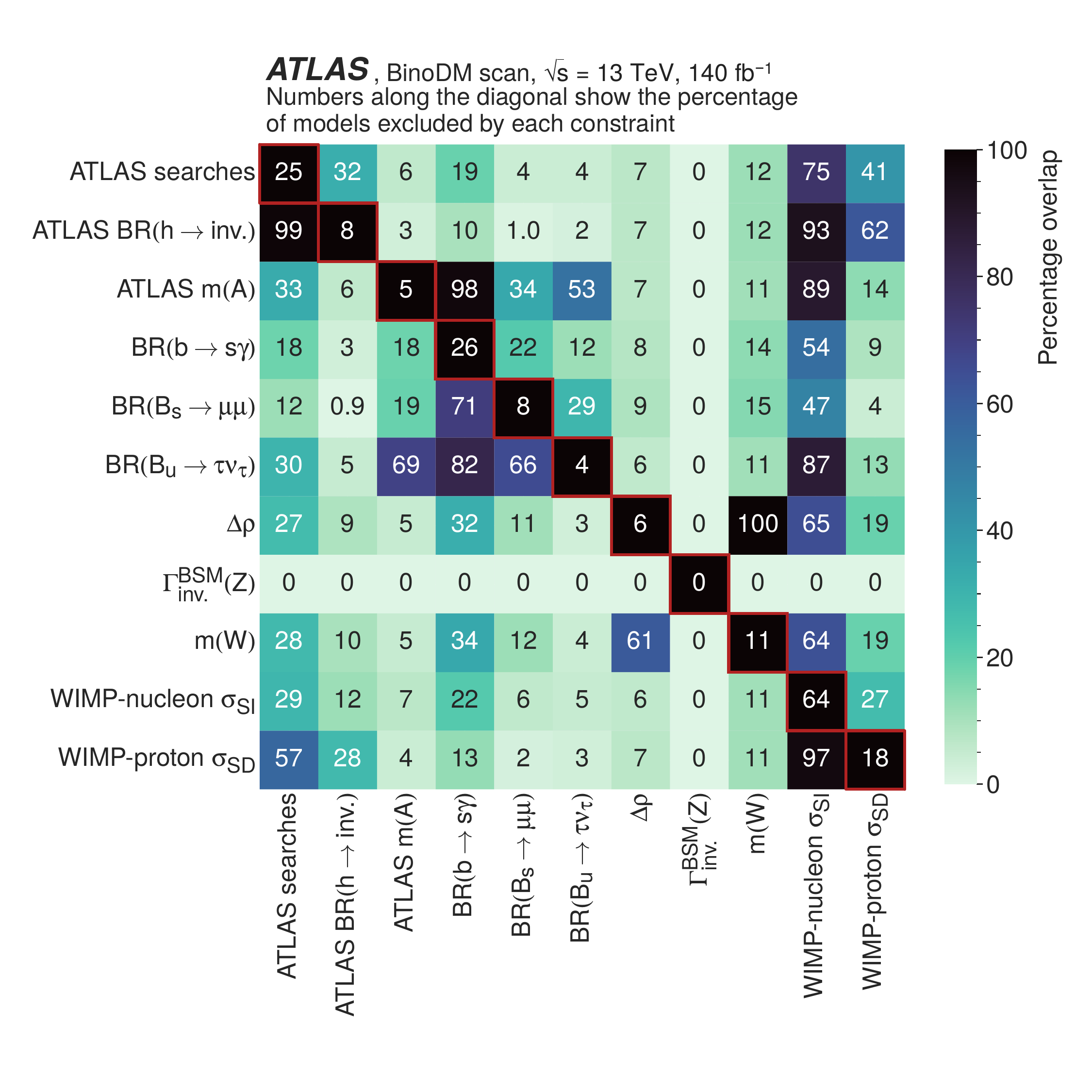}\label{constraint_overlap_bino}}
\caption{Percentage overlaps in the models excluded by each search and constraint. For the diagonal entries, the number shows the percentage of all considered models excluded by that search/constraint.
For each off-diagonal square, the number and bin colour indicates the percentage of the models excluded by the result on the $y$~axis that were also excluded by the result on the $x$~axis, i.e. $100\times N_{xy}/N_{y}$ where $N_{xy}$ is the number of models excluded by the results on the $x$ and $y$~axes and $N_{y}$ is the number of models excluded by the result on the $y$~ axis.
The figures in (a) and (b) show the overlap of each search, while (c) and (d) show the overlap of each constraint. Figure (a) and (c) show the EWKino scan, while (b) and (d) are for the BinoDM scan. Where no models are excluded, the overlap is shown as zero.}
\label{search_overlap}
\end{figure}

In assessing the exclusion, the CL$_{\text{s}}$ value is taken from the ATLAS analysis with the best expected sensitivity, however for many models multiple analyses would have the required sensitivity to claim exclusion. The overlap in sensitivity between the analyses and constraints is displayed in Figure \ref{search_overlap}. For the diagonal entries, the number in bold shows the percentage of all considered models excluded by that search/constraint. For the off-diagonal entries the number in each box indicates the percentage of the models excluded by the analysis/constraint in that row that were also excluded by the analysis/constraint on the column label. In general, the 2L2J and 3L searches have high overlaps with all searches except the disappearing-track search for both the EWKino and BinoDM scans, due to their abilities to probe both associated chargino--neutralino production with gauge boson-mediated decays as well as leptonic final states produced in cascade decays involving heavier electroweakinos.
 
The ATLAS searches exclude 100\% and 99\% of the models excluded by the ATLAS $\mathcal{B}(h\rightarrow\textup{inv})$ constraint in the EWKino and BinoDM scans respectively.
This is because models with high $\mathcal{B}(h\rightarrow\textup{inv}$) lie in the Higgs-funnel region, which corresponds to mass ranges that the ATLAS searches considered are most sensitive to.
 
When examining the overlaps between the ATLAS searches and external constraints, the highest overlaps with the ATLAS searches come from the spin-independent cross-section measurement and relic density measurement for the EWKino scan, but the numbers are still relatively low (30\% and 24\% respectively). Alternatively, for the BinoDM scan the spin-independent cross-section measurement has a high overlap with all relevant searches and constraints.
 
Finally, Table~\ref{tab:LSP_excl} contains a summary of the percentage of models excluded for each LSP type, for each search and constraint individually and overall. The final row shows that over 77\% of models for each LSP type and scan are excluded when considering both ATLAS Run~2 and external constraints. However as these fractions are calculated from the entire model set they depend on the priors (scan ranges and strategy) used. It is therefore more useful for planning future searches to
consider the models that remain unexcluded with low-mass electroweakinos that are discussed in the next section. For the EWKino scan, of the different LSP types, wino-like LSPs have the highest ATLAS Run 2 exclusion, which is almost entirely driven by the disappearing-track analysis. For the BinoDM scan, the largest individual exclusion comes from the 2L2J and 3L analyses, which, as mentioned above, have degrees of overlap. The external constraints have a large impact on both bino LSP models in the
EWKino scan and the BinoDM scan, which is mainly driven by the DM constraints. The non-DM external constraints (flavour and precision electroweak) have a similar impact across the LSP types and scans, though flavour constraints have a weaker impact on models with a higgsino-like LSP.

\begin{table}[h!]
\caption{Percentage of models excluded by each search and constraint, for each LSP type. }
\label{tab:LSP_excl}
\centering
\begin{tabular}{| c | r r r | r |}
\hline
& \multicolumn{3}{ c |}{EWKino scan LSP type} &  BinoDM \\
Search & Bino-like & Higgsino-like & Wino-like & scan \\
\hline
\hline
FullHad 			& 5.3 \% & 2.6 \% & 3.0 \% & 0.2 \% \\
1Lbb 				& 1.2 \% & 0.0 \% & 0.0 \% & 0.1 \% \\
Compressed 			& 0.2 \% & 1.4 \% & 0.0 \% & 1.9 \% \\
2L0J 				& 2.2 \% & 0.0 \% & 0.0 \% & 0.8 \% \\
2L2J 				& 5.6 \% & 2.9 \% & 3.1 \% & 15.9 \% \\
3L (off-shell) 			& 1.4 \% & 0.6 \% & 0.0 \% & 11.7 \% \\
3L (on-shell)        & 4.7 \% & 0.8 \% & 1.2 \% & 17.7 \% \\
4L 					& 0.3 \% & 0.1 \% & 0.0 \% & 2.4 \% \\
Disappearing-track  & 0.1 \% & 0.1 \% & 45.7 \% & 1.9 \% \\
\hline
\textbf{Overall ATLAS Run~2 SUSY searches} & \textbf{11.2 \%} & \textbf{5.7 \%} & \textbf{48.5 \%} & \textbf{25.0 \%} \\
\hline
$m(A)>480~\GeV$ 					& 0.4 \% & 1.6 \%  & 0.4 \% & 4.8 \% \\
$\mathcal{B}(h\rightarrow\mathrm{inv.})\leq0.107$ 	& 0.4 \% & 0.0 \% & 0.0 \% & 8.1 \% \\
\hline
\textbf{Overall ATLAS Run~2} & \textbf{11.6 \%} & \textbf{7.2 \%} & \textbf{48.7 \%} & \textbf{28.3\%} \\
\hline
Flavour constraints & 33.8 \% & 22.8 \% & 32.5 \% & 28.8 \% \\
Electroweak precision constraints & 9.7 \% & 9.8 \% & 9.7 \% & 10.5 \% \\
Relic density constraint & 96.0 \% & 12.8 \% & 0.3 \% & N/A \\
Direct detection constraints & 83.8 \% & 67.7 \% & 24.4 \% & 64.7 \% \\
All non-DM external constraints & 39.3 \% & 29.3 \% & 37.8 \% & 35.4 \% \\
All external constraints 	& 98.8 \% & 75.8 \% & 54.0 \% & 80.1 \% \\
 
\hline
\textbf{Overall ATLAS Run~2 + non-DM external} & \textbf{46.7 \%} & \textbf{33.9 \%} & \textbf{67.5 \%} & \textbf{53.4 \%} \\
\textbf{Overall ATLAS Run~2 + external} & \textbf{98.9 \%} & \textbf{77.3 \%} & \textbf{83.0 \%} & \textbf{84.1 \%} \\
\hline
\end{tabular}
\end{table}
 
\FloatBarrier

\begin{figure}[h]
\centering
 
\subfloat[][EWKino scan]{\includegraphics[width=0.47\textwidth]{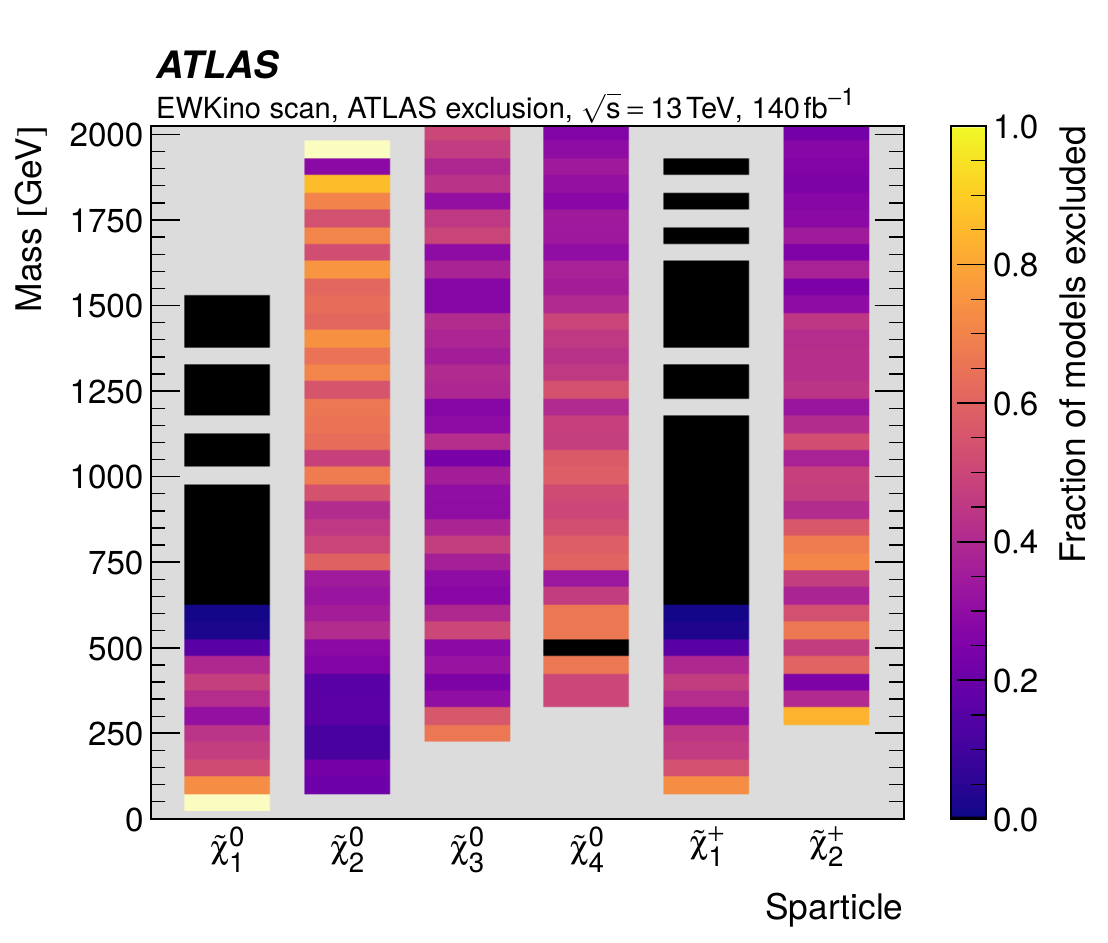}\label{summary_ATLAS_EWK_postExt}}
\qquad
\subfloat[][BinoDM scan]{\includegraphics[width=0.47\textwidth]{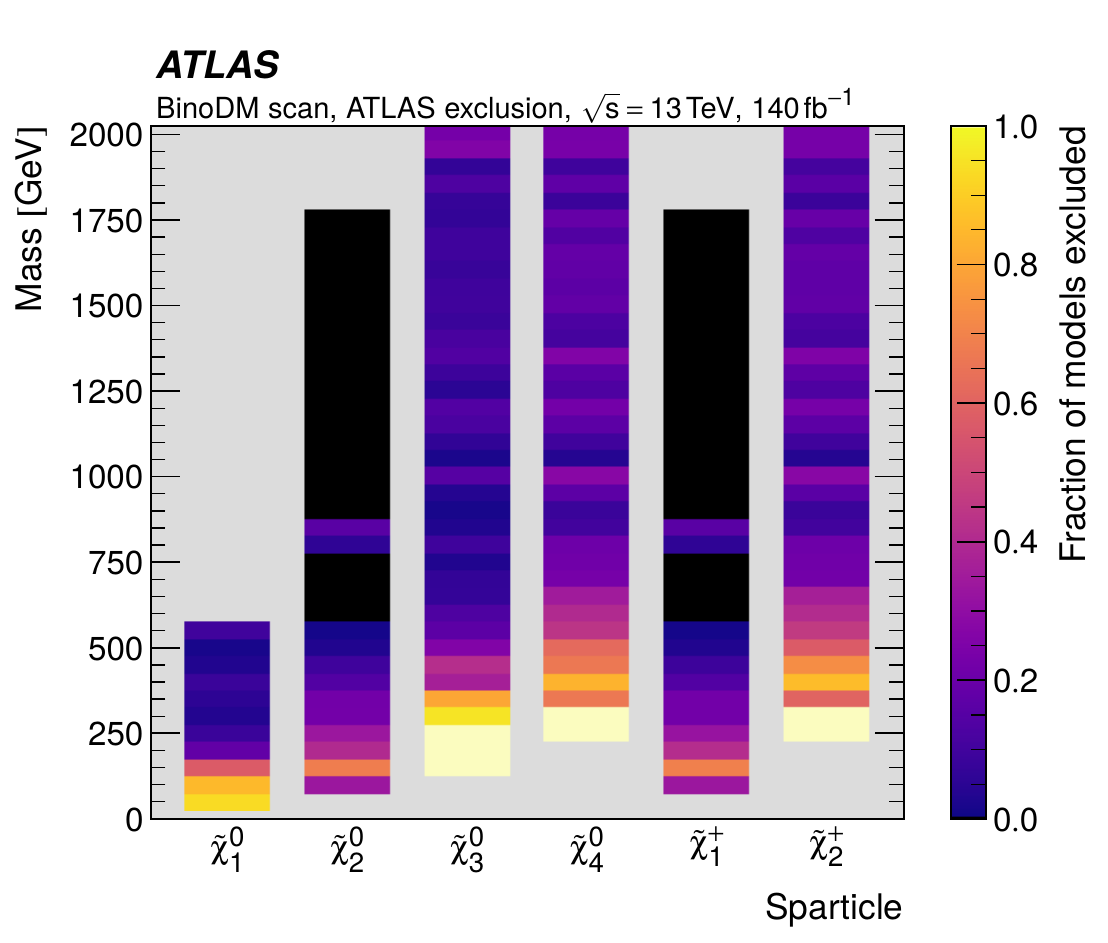}\label{summary_ATLAS_Bino_postExt}}
\caption{Summary plots showing the fraction of models excluded by ATLAS across the masses of each electroweakino for models that satisfy external constraints. \protect\subref{summary_ATLAS_EWK_postExt} shows the EWKino scan and \protect\subref{summary_ATLAS_Bino_postExt} the BinoDM scan. Bins in grey have no models to consider, while for bins in cream (black) all models are (not) excluded.}
\label{fig:summary_postExt}
\end{figure}

Figure~\ref{fig:summary_postExt} provides a visual summary of the constraints by showing the fraction of models excluded by ATLAS as a function of the masses of each electroweakino for models satisfying all of the external constraints. For both the EWKino and BinoDM scans, the bar for the LSP shows predictable behaviour where the fraction of models excluded decreases with increasing LSP mass. A similar trend is observed for all electroweakinos in the BinoDM scan. For the EWKino scan, there is
an interesting effect for the \N{2} where models at higher mass show high exclusion fractions. This is partly driven by wino-like LSP models, where $\Delta m(\Ch{1},\N{1})$ and the $\Ch{1}$ lifetime, and therefore the sensitivity of the disappearing-track analysis, are correlated with $m(\N{2})$.
 
\FloatBarrier
\subsection{Viable pMSSM models surviving the ATLAS Run~2 constraints}
\label{results:viablemodels}

One of the benefits of this study is understanding that pMSSM models have been missed by previous searches. This section presents six benchmark models with a bino-like or higgsino-like LSP that satisfy all the external constraints but are not excluded by the  ATLAS Run~2 analyses considered. These have SUSY spectra that deviate from those typically encountered in ATLAS simplified models, either through having different hierarchies between the bino, wino and higgsino mass parameters, mixed decay modes, or relatively light \N{3}, \Ch{2} and \N{4} that are not within the sensitivity of the FullHad or 2L2J searches. These can be used to optimise dedicated new searches for Run~3 of the LHC and beyond. Spectra are not presented here for wino-like LSPs as the main analysis targeting this scenario, the disappearing-track search, was applied as cross-section upper limits, meaning that the reason many wino-like models were not excluded is simply that they did not have a sufficiently large production cross-section.
 
\begin{figure}[h]
\centering
\subfloat[][$Z$/$h$ funnel region]{\includegraphics[width=0.5\textwidth]{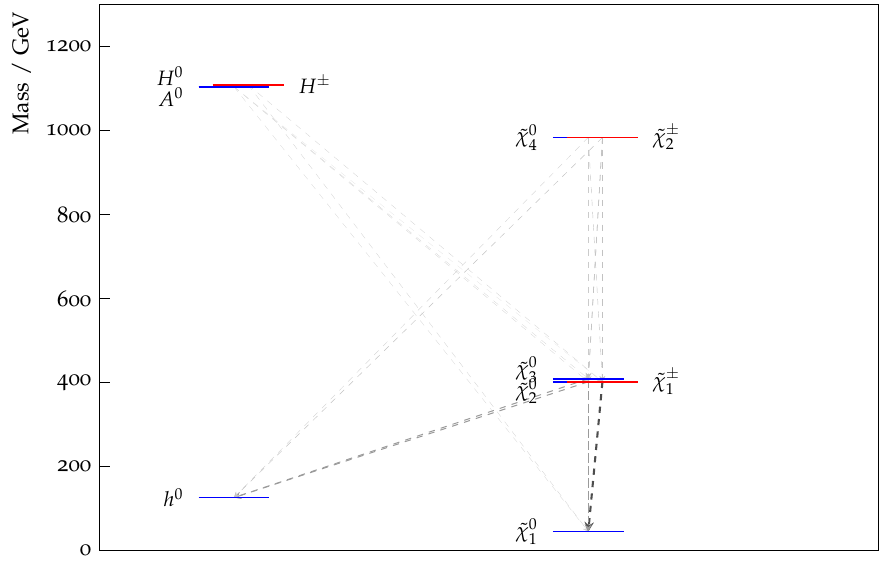}\label{slha:Zhfunnel}}
\subfloat[][$A/H$ funnel region]{\includegraphics[width=0.5\textwidth]{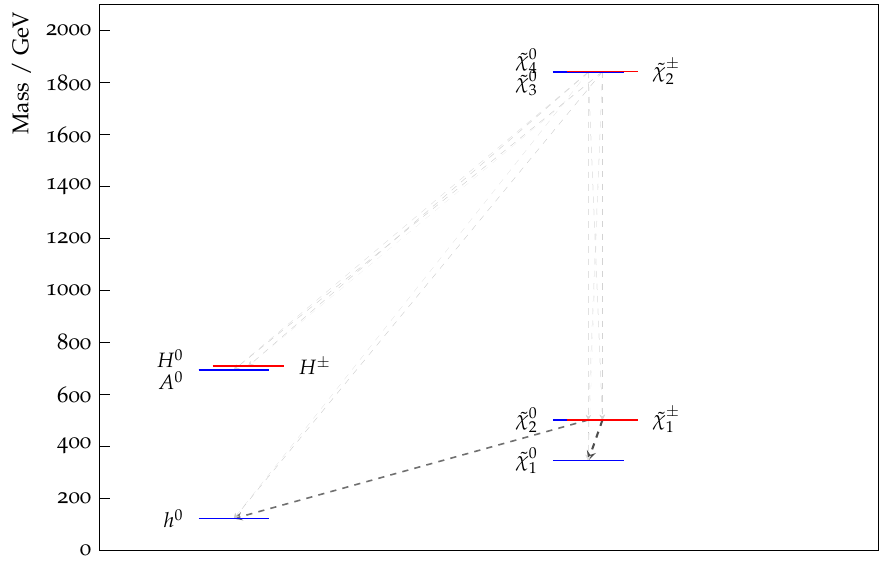}\label{slha:AHfunnel}}
\qquad
\subfloat[][Compressed region]{\includegraphics[width=0.5\textwidth]{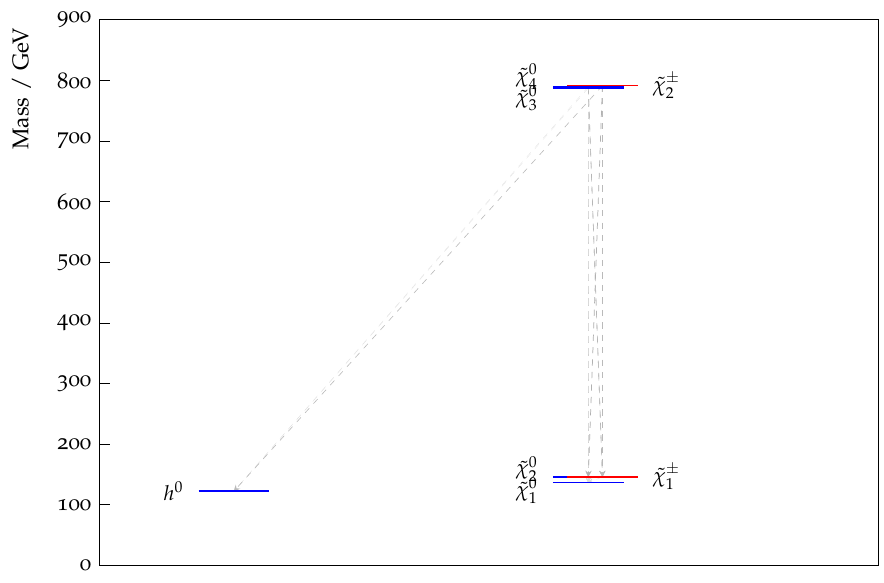}\label{slha:comp1}}
\subfloat[][Compressed region]{\includegraphics[width=0.5\textwidth]{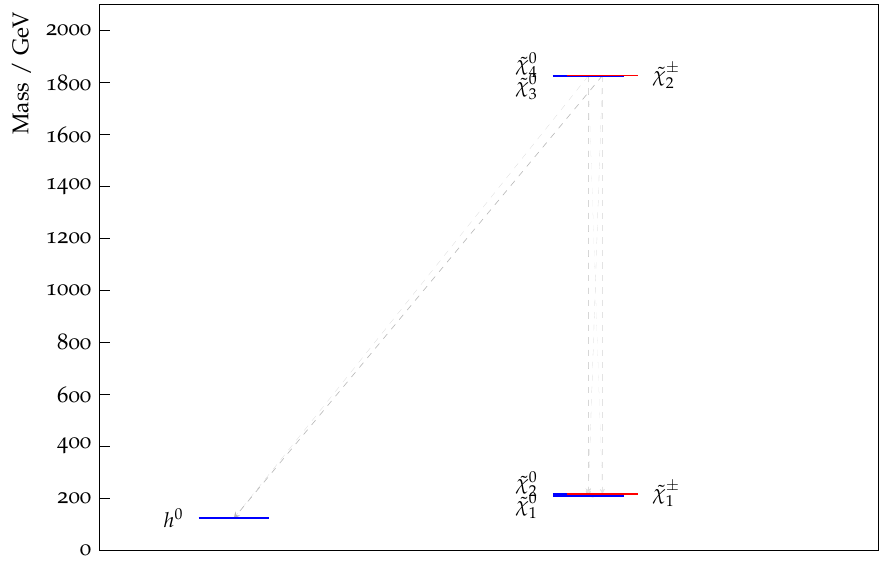}\label{slha:comp2}}
\caption{Mass spectrum for four benchmark models with a bino-like LSP that satisfy all constraints and are not excluded despite having a mass-spectrum within published ATLAS simplified model contours. Produced using PySLHA~\cite{Buckley:2013jua}.}
\label{binoDM_benchmarks}
\end{figure}

Figure \ref{binoDM_benchmarks} shows the mass spectrum for four benchmark models with a bino-like LSP that satisfy all constraints and are not excluded by ATLAS Run~2 analyses.
The \N{1}, \Ch{1} and \N{2} masses of all of these models lie within the published ATLAS simplified model contours for at least one of the SUSY searches considered in this analysis.
 
Figure \ref{slha:Zhfunnel} shows a model in the $Z$/$h$ funnel region.
This model differs from typical simplified models as the \Ch{1} and \N{2} higgsino fraction is greater than 98\%.
Additionally, this scenario has $\mathcal{B}(\N{2}\rightarrow\N{1} h)\approx\mathcal{B}(\N{2}\rightarrow\N{1} Z)\approx50\%$.
Scenarios such as this will therefore have smaller signal yields than typical simplified models due to the higgsino-like \Ch{1}/\N{2} production cross-sections being smaller than for wino-like particles and the branching ratios to the typical \N{2} decay modes being half the usual simplified model assumption of 100\%.
 
Figure \ref{slha:AHfunnel} shows a model with a wino-like \Ch{1} and \N{2} of similar mass with a 156~\GeV\ mass splitting to the bino-like LSP\@.
This scenario has relatively light $A$ and $H$ bosons at twice the LSP mass such that the DM relic density is reduced via the $A/H$ funnel self-annihilation mechanisms, $\N{1}\N{1}\rightarrow A/H$.
Similar to the $Z$/$h$ funnel model, this model has mixed $\N{2}$ decay modes, with $\mathcal{B}(\N{2}\rightarrow\N{1} h)=77\%$ and $\mathcal{B}(\N{2}\rightarrow\N{1} Z)=23\%$.
The model lies just beyond the 1Lbb region of sensitivity, so despite the \N{2} decay into $h$ being dominant, the 3L search (specifically the signal regions that targeted the \N{2} decay via $Z$) has the best sensitivity.
 
Finally, two scenarios with small mass splittings between the \Ch{1}, \N{2} and \N{1} are shown in Figures \ref{slha:comp1} and \ref{slha:comp2}. These scenarios have a smaller branching fraction for the $\N{2}\rightarrow\N{1}\ell^+\ell^-$ three-body decay than the simplified models used in the original compressed analysis. They are instead dominated by the radiative decay $\N{2}\rightarrow\N{1}\gamma$.
Therefore these scenarios are not excluded despite having masses within the compressed analysis exclusion contour.
In the case of Figure~\ref{slha:comp1}, the \N{3}, \Ch{2} and \N{4} are light enough to be within reach of searches such as FullHad and 2L2J, but the production cross-section is lower due to
those being higgsino-like and they all have decays through $W$, $Z$ and $h$ bosons at similar branching fractions.
 
\begin{figure}[h]
\centering
\subfloat[][]{\includegraphics[width=0.5\textwidth]{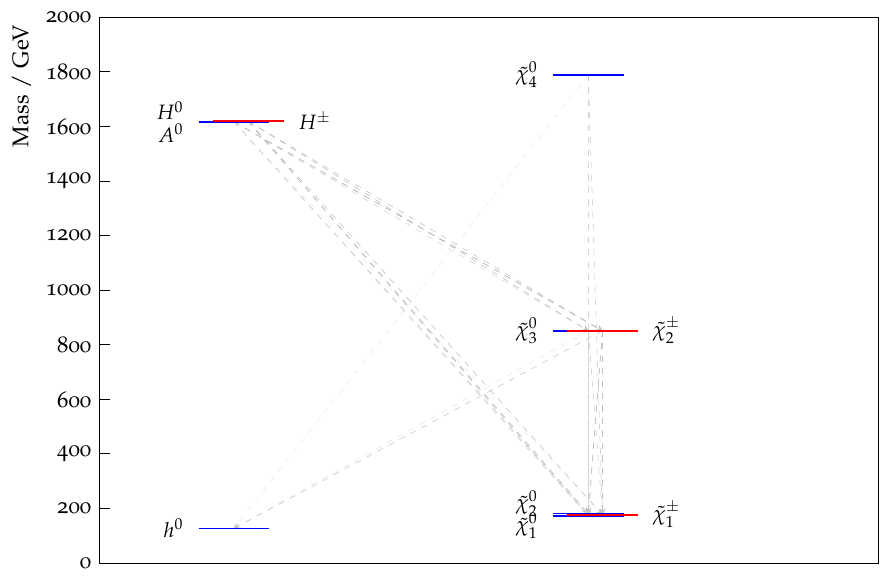}\label{slha:hino1}}
\subfloat[][]{\includegraphics[width=0.5\textwidth]{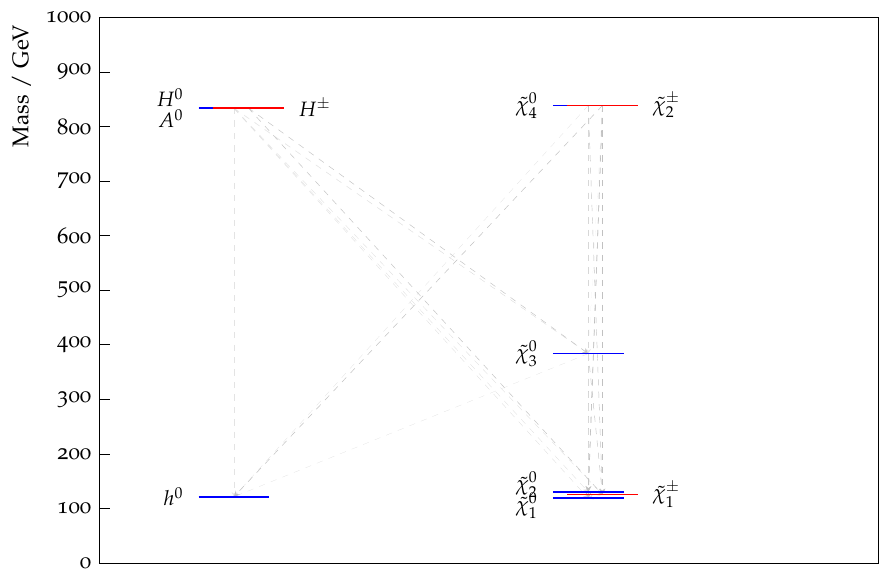}\label{slha:hino2}}
\caption{Mass spectrum for two benchmark models with a higgsino-like LSP that satisfy all constraints and are not excluded despite having a mass-spectrum within published ATLAS simplified model contours.}
\label{hino_benchmarks}
\end{figure}

Figure~\ref{hino_benchmarks} shows the mass spectrum of two benchmark models, each with a higgsino-like LSP that satisfy all constraints and are not excluded by ATLAS Run~2 analyses.
The \N{1} mass and \N{2} mass of these models is within the published simplified-model contours for the ATLAS 3L (off-shell) or compressed searches.
As with the compressed bino LSP models discussed previously, the $\N{2}\rightarrow\N{1}\ell^+\ell^-$ decay mode has a smaller branching fraction for these models than the compressed analysis simplified models, and the radiative decay is the dominant decay mode.
Otherwise, the bino, wino and higgsino mass parameters are well separated in these models, making the resulting electroweakinos almost pure bino, wino or higgsinos states as would be expected in simplified models.
The \N{3} and \Ch{2} each have masses under a \TeV\ and large mass-splittings with the other electroweakinos for both models. Therefore future searches could achieve sensitivity to the production of these heavier electroweakinos and their high \pT\ decay products.

\FloatBarrier
 
\section{Conclusion}
\label{sec:conclusion}

This paper presents constraints from eight ATLAS Run~2 searches for electroweak SUSY, the invisible Higgs boson width and searches for additional Higgs bosons, on the phenomenological minimal SUSY standard model, or pMSSM\@. The selected searches are considered most relevant for the model space being considered. Two scans are performed to evaluate the sensitivity of the ATLAS Run 2 searches. The scans use flat priors across the ranges of pMSSM parameters relevant to the electroweakino sector of the pMSSM, and external constraints from electroweak, flavour and
dark-matter related measurements are applied. The EWKino scan considers models where the lightest neutralino (assumed to be the dark matter candidate) can be either predominantly bino-, wino- or higgsino-like. As models with a bino-like LSP typically over-estimate the dark matter relic density unless some additional annihilation mechanisms are present, a second BinoDM scan is presented that aims to oversample the region with a
bino-like LSP less than 500~\GeV.
 
The impact of the ATLAS constraints is evaluated by considering the fraction of models that satisfy external constraints that are excluded by the selected ATLAS results. The results are presented as a function of electroweakino masses, and additional observables related to dark matter phenomenology are also presented that highlight the complementarity between collider and non-collider dark matter searches and measurements. In general, the bounds on electroweakino masses from the ATLAS searches are weaker in the pMSSM when assumptions entering the simplified models are relaxed. For some models where heavier electroweakinos are also experimentally accessible and can be constrained by the ATLAS searches, constraints beyond the simplified model contours are obtained. Highlights of the ATLAS Run~2 constraints include almost complete exclusion in the $Z$/$h$ `funnel regions' where a light bino-like LSP can avoid oversaturating the relic dark matter density through annihilating through a $Z$ or Higgs boson, and increased sensitivity relative to Run~1 for models with more compressed mass splittings ($\Delta m(\N{1},\N{2})$ or $\Delta m(\N{1},\Ch{1})$). Example spectra for surviving SUSY models with light charginos and neutralinos are also presented that can be used to optimise future searches.


\section*{Acknowledgements}


We thank CERN for the very successful operation of the LHC and its injectors, as well as the support staff at
CERN and at our institutions worldwide without whom ATLAS could not be operated efficiently.
 
The crucial computing support from all WLCG partners is acknowledged gratefully, in particular from CERN, the ATLAS Tier-1 facilities at TRIUMF/SFU (Canada), NDGF (Denmark, Norway, Sweden), CC-IN2P3 (France), KIT/GridKA (Germany), INFN-CNAF (Italy), NL-T1 (Netherlands), PIC (Spain), RAL (UK) and BNL (USA), the Tier-2 facilities worldwide and large non-WLCG resource providers. Major contributors of computing resources are listed in Ref.~\cite{ATL-SOFT-PUB-2023-001}.
 
We gratefully acknowledge the support of ANPCyT, Argentina; YerPhI, Armenia; ARC, Australia; BMWFW and FWF, Austria; ANAS, Azerbaijan; CNPq and FAPESP, Brazil; NSERC, NRC and CFI, Canada; CERN; ANID, Chile; CAS, MOST and NSFC, China; Minciencias, Colombia; MEYS CR, Czech Republic; DNRF and DNSRC, Denmark; IN2P3-CNRS and CEA-DRF/IRFU, France; SRNSFG, Georgia; BMBF, HGF and MPG, Germany; GSRI, Greece; RGC and Hong Kong SAR, China; ISF and Benoziyo Center, Israel; INFN, Italy; MEXT and JSPS, Japan; CNRST, Morocco; NWO, Netherlands; RCN, Norway; MEiN, Poland; FCT, Portugal; MNE/IFA, Romania; MESTD, Serbia; MSSR, Slovakia; ARRS and MIZ\v{S}, Slovenia; DSI/NRF, South Africa; MICINN, Spain; SRC and Wallenberg Foundation, Sweden; SERI, SNSF and Cantons of Bern and Geneva, Switzerland; MOST, Taipei; TENMAK, T\"urkiye; STFC, United Kingdom; DOE and NSF, United States of America.
 
Individual groups and members have received support from BCKDF, CANARIE, CRC and DRAC, Canada; CERN-CZ, PRIMUS 21/SCI/017 and UNCE SCI/013, Czech Republic; COST, ERC, ERDF, Horizon 2020, ICSC-NextGenerationEU and Marie Sk{\l}odowska-Curie Actions, European Union; Investissements d'Avenir Labex, Investissements d'Avenir Idex and ANR, France; DFG and AvH Foundation, Germany; Herakleitos, Thales and Aristeia programmes co-financed by EU-ESF and the Greek NSRF, Greece; BSF-NSF and MINERVA, Israel; Norwegian Financial Mechanism 2014-2021, Norway; NCN and NAWA, Poland; La Caixa Banking Foundation, CERCA Programme Generalitat de Catalunya and PROMETEO and GenT Programmes Generalitat Valenciana, Spain; G\"{o}ran Gustafssons Stiftelse, Sweden; The Royal Society and Leverhulme Trust, United Kingdom.
 
In addition, individual members wish to acknowledge support from Chile: Agencia Nacional de Investigaci\'on y Desarrollo (FONDECYT 1190886, FONDECYT 1210400, FONDECYT 1230987); China: National Natural Science Foundation of China (NSFC - 12175119, NSFC 12275265); European Union: European Research Council (ERC - 948254), Horizon 2020 Framework Programme (MUCCA - CHIST-ERA-19-XAI-00), Italian Center for High Performance Computing, Big Data and Quantum Computing (ICSC, NextGenerationEU), Marie Sklodowska-Curie Actions (EU H2020 MSC IF GRANT NO 101033496); France: Agence Nationale de la Recherche (ANR-20-CE31-0013, ANR-21-CE31-0013, ANR-21-CE31-0022), Investissements d'Avenir Idex (ANR-11-LABX-0012), Investissements d'Avenir Labex (ANR-11-LABX-0012); Germany: Baden-Württemberg Stiftung (BW Stiftung-Postdoc Eliteprogramme), Deutsche Forschungsgemeinschaft (DFG - CR 312/5-1); Italy: Istituto Nazionale di Fisica Nucleare (FELLINI G.A. n. 754496, ICSC, NextGenerationEU); Japan: Japan Society for the Promotion of Science (JSPS KAKENHI 22H01227, JSPS KAKENHI JP21H05085, JSPS KAKENHI JP22H04944); Netherlands: Netherlands Organisation for Scientific Research (NWO Veni 2020 - VI.Veni.202.179); Norway: Research Council of Norway (RCN-314472); Poland: Polish National Agency for Academic Exchange (PPN/PPO/2020/1/00002/U/00001), Polish National Science Centre (NCN 2021/42/E/ST2/00350, NCN OPUS nr 2022/47/B/ST2/03059, NCN UMO-2019/34/E/ST2/00393, UMO-2020/37/B/ST2/01043, UMO-2021/40/C/ST2/00187); Slovenia: Slovenian Research Agency (ARIS grant J1-3010); Spain: BBVA Foundation (LEO22-1-603), Generalitat Valenciana (Artemisa, FEDER, IDIFEDER/2018/048), La Caixa Banking Foundation (LCF/BQ/PI20/11760025), Ministry of Science and Innovation (RYC2019-028510-I, RYC2020-030254-I), PROMETEO and GenT Programmes Generalitat Valenciana (CIDEGENT/2019/023, CIDEGENT/2019/027); Sweden: Swedish Research Council (VR 2022-03845, VR 2022-04683), Knut and Alice Wallenberg Foundation (KAW 2017.0100, KAW 2018.0157, KAW 2019.0447); Switzerland: Swiss National Science Foundation (SNSF - PCEFP2\_194658); United Kingdom: Leverhulme Trust (Leverhulme Trust RPG-2020-004); United States of America: Neubauer Family Foundation.


\clearpage

\printbibliography
 
\clearpage
 
\begin{flushleft}
\hypersetup{urlcolor=black}
{\Large The ATLAS Collaboration}

\bigskip

\AtlasOrcid[0000-0002-6665-4934]{G.~Aad}$^\textrm{\scriptsize 102}$,
\AtlasOrcid[0000-0002-5888-2734]{B.~Abbott}$^\textrm{\scriptsize 120}$,
\AtlasOrcid[0000-0002-1002-1652]{K.~Abeling}$^\textrm{\scriptsize 55}$,
\AtlasOrcid[0000-0001-5763-2760]{N.J.~Abicht}$^\textrm{\scriptsize 49}$,
\AtlasOrcid[0000-0002-8496-9294]{S.H.~Abidi}$^\textrm{\scriptsize 29}$,
\AtlasOrcid[0000-0002-9987-2292]{A.~Aboulhorma}$^\textrm{\scriptsize 35e}$,
\AtlasOrcid[0000-0001-5329-6640]{H.~Abramowicz}$^\textrm{\scriptsize 151}$,
\AtlasOrcid[0000-0002-1599-2896]{H.~Abreu}$^\textrm{\scriptsize 150}$,
\AtlasOrcid[0000-0003-0403-3697]{Y.~Abulaiti}$^\textrm{\scriptsize 117}$,
\AtlasOrcid[0000-0002-8588-9157]{B.S.~Acharya}$^\textrm{\scriptsize 69a,69b,m}$
% ...
\bigskip
\\

% ...
$^{29}$Physics Department, Brookhaven National Laboratory, Upton NY; United States of America.\\
% ...
$^{35}$$^{(a)}$Facult\'e des Sciences Ain Chock, R\'eseau Universitaire de Physique des Hautes Energies - Universit\'e Hassan II, Casablanca;$^{(b)}$Facult\'{e} des Sciences, Universit\'{e} Ibn-Tofail, K\'{e}nitra;$^{(c)}$Facult\'e des Sciences Semlalia, Universit\'e Cadi Ayyad, LPHEA-Marrakech;$^{(d)}$LPMR, Facult\'e des Sciences, Universit\'e Mohamed Premier, Oujda;$^{(e)}$Facult\'e des sciences, Universit\'e Mohammed V, Rabat;$^{(f)}$Institute of Applied Physics, Mohammed VI Polytechnic University, Ben Guerir; Morocco.\\
% ...
$^{49}$Fakult\"{a}t Physik , Technische Universit{\"a}t Dortmund, Dortmund; Germany.\\
% ...
$^{55}$II. Physikalisches Institut, Georg-August-Universit\"{a}t G\"ottingen, G\"ottingen; Germany.\\
% ...
$^{69}$$^{(a)}$INFN Gruppo Collegato di Udine, Sezione di Trieste, Udine;$^{(b)}$ICTP, Trieste;$^{(c)}$Dipartimento Politecnico di Ingegneria e Architettura, Universit\`a di Udine, Udine; Italy.\\
% ...
$^{102}$CPPM, Aix-Marseille Universit\'e, CNRS/IN2P3, Marseille; France.\\
% ...
$^{117}$Department of Physics, New York University, New York NY; United States of America.\\
% ...
$^{120}$Homer L. Dodge Department of Physics and Astronomy, University of Oklahoma, Norman OK; United States of America.\\
% ...
$^{150}$Department of Physics, Technion, Israel Institute of Technology, Haifa; Israel.\\
$^{151}$Raymond and Beverly Sackler School of Physics and Astronomy, Tel Aviv University, Tel Aviv; Israel.\\
% ...
$^{a}$ Also Affiliated with an institute covered by a cooperation agreement with CERN.\\
$^{b}$ Also at An-Najah National University, Nablus; Palestine.\\
$^{c}$ Also at Borough of Manhattan Community College, City University of New York, New York NY; United States of America.\\
$^{d}$ Also at Center for High Energy Physics, Peking University; China.\\
$^{e}$ Also at Center for Interdisciplinary Research and Innovation (CIRI-AUTH), Thessaloniki; Greece.\\
$^{f}$ Also at Centro Studi e Ricerche Enrico Fermi; Italy.\\
% ...
$^{m}$ Also at Department of Physics, King's College London, London; United Kingdom.\\
% ...

\end{flushleft}


\end{document}